\documentclass[onecolumn,%
sort&compress,numbers%
]{els-mrw} 

\usepackage{amsmath,amssymb,amsfonts,amsthm,graphicx,slashed,dsfont}
\usepackage{txfonts}
\usepackage{helvet}
\usepackage{xspace,bm}

\usepackage[colorlinks=true,pdfstartview=FitV,linkcolor=blue,citecolor=magenta,urlcolor=blue,bookmarks=true,bookmarksnumbered=true]{hyperref}

\newcommand{\bcite}[1]{\cite{#1}}

\newcommand{\ev}[1]{\langle #1 \rangle}

\newcommand{\chipt}{$\chi$PT\xspace}
\newcommand{\SU}{\text{SU}}
\newcommand{\U}{\text{U}}

\newcommand{\eb}{eB}
\newcommand{\B}{\mathcal{B}}
\newcommand{\I}{\mathcal{I}}
\newcommand{\Q}{\mathcal{Q}}
\renewcommand{\S}{\mathcal{S}}
\newcommand{\Utauthree}{U_Q(1)}
\newcommand{\Z}{\mathcal{Z}}
\newcommand{\D}{\mathcal{D}}
\newcommand{\A}{\mathcal{A}}
\newcommand{\Ds}{\slashed{D}}
\newcommand{\M}{\mathcal{M}}

\newcommand{\muB}{\mu_\B}
\newcommand{\muI}{\mu_\I}
\newcommand{\muQ}{\mu_\Q}
\newcommand{\muS}{\mu_\S}

\begin{document}

\chapter{Thermodynamics of magnetized matter in hot and dense QCD}\label{chap1}

\author[1]{Bastian B.\ Brandt}%

\address[1]{
\orgname{Faculty of Physics, Bielefeld University},
\orgaddress{Universit\"{a}tsstra{\ss}e 25, D-33615 Bielefeld, Germany}
}

\author[2]{Gergely Endr\H{o}di}%

\address[2]{
\orgname{Institute of Physics and Astronomy,
ELTE E\"otv\"os Lor\'and University},
\orgaddress{\\ \;\;\,P\'azm\'any P.\ s\'et\'any 1/A, H-1117 Budapest, Hungary}
}

\maketitle

\begin{abstract}[Abstract]
This chapter, to appear in the section on QCD under extreme conditions within the Encyclopedia of Nuclear Physics, aims to provide a pedagogical introduction to the physics of quarks and gluons in the presence of high temperature, nonzero (isospin) density and strong background electromagnetic fields. Extreme conditions of these types are relevant for the description of high-energy heavy-ion collisions, neutron stars and their mergers, as well as the evolution of the early Universe in its first microsecond. Most of the existing results on this topic have been obtained by means of first-principles simulations of the discretized theory of the strong interactions, lattice Quantum Chromodynamics (QCD). This lays the focus of this review chapter, although various calculations within effective theories of QCD -- most notably chiral perturbation theory -- are also discussed. Furthermore, we provide an outlook concerning open questions and yet uncharted parameter regions within this fascinating system.
\end{abstract}

\section{Introduction}
\label{sec-EB:intro}

Under the typical conditions present in today's Universe, the strong interaction confines quarks and gluons in baryons, such as protons and neutrons, and mesons, such as pions and kaons. Within the Standard Model of particle physics, the strong interaction is formulated in terms of a non-abelian $\SU(3)$ gauge theory, quantum chromodynamics (QCD), which is understood to encode the confinement of quarks and gluons through the gluonic self-interactions and the associated growing coupling strength at large distances. When strongly interacting matter is subjected to extreme temperatures, as in the early Universe shortly after the Big Bang, quarks and gluons effectively become free, or deconfined, forming a new state of matter, the so-called quark gluon plasma (QGP). Similarly, in the presence of very high baryon densities, as present in neutron stars in today's Universe, a similar change to a new phase of strongly interacting matter is expected, potentially even to a phase including color superconductivity (CSC) in which the degrees of freedom are Cooper pair-like diquarks. The region of high temperatures and intermediate baryon densities is probed by collisions of ultra-relativistic heavy ions (HICs) at facilities like CERN and Brookhaven National Laboratory, as well as the future FAIR facility, where the QGP is also expected to form. Apart from non-zero baryon densities, all of the above mentioned systems also feature a non-zero density of electromagnetic charge, which affects the properties of the system. While the baryon density is typically dominant, there might be physical systems where the charge density governs the properties of the system, such as in the early Universe in the presence of large lepton flavor asymmetries~\bcite{Oldengott:2017tzj}.

In addition to non-zero temperature and density, the aforementioned systems can also feature electromagnetic fields with magnitudes comparable to the QCD scale. These are expected to be present in the core of magnetars~\bcite{Duncan:1992hi}, magnetized neutron stars, in off-central HICs~\bcite{Skokov:2009qp}, and might have been created at early times during the evolution of the Universe~\bcite{Grasso:2000wj} prior to the QCD transition. Electromagnetic fields of this magnitude strongly affects the quarks and, through them, the thermodynamic properties of strongly interacting matter. For a consistent phenomenological treatment, the effects of these fields therefore has to be taken into account.

The investigation of strongly interacting matter under extreme conditions is one of the major topics of modern theoretical and experimental high-energy nuclear and particle physics. Not only does it help us understand the evolution of the early Universe and compact stars in our cosmos, it also allows to directly study the properties of quarks and gluons, which usually can only be studied indirectly through their properties in hadronic bound states. While extremely high temperatures and densities are accessible within perturbation theory, allowing for analytical computations, in the region interesting for the aforementioned physical systems QCD remains strongly coupled, so that non-perturbative methods are necessary for theoretical investigations of the thermodynamics. The two non-perturbative ab initio approaches for the study of QCD are functional methods and numerical simulations of lattice QCD. While the former have shown considerable progress in methodology in the past decade (see~\bcite{Fischer:2026uni} for a recent review) lattice QCD still remains the most widely used method to study the equilibrium thermodynamics of strongly interacting matter and allows for a systematic improvement of the results.

In the phenomenologically interesting region of large baryon density and intermediate temperatures, lattice QCD is plagued by the complex action problem, to which we get back to below, prohibiting direct simulations. The latter are only possible at vanishing density, both for zero or non-zero external magnetic fields, or for a particular combination of quark densities corresponding to the so-called isospin-asymmetric setting. The study of the non-perturbative properties of QCD in other regions of the parameter space is only possible using indirect methods, such as Taylor expansions~\bcite{Allton:2002zi}, for instance. In this review chapter we will focus on the regions where direct simulations are possible, allowing for results with full control over the main systematic effects. Other chapters of this encyclopedia are devoted to the more detailed discussion of indirect methods and the results obtained for the region of large non-zero baryon densities in particular.

Following a more in-depth discussion of the relevant parameter space and the QCD phase diagram, we will briefly introduce in Sec.~\ref{sec-EB:approaches} the theoretical approaches that are available in this context. Sec.~\ref{sec-EB:lattice} is devoted to a short introduction to lattice QCD and a discussion of the complex action problem, whereas in Sec.~\ref{sec-EB:chipt-form} we will consider chiral perturbation theory (\chipt). We then proceed to reviewing the properties of the phase diagram and the EoS up to large values of $\muI$ in Sec.~\ref{sec-EB:isospin}. Finally, Sec.~\ref{sec-EB:magnetic} is devoted to the discussion of results at nonzero electromagnetic fields, with a focus on the phase diagram, electromagnetic susceptibilities and the interesting features of dense and magnetized QCD matter. The review is concluded in Sec.~\ref{sec-EB:applications} with a summary and list of open questions, as well as possible directions for future research of these fascinating systems.

\subsection{Phase diagram and equation of state}
\label{sec-EB:intro_pd_eos}

In the context of the physical systems mentioned above, it is a good approximation to take into account the three lightest quark flavors only, i.e. up ($u$), down ($d$) and strange quarks ($s$).\footnote{We note, however, that the effect of the charm quark might still play a role for a consistent description of some of the physical systems mentioned above.} The relevant parameter space is therefore spanned by the densities $n_u$, $n_d$ and $n_s$ in addition to the temperature $T$. In HICs, the system is thought to be in equilibrium with respect to the strong interactions after initial thermalization, but not with respect to the weak force and electromagnetism, thus the density $n_f$ of each quark flavor is effectively conserved. In the QCD epoch of the early Universe, weak and electromagnetic equilibrium is reached, but neutrino oscillations are absent above temperatures of $T\approx10\textmd{ MeV}$. In this case, baryon number, electric charge and the individual lepton flavor numbers are conserved~\bcite{Wygas:2018otj}. Finally, in cold neutron stars, where neutrinos have already escaped, matter is in equilibrium with respect to all three fundamental forces and only baryon number and electric charge are conserved.

Keeping these in mind, one can consider the grand canonical ensemble to describe this statistical physical system, trading the conserved densities $n$ for the associated chemical potentials $\mu$. Motivated by the HIC scenario, below we consider all three light quark chemical potentials $\mu_f$.
Through a change of basis, we can equivalently describe dense three-flavor QCD matter via three quantum numbers: the baryon number $\B$, the third component $\I$ of isospin controlling the above mentioned isospin-asymmetry, as well as the strangeness $\S$,
\begin{equation}
\mu_u=\frac{\muB}{3}+\frac{\muI}{2}, \qquad
\mu_d=\frac{\muB}{3}-\frac{\muI}{2}, \qquad
\mu_s=\frac{\muB}{3}-\muS\,.
\label{eq:BIS_basis}
\end{equation}
Another alternative is the basis spanned by baryon number, strangeness and the electric charge $\Q=\I+(\B+\S)/2$.\footnote{\label{fn:convention}We emphasize that occasionally, a different convention has also been used in the literature, with the isospin quantum number $\bar\I=2\I$, implying $\mu_{\bar\I}=\muI/2$. Throughout this chapter, we stick to the convention~\eqref{eq:BIS_basis}.}

\begin{figure}[t]
\centering
\includegraphics[width=.48\textwidth]{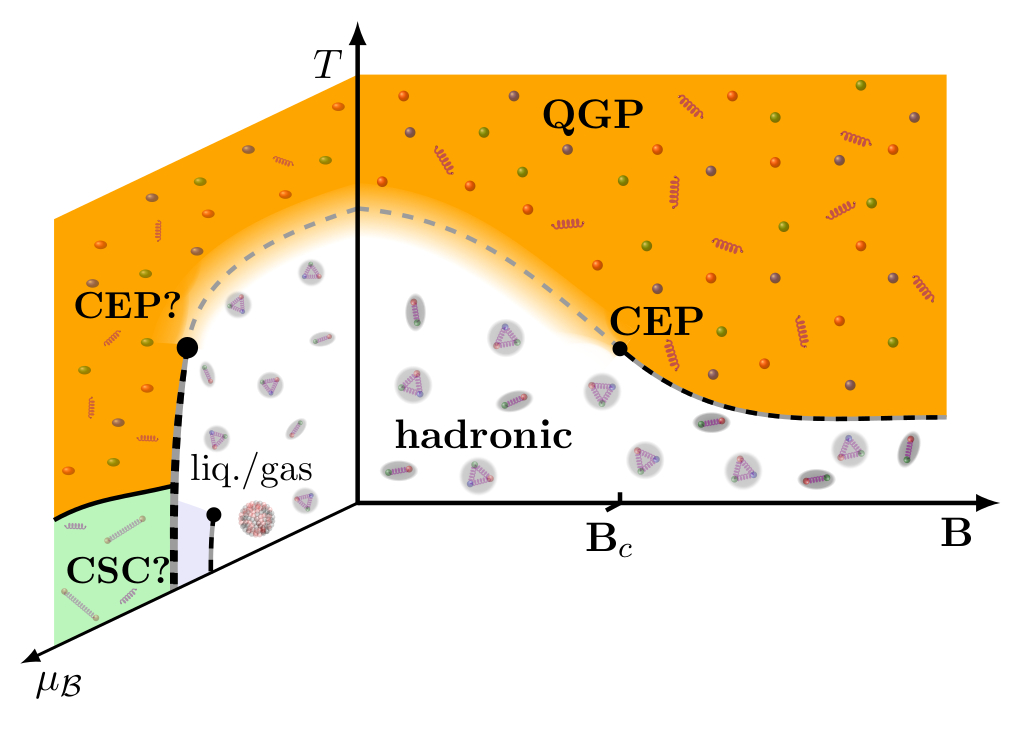}\qquad
\includegraphics[width=.48\textwidth]{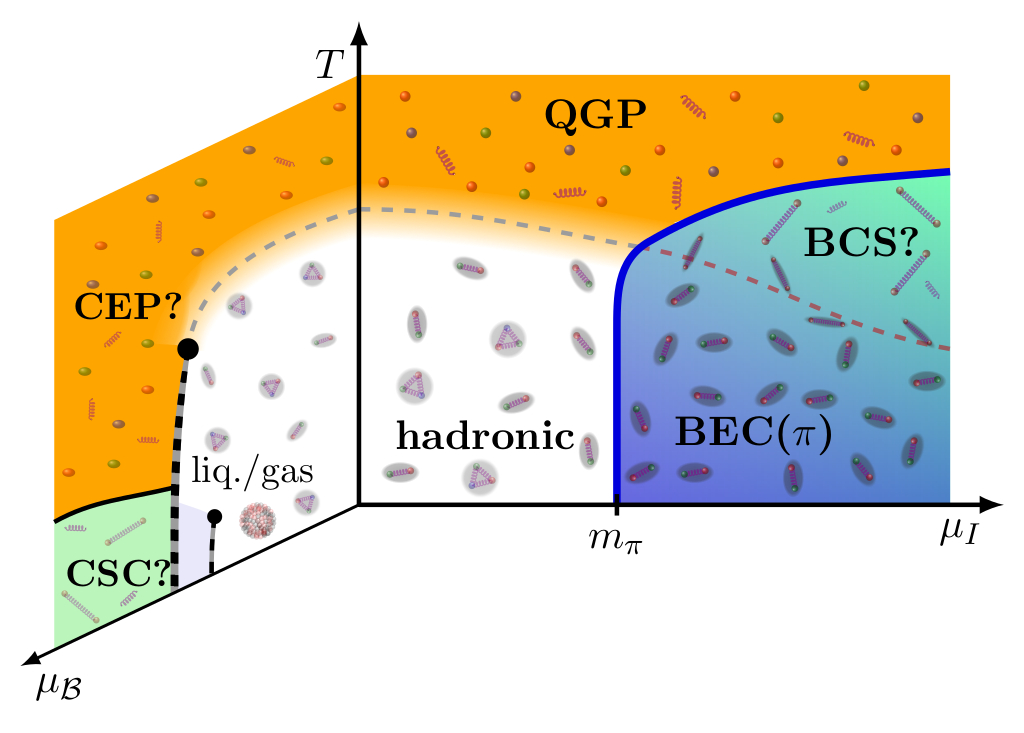}
\caption{Three-dimensional slices of the QCD phase diagram in the $T-\muB-B$ (left panel) and the $T-\muB-\muI$ space (right panel). The hadronic, QGP, pion condensed (BEC), superconducting (BCS) and color superconducting (CSC) phases (indicated by the different shaded regions) are separated by phase transitions: crossovers (gray and red dashed line), second-order phase transitions (solid blue) and first-order phase transitions (gray-black dashed), ending at critical endpoints (CEP). Question marks indicate features that have been conjectured but not confirmed by first-principles calculations.}
\label{fig-EB:extended phasediags}
\end{figure}

Thus, our aim is to study the behavior of strongly interacting matter as a function of the parameters $T$, $\muB$, $\muI$, $\muS$ as well as the electromagnetic fields $B$ and $E$.
The thermodynamics in this parameter space may be summarized compactly in the QCD phase diagram. Sketches of the phase diagram in various three-dimensional spaces are shown in Fig.~\ref{fig-EB:extended phasediags}. These diagrams show the different characteristic phases of strongly interacting matter, differing from each other in the realized pattern of chiral symmetry breaking. 
At low energies, the system is in the confined phase, where the effective degrees of freedom are hadrons. In addition, the confined phase features explicit and spontaneous breaking of the chiral symmetry of the massless QCD action by the expectation value $\ev{\bar\psi\psi}$ of the chiral condensate. This symmetry breaking has a characteristic influence on various properties of hadrons. In turn, at high energies, chiral symmetry is effectively restored and quarks and gluons become effectively deconfined in the QGP phase. Additionally, at high isospin chemical potential $\muI$, a superconducting phase with Bose-Einstein condensation (BEC) of charged pions emerges. In this phase
a subgroup of chiral symmetry is broken spontaneously by the pion condensate $\ev{\pi}$. At large $\muI$, a BCS-type phase with pseudoscalar Cooper pairs composed of up and anti-down quark pairs is expected to appear.

These phases are separated from each other by phase transition lines. The transition between confined and deconfined matter at zero density and zero electromagnetic field is an analytic crossover~\bcite{Aoki:2006we,Bhattacharya:2014ara} and takes place at a temperature of about $T_c\approx150\textmd{ MeV}$~\bcite{Aarts:2023vsf}. At magnetic fields above $B_c\approx(4-9)\textmd{ GeV}^2$, this transition is known to turn into a first-order phase transition through a second-order $\mathrm{Z}(2)$ critical endpoint~\bcite{Endrodi:2015oba,DElia:2021yvk}. In turn, the transition to the BEC phase at nonzero $\muI$ is known to occur at zero temperature at $\muI = m_\pi \approx 135 \textmd{ MeV}$ via a second-order phase transition in the $\mathrm{O}(2)$ universality class~\bcite{Brandt:2017oyy}.

The above features of the phase diagram have been obtained by first-principles lattice QCD simulations. As mentioned above, away from the $\muB=\muS=0$ axis, direct lattice simulations are not possible due to the complex action problem and one is restricted to indirect methods with a limited range of validity.
In the bulk of the phase diagram, one must therefore rely on alternative approaches. Low-energy models of QCD suggest that the crossover at high temperature may turn into a first-order phase transition through a $\mathrm{Z}(2)$ critical endpoint at large $\muB$. Within functional approaches to QCD, indications are also observed for a critical endpoint at around $\muB\approx 600\textmd{ MeV}$~\bcite{Fischer:2026uni}. In a different corner of the phase diagram, nuclear effective theory predicts a nuclear liquid-gas transition to take place at low temperatures and large $\muB$, for a recent review see~\bcite{Kaiser:2026msy}. A color-superconducting phase is expected to emerge at even larger values of $\muB$, see~\bcite{Alford:2007xm} for a comprehensive review. Further exotic scenarios including inhomogeneous phases, for a review see~\bcite{Buballa:2014tba}, moat regimes~\bcite{Fu:2019hdw} or quarkyonic phases~\bcite{McLerran:2007qj} have been predicted at nonzero $T$ and $\muB$. 
Currently, lattice simulations based on Taylor expansions around $\muB=0$ or analytic continuations from imaginary $\muB$ have not found indications for such phases, nor clear evidence for a critical endpoint~\bcite{Aarts:2023vsf}.

Another feature, relevant for all of the physical systems mentioned in the introduction, is the equation of state (EoS), i.e.\ the relationship between the pressure, the energy density, the entropy density and further thermodynamic state functions. First, the EoS is needed to solve the Tolman-Oppenheimer-Volkoff equations, which determine the mass and radius of gravitationally stable neutron stars~\cite{Lattimer:2000nx}. Second, the EoS enters the Friedmann equation, which governs the isentropic expansion of the early Universe~\cite{Boyanovsky:2006bf} and dictates the power spectrum of primordial gravitational waves. Finally, the EoS is required within the hydrodynamic evolution of the quark-gluon plasma and the subsequent hadronic freeze-out~\cite{Teaney:2001av,Kolb:2003dz}.
The equation of state at nonzero $\muI$
has been determined via lattice simulations both at low temperature~\bcite{Brandt:2018bwq,Abbott:2023coj,Abbott:2024vhj} and over a broad region of the $T-\muI$ plane~\bcite{Vovchenko:2020crk,Brandt:2022hwy}. Similarly, the effect of the magnetic field on the pressure and other thermodynamic observables is also well understood~\bcite{Endrodi:2024cqn}. In turn, the determination of the impact of $\muB$, $\muS$ and $E$ on the EoS is difficult due to the complex action problem. Our knowledge from lattice simulations is therefore limited to the weak-field and weak-chemical potential regions.

\section{Theoretical approaches}
\label{sec-EB:approaches}

In this review we will focus on results obtained with lattice simulations and \chipt. 
We mention however that further methods to investigate QCD thermodynamics include low-energy QCD models like the hadron-resonance gas, the quark-meson or the Nambu-Jona-Lasinio models, as well as the aforementioned functional approaches, such as the functional renormalization group and Dyson-Schwinger equations.
The latter methods have been used extensively for studies of the QCD phase diagram at nonzero $\muB$,
but less in the case of $\muI\neq0$ (see, however, the calculation within the quark-meson model~\bcite{Kamikado:2012bt}). Furthermore, at large chemical potentials or high temperatures, perturbation theory is applicable. In particular, the latter approach can provide important insight into the nature of the expected BCS-type color-superconducting phase at large $\muB$ as well as the analogous BCS superconducting phase at large $\muI$ (e.g.~\bcite{Fujimoto:2023mvc} and references therein).

\subsection{Lattice QCD simulations}
\label{sec-EB:lattice}

Lattice QCD is a systematic approach that enables us to understand the non-perturbative thermodynamic properties of QCD solely from the first principles. The quark and gluon fields are discretized on a Euclidean space-time grid with geometry $N_s^3\times N_t$ and lattice spacing $a$. The spatial volume and the temperature of the system read $V=(N_s a)^3$ and $T=1/(N_ta)$. To ensure the correct particle statistics, the gluon (quark) fields satisfy periodic (antiperiodic) boundary conditions in the imaginary time direction and it is convenient to employ periodic boundary conditions in space. To recover the continuum theory, one needs to carry out the $a\to0$ extrapolation, the so-called continuum limit, which at fixed temperature amounts to $N_t\to\infty$.
A detailed discussion of lattice QCD and the associated numerical simulations is beyond the scope of this review (see for instance the book~\bcite{Gattringer:2010zz}). Here we merely intend to point out a few features of lattice simulations that will be important later on.

The central object that encodes the impact of the parameters mentioned in Sec.~\ref{sec-EB:intro_pd_eos} on strongly interacting matter is the partition function of the system. Represented by the Euclidean path integral over the fields, it reads 
\begin{equation}
\Z=
\int \D\bar\psi\,\D\psi\,\D \A_\nu \, e^{-S[\A,\bar\psi,\psi]} =
\int \D U_\nu \, e^{-S_G[U]}\, \prod_f \det \left[ \Ds_f+m_f \right]\,,
\label{eq:partfunc}
\end{equation}
where $S$ is the total and $S_G$ the gluon action and in the second step the path integral over quark and antiquark fields has been carried out, resulting in the Dirac determinant and a variable substitution from the gluon field $\A_\nu$ to the gluon links $U_\nu=\exp(ia \A_\nu)$ was made.
The determinant contains the quark masses $m_f$, as well as the Dirac operator $\Ds_f$ for the flavor $f$, through which the chemical potentials and background electromagnetic fields $F_{\nu\rho}$ enter. The latter are generated by an electromagnetic vector potential $A_\nu$ and couple to the quark electric charges $q_f$. 
We note that the background nature of the field reflects the fact that $B$ and $E$ are generated by electromagnetic currents outside of the system -- 
we are therefore only interested in the impact of the field on the medium and
the back-reaction on the electromagnetic field is excluded.
When expressing the dependence of physical observables on the field, it is natural to combine the field strength with the elementary charge $e$ in the renormalized combination $eF_{\nu\rho}$, so multiplicative renormalization constants are not required, see e.g.~\bcite{Endrodi:2024cqn}.

Below we will specialize to pure magnetic $B=F_{12}$ and pure electric fields $E=F_{34}$, assumed to point in the positive $x_3$ direction without loss of generality. In the infinite volume, homogeneous fields are respectively realized by the Euclidean vector potentials of the form $A_\nu(x)=Bx_1\delta_{\nu2}$ and $A_\nu(x)=iEx_3\delta_{\nu4}$, for example.\footnote{Notice that after the Wick rotation to Euclidean space-time, $A_4$ contains the imaginary electric field.} In turn, for the finite periodic volumes used in the simulations, the vector potential is slightly more complicated and the flux of the magnetic and (imaginary) electric fields become quantized~\bcite{tHooft:1979rtg},
\begin{equation}
a^2N_s^2 \,q_dB = 2\pi N_b, \qquad
a^2N_sN_t \, iq_dE = 2\pi N_e, \qquad N_b, N_e\in \mathds{Z}\,,
\label{eq:quantization}
\end{equation}
in units of the smallest electric charge (that of the down quark, $q_d=-e/3$). We will
get back to this point in Sec.~\ref{sec-EB:chixi} when we discuss electromagnetic susceptibilities.

The central idea of lattice QCD is to sample gluon configurations from~\eqref{eq:partfunc} using the integrand as probability weight. A prerequisite for such importance sampling-based simulation algorithms is that this integrand is real and positive.
The reality of the determinant can be shown by finding a unitary matrix $\Gamma$ that satisfies the Hermiticity relation $\Gamma \Ds_f \Gamma^\dagger=\Ds_f^\dagger$, since in that case
\begin{equation}
\det \left[ \Ds_f +m_f\right]
=\det \left[ \Gamma^\dagger\Gamma(\Ds_f +m_f)\right]
=\det \left[ \Gamma(\Ds_f+m_f)\Gamma^\dagger\right]
=\det \left[ \Ds_f^\dagger+m_f\right]
=\det \left[ \Ds_f+m_f\right]^*\,,
\end{equation}
where we used the cyclicity of the trace. 
Positivity typically follows when one considers two degenerate quark flavors.
At nonzero temperature, magnetic fields as well as imaginary-valued chemical potentials and imaginary-valued electric fields, one finds $\Gamma=\gamma_5$, ensuring that standard, importance sampling-based Monte-Carlo algorithms work.
In contrast, no such matrix can be found whenever $\muB$, $\muS$ or $E$ are nonzero and real. This is the so-called complex action or sign problem, which hinders direct simulations. Such cases require indirect methods, in particular Taylor expansions~\bcite{Allton:2002zi} around $\muB=\muS=E=0$ or analytic continuations~\bcite{deForcrand:2003vyj} from imaginary values of these parameters. 

For the case of the isospin chemical potential $\muI$, it is instructive to combine the up and down quark Dirac operators into the light quark matrix $\M_\ell=\textmd{diag}(\Ds_u+m_u,\Ds_d+m_d)$. For degenerate masses, $m_u=m_d\equiv m_\ell$, the matrix we are seeking takes the form $\Gamma=\gamma_5\tau_1$~\bcite{Alford:1998sd}, acting not just in Dirac but also in flavor space, $\Gamma \M_\ell \Gamma^\dagger=\M_\ell^\dagger$. 
Thus, standard algorithms are again applicable, however, the lattice simulations in this setup involve further peculiarities.
In particular, within the pion condensed phase mentioned above, $\M_\ell$ develops near-zero modes (more precisely, near-zero singular values) that lead to numerical issues in the treatment of the determinant. These small singular values stem from the Goldstone mode associated with the spontaneous breaking of isospin symmetry by the pion condensate. To protect the simulations from this infrared singular behavior, a pion source parameter $\lambda$ is included~\bcite{Kogut:2002zg}, which shifts the singular values away from zero and renders the problem infrared-safe,
\begin{equation}
\M_\ell=\begin{pmatrix}
\Ds_u+m_\ell & \lambda\gamma_5 \\
-\lambda\gamma_5 & \Ds_d+m_\ell \\
\end{pmatrix}, \qquad
\det \M_\ell = \det \left[ (\Ds_u+m_\ell)^\dagger (\Ds_u+m_\ell) + \lambda^2\right]\,.
\label{eq:lattice-pionsource}
\end{equation}
In fact, the inclusion of $\lambda\neq0$ is necessary to observe spontaneous symmetry breaking by the pion condensate in finite volumes, to which we get back to in Sec.~\ref{sec-EB:pionisospin}. This is tantamount to how a nonzero quark mass is necessary for ordinary chiral symmetry breaking to occur in the vacuum of QCD in finite volumes. The pion source parameter needs to be extrapolated to zero at the end of the analysis.
We note that when $\muI\neq0$ and electromagnetic fields coupled to the quark electric charges are included as well, no $\Gamma$ matrix exists and one yet again faces a complex action problem.

\subsection{Chiral perturbation theory}
\label{sec-EB:chipt-form}

In the low-energy regime, the dynamics encoded in~\eqref{eq:partfunc} is governed by the lightest hadronic excitations i.e.\ pions. The latter arise as the Goldstone bosons of the spontaneous breaking of chiral symmetry by the chiral condensate, rendered pseudo-Goldstone bosons at nonzero quark mass. The effective theory that describes pions consistently is chiral perturbation theory~\bcite{Gasser:1983yg} (for an introductory lecture see~\bcite{Scherer:2002tk}). The behavior of QCD at low values of our parameters can be found analytically in this framework, serving as a useful benchmark for lattice QCD results at low energies. In particular, we will discuss the \chipt  predictions for the behavior of QCD matter at low $T$, low $\muI$ and low $B$ below.

The thermodynamic properties are to a large extent governed by the global symmetries of the theory. In QCD, the most important one is the approximate symmetry between the light quark flavors, i.e.\ chiral symmetry.  For the purpose of this chapter it is sufficient to consider the symmetry for the two lightest ($u$ and $d$) quark flavors joined together in the doublet $\psi=(u,d)^\top$. The symmetry group is then given by
\begin{equation}
    \label{eq-EB:chiral-symm}
    \SU_L(2) \times \SU_R(2) \times \U_V(1) \times \U_A(1) \,,
\end{equation}
where the subscripts $L$ and $R$ refer to generators acting on left- and right-handed spinors, $V$ indicates a vector and $A$ an axial transformation. At low energies, the $\SU_L(2) \times \SU_R(2)$ symmetry is broken spontaneously to $\SU_V(2)$ by the light-quark chiral condensate $\ev{\bar\psi\psi}=\ev{\bar{u}u}+\ev{\bar{d}d}$.
This gives rise to three Goldstone bosons via Goldstone's theorem, the pions $\pi^{\pm,0}$. The $\U_A(1)$ symmetry is anomalously broken at the quantum level, responsible for rendering the $\eta'$ meson more massive than the pions~\bcite{Witten:1979vv,Veneziano:1979ec}, and also playing a key role in anomalous transport phenomena to which we get back to in Sec.~\ref{sec-EB:magnetic}. The remaining $\SU_V(2)$ symmetry is dubbed isospin symmetry and will be important for the phenomena discussed in the following. The global $\U_V(1)$ is associated with baryon number conservation and remains conserved in the complete parameter space.

For nonzero (degenerate) light quark masses $m_\ell$, only the $\SU_V(2)$ subgroup of chiral symmetry remains intact. Furthermore, a difference between the $u$ and $d$ quark masses breaks the $\SU_V(2)$ isospin symmetry explicitly to a residual $\Utauthree$ symmetry, with the third Pauli matrix $\tau_3$ as generator.
The symmetry breaking due to non-degenerate light quarks is typically expected to be a subleading effect -- it does not significantly affect the QCD transition temperature~\bcite{Gavai:2002fi}, for instance -- and thus typically neglected when studying the thermodynamics and phase diagram of QCD.
The same explicit breaking, however, occurs at non-zero isospin chemical potentials, appearing in the chemical potential matrix $\bm \mu$ or electromagnetic fields coupled to the electric charge $\Q$, written as $2\times2$ matrices in flavor space,
\begin{equation}
  \label{eq-EB:isobreak-mass}
   \bm\mu = \frac{\muB}{3} \mathds{1} + \frac{\muI}{2} \tau_3, \qquad
\Q=\frac{\mathds{1}}{6}+\frac{\tau_3}{2}\,.
\end{equation}
In this generalized form, the symmetry breaking fields couple to the quarks via
\begin{equation}
    \label{eq-EB:coupling terms}
    \mathcal{L}_{\rm SB} = \bar{\psi} \gamma^\nu \mathcal{V}_\nu \psi \,, \qquad \text{with} \quad \mathcal{V}_0 = \bm\mu - \Q\,A_0(x) \quad \text{and} \quad \mathcal{V}_i = - \Q\, A_i(x) \,,
\end{equation}
where $A_\nu(x)$ is the electromagnetic vector potential generating the background fields. 

In chiral perturbation theory, the pseudoscalar meson (pion) fields are collected in the Lie algebra-valued field $\phi=\sum_{i=1}^{3} \phi_i \tau_i/2$ and the effective theory framework is formulated using the exponentiated meson fields $\Sigma=\exp(i\phi)$, which transform as $\Sigma \to L\Sigma R^\dagger$ under $\SU_L(2) \times \SU_R(2)$ transformations with $L\in\SU_L(2)$ and $R\in\SU_R(2)$. 
The physical (charged and neutral) pion fields are written as the linear combinations $\pi^\pm = (\phi_1 \pm i \phi_2)/\!\sqrt{2}$ and $\pi^0 = \phi_3$.
The Lagrangian density of \chipt contains all terms allowed by symmetry up to a given order in the expansion of the Goldstone boson momenta, assuming $p\ll \Lambda_\chi$. Here $\Lambda_\chi\sim 1$~GeV is the scale associated with the breakdown of \chipt. To leading order, this Lagrangian density is given by
\begin{equation}
    \label{eq-EB:lag-chipt}
    \mathcal{L}_{\text{\chipt}} = \frac{f^2}{4} \text{Tr}\,\left( D^\mu \Sigma^\dagger D_\mu \Sigma \right) + \frac{f^2 B}{2} \text{Tr}\,\left( \Sigma^\dagger M + M^\dagger \Sigma \right) \,,
\end{equation}
where all symmetry breaking fields from~\eqref{eq-EB:coupling terms} enter through the covariant derivative, defined by
\begin{equation}
    \label{eq-EB:chipt-codev}
    D_\mu \Sigma = \partial_\mu \Sigma - i \,\big[ \mathcal{V}_\mu , \, \Sigma \big] \,,
\end{equation}
$M= m_\ell \mathds{1}$ is the quark mass matrix for degenerate $u$ and $d$ quarks and $f$ and $B$ are low-energy constants related to the chiral limit of the pion decay constant and the chiral condensate, $\ev{\bar{\psi}\psi}=-f^2 B$, respectively. $B$ is also directly connected to the vacuum pion mass as $m_\pi^0=\sqrt{2Bm_\ell}$. Note that in the commutator, terms proportional to the unit matrix drop out, so that in \chipt a baryon chemical potential has no effect and only the charge difference between $u$ and $d$ quarks is of relevance.

We note at this point the qualitative similarity between this setup and two-color QCD, i.e., with gauge group $\SU(2)$, for two degenerate quarks at non-zero baryon chemical potential. Indeed, the latter system also allows for direct simulations~\bcite{Dagotto:1986gw,Dagotto:1986ms,Dagotto:1986xt,Hands:1999md}. The similarity becomes most evident in the formulation of two-color QCD in the effective field theory equivalent to \chipt~\bcite{Kogut:1999iv,Kogut:2000ek,Splittorff:2000mm}, which can be brought to a form similar to Eq.~\eqref{eq-EB:lag-chipt}. The main difference concerns the interpretation of the effective degrees of freedom, which are given by di-quarks in the two-color theory instead of the pseudoscalar mesons in QCD. A review of the results in two-color QCD, however, is beyond the scope of this chapter. For more recent results see e.g. Refs.~\bcite{Begun:2022bxj,Iida:2022hyy,Hands:2024nkx} and references therein.

\section{Thermodynamic properties of isospin-asymmetric QCD matter}
\label{sec-EB:isospin}

Having introduced lattice simulations and \chipt, we continue by discussing results concerning the impact of external magnetic fields and isospin chemical potentials on the thermodynamic properties of QCD using these approaches. In both cases, a natural starting point is to investigate how these external parameters change the global symmetries and the response of the low-energy degrees of freedom, the pions, at low temperatures. Then we proceed with presenting results obtained via first-principles lattice QCD simulations. In this section, we concentrate on the case of a nonzero isospin-asymmetry.

\subsection{Pion condensation at non-zero isospin chemical potential and low energies}
\label{sec-EB:pionisospin}

\begin{figure}[b]
\centering
\includegraphics[width=.4\textwidth]{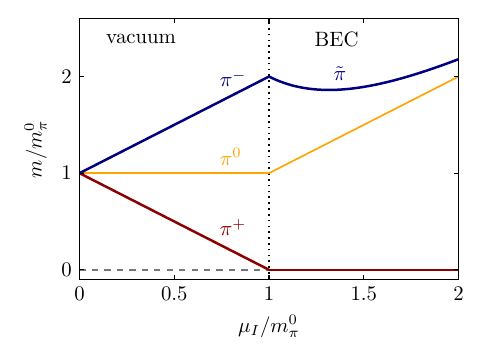}
\includegraphics[width=.42\textwidth]{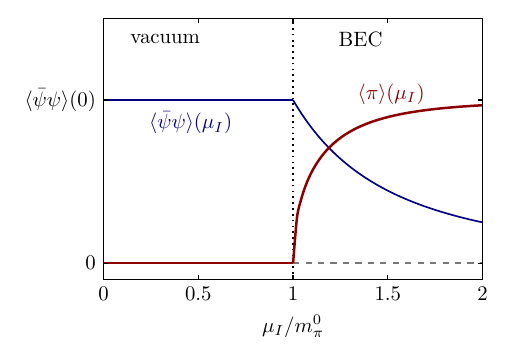}
\caption{{\it Left panel:} Masses of the lightest pseudoscalar excitations at non-zero isospin chemical potential $\muI$ from \chipt at leading order, following~\cite{Son:2000by}. $\tilde{\pi}$ is the radial pseudoscalar meson excitation in the phase with Bose-Einstein condensation (BEC) of charged pions and $m_\pi^0$ the vacuum pion mass.
{\it Right panel:} Expectation values of chiral and pion condensates from \chipt to leading order at $\muI\neq0$.}
\label{fig-EB:chipt-pmasses}
\end{figure}

We start our discussion of the low-energy phenomena of isospin-asymmetric QCD by considering $T=\muB=\muS=B=E=0$ but nonzero $\muI$. For low values of the isospin chemical potential, its main effect is the change of the hadron masses with non-zero isospin quantum numbers. Thermodynamic quantities are not affected following the Silver-Blaze phenomenon~\bcite{Cohen:2003kd}.
In particular, the change of the pion masses with $\muI$ can be computed in \chipt by expanding the Lagrangian density of Eq.~\eqref{eq-EB:lag-chipt} in the pion fields. For $0<\muI<m_\pi^0$, the result is a decreasing mass of $\pi^+$ and an increasing mass of $\pi^-$ with $\muI$, as shown in the left panel of Fig.~\ref{fig-EB:chipt-pmasses}, while $\pi^0$ -- being the $\I=0$ component of the pion triplet -- is not affected by $\muI$~\bcite{Son:2000xc,Son:2000by,Kogut:2001id,Mammarella:2015pxa,Adhikari:2019mdk}. At the critical chemical potential $\muI=\muI^c=m_\pi^0$, the mass of $\pi^+$ reaches zero, signalling the change of the ground state of the system. Here charged pions, being bosonic degrees of freedom, all populate the same zero-momentum state, generating a macroscopic Bose-Einstein condensate (BEC), the so-called pion condensate. While this result, shown in Fig.~\ref{fig-EB:chipt-pmasses}, follows from \chipt at leading order, inclusion of the next order does not change the location of this phase transition~\bcite{Carignano:2016lxe}.

Within the low-energy effective theory, in the new ground state $\overline{\Sigma}$ a linear combination of the fields $\phi_{i=1,2}$ related to the charged pions acquires a non-zero expectation value. Following~\cite{Son:2000by}, this ground state can be parameterized as
\begin{equation}
    \label{eq-EB:bec-gstate}
    \overline{\Sigma} = \mathds{1}\cos(\alpha) + i\,\big( \tau_1 \cos\varphi + \tau_2 \sin\varphi \big) \sin\alpha \,,
\end{equation}
where $\alpha$ and $\varphi$ represent the angles of the ground state in the three-dimensional space spanned by the expectation values of the chiral condensate and the pion condensates with opposite charges. Their value can be obtained by maximizing the static Lagrangian density. This maximum is at $\cos\alpha=(m_\pi^0)^2/\muI^2$, independent of $\varphi$, for which the value is chosen spontaneously. The hadron masses in the medium can then be computed via the fluctuation field with respect to the new ground state~\bcite{Son:2000xc,Son:2000by} (see also~\cite{Kogut:1999iv,Kogut:2000ek} for the prototype computations in the very similar two-color QCD setting at non-zero baryon chemical potential as well as~\bcite{Kogut:2001id,Adhikari:2019mdk} for more details). The spontaneous choice of the angle $\varphi$ indicates spontaneous symmetry breaking of the residual $\Utauthree$ symmetry. This (a) implies the existence of one massless Goldstone boson, following Goldstone's theorem, which can be seen as the massless pseudoscalar state in the BEC phase in Fig.~\ref{fig-EB:chipt-pmasses}, and (b) a second-order transition at the onset of the BEC phase in the $\mathrm{O}(2)$ universality class~\bcite{Son:2000xc}. The other two states are massive and can be identified with $\pi^0$ and the massive excitation $\tilde{\pi}$. Note that due to the fact that the pion condensate breaks the global symmetry associated with the conservation of electric charge, the massless pion and the $\tilde\pi$ states are not charge eigenstates.

As we have just seen, the BEC is characterized by a non-zero expectation value $\ev{\pi}$ of a condensate with pseudoscalar quantum numbers given by a linear combination in the $\pi^\pm$ plane. Pion and chiral condensates can also be computed in \chipt and the leading order result is given by~\bcite{Son:2000xc,Son:2000by},
\begin{equation}
    \label{eq-EB:condensates}
    \ev{\bar{\psi}\psi}_{\muI} = \ev{\bar{\psi}\psi}_{\muI=0} \cdot \cos\alpha \quad \text{and} \quad \ev{\pi}_{\muI} = \ev{\bar{\psi}\psi}_{\muI=0} \cdot \sin\alpha \,,
\end{equation}
so that the chiral condensate rotates into the pion condensate, as visualized in the right panel of Fig.~\ref{fig-EB:chipt-pmasses}. As shown in~\cite{Adhikari:2020ufo}, this prediction only receives small corrections at next-to-leading order and for moderate values of $\muI$, as well as from the introduction of a third quark flavor~\bcite{Adhikari:2020qda}. We note as a side remark that -- together with the pion condensate -- an axial condensate $\ev{\sigma_A}=\ev{\bar{u}\gamma_0\gamma_5d+\bar{d}\gamma_0\gamma_5u}/2$ develops~\bcite{Brauner:2016lkh,Adhikari:2020qda} as well (see also~\bcite{Brandt:2018bwq}) due to the mixing between the pseudoscalar and the temporal component of the axial vector channels, responsible for weak decays of the pion condensate when electroweak interactions are included~\bcite{Brandt:2018bwq}.

\begin{figure}[t]
\centering
\raisebox{.4cm}{\includegraphics[width=.4\textwidth]{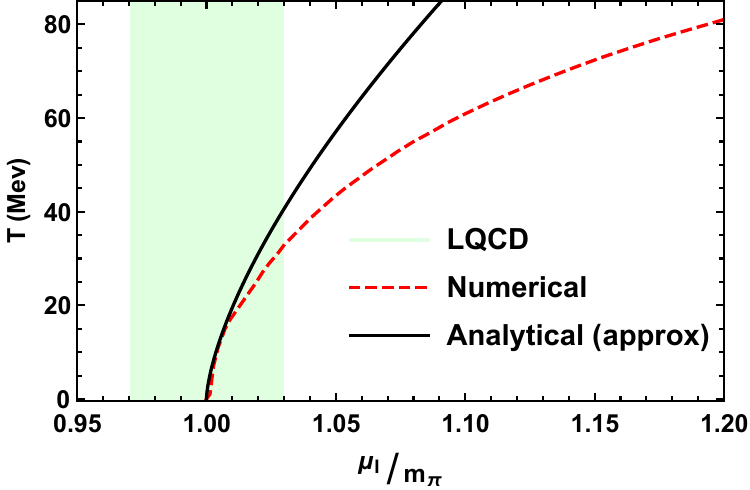}} \qquad
\includegraphics[width=.36\textwidth]{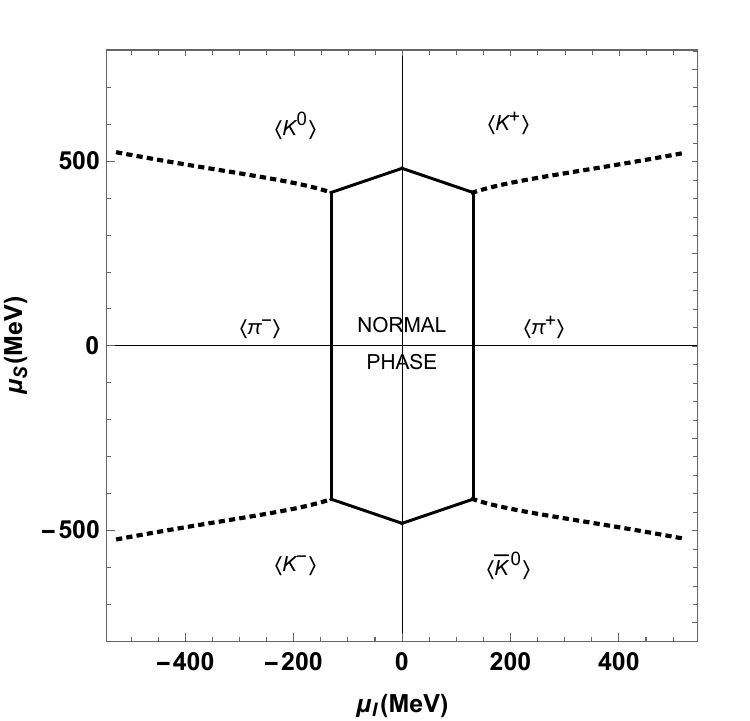}
\caption{{\it Left panel:} Prediction for the BEC phase boundary in the $\muI-T$ plane from \chipt~\bcite{Adhikari:2020kdn}. The full result is the curve referenced as ``Numerical'', while the one denoted as ``Analytical'' is the result of the approximation from Ref.~\bcite{Splittorff:2002xn}. The transition to the BEC phase is expected to be a second order transition in the $\mathrm{O}(2)$ universality class~\bcite{Son:2000xc}. In addition, the green band shows the low-temperature region of the phase boundary obtained from lattice QCD~\bcite{Brandt:2017zck}.
{\it Right panel:} Zero temperature phase diagram in the plane of isospin $\muI$ and strangeness $\muS$ chemical potentials as obtained from \chipt at leading order, taken from Ref.~\bcite{Adhikari:2019zaj}. The region denoted as ``normal phase'' is the vacuum, while the other regions are labelled by the non-vanishing condensates, associated to the pseudoscalar mesons which become massless.}
\label{fig-EB:chipt-pdiag}
\end{figure}

The effects of non-zero temperatures can also be included in \chipt computations. As first anticipated in Refs.~\bcite{Son:2000xc,Son:2000by}, the critical isospin chemical potential $\muI^c$, determined from the point where the charged pion mass vanishes, increases with growing temperature~\bcite{Loewe:2002tw,Loewe:2004mu}. In other words, the critical temperature, where the pion condensate melts is found to be a monotonously increasing function of $\muI$. This result was later also confirmed using the Ginzburg-Landau expansion of the effective potential~\bcite{Adhikari:2020kdn}. The result for the BEC phase boundary of Ref.~\bcite{Adhikari:2020kdn} is shown in the left panel of Fig.~\ref{fig-EB:chipt-pdiag}. Making contact with the regime of asymptotically large $\muI$, it has been proposed that the transition temperature could be a monotonous function for all $\muI$.

\chipt can also be used to compute the EoS, i.e.\ the dependence of the pressure on $\muI$. Results at $T=0$, as well as for low temperatures have been obtained in \chipt with two~\bcite{Adhikari:2019zaj,Adhikari:2020kdn,Andersen:2023ivj} and three~\bcite{Carignano:2016rvs,Adhikari:2019zaj,Andersen:2023ofv} flavors. For the latter, non-zero strangeness chemical potentials have also been considered. Due to the Silver-Blaze property, at $T=0$ bulk thermodynamic observables are only affected by $\muI$ beyond the onset of pion condensation at $\muI=m_\pi$. The EoS develops several peculiar features, such as a strong increase of the speed of sound with $\muI$ (e.g.,~\bcite{Adhikari:2019zaj,Adhikari:2020kdn,Andersen:2023ofv}), soon crossing the conformal value~\bcite{Cherman:2009tw} $c_s^2=c/3$, and a strong decrease of the interaction measure. These results will be discussed further together with the lattice results in Sec.~\ref{sec-EB:isospin-zeroT}.

In addition to a non-vanishing $\muI$, the \chipt treatment of three light quark flavors also readily allows for the introduction of a non-zero strangeness chemical potential $\muS$~\bcite{Kogut:2001id,Mammarella:2015pxa,Adhikari:2019mlf}. For large enough $\muS$ and $\muI<m_\pi^0$, a similar BEC of pseudoscalar mesons -- kaons in this case -- has been observed. While the phase boundary between the vacuum and the phase with a kaon condensate is a second-order transition, just like for the BEC phase with condensing pions, the boundary between pion and kaon condensates constitutes a first-order phase transition, where the pion and kaon condensates develop a jump. The resulting phase diagram is shown in the right panel of Fig.~\ref{fig-EB:chipt-pdiag} and we refer the interested reader to the review~\bcite{Mannarelli:2019hgn} for more details. As in the case of non-zero $\muI$, outside the pion condensed phase thermodynamic observables only respond to $\muS$ at the onset of kaon condensation due to the Silver-Blaze property.

We note that the above realization of spontaneous symmetry breaking, involving a ground state with meson condensation, is valid in the infinite volume. In contrast, in finite volumes spontaneous symmetry breaking does not occur, so that the expectation value of the pion condensate vanishes identically. Instead, a small explicit breaking is necessary to trigger a nonzero condensate. In the case of pion condensation, this is achieved by a term $M\to M+\lambda i\tau_2$ in the mass matrix of~\eqref{eq-EB:lag-chipt}, which picks the direction $\varphi=\pi/2$ in the pion condensate plane. This is exactly the choice made in Eq.~\eqref{eq:lattice-pionsource} for the Dirac operator on the lattice. The impact of the explicit breaking with $\lambda\neq0$ can also be studied in \chipt, see Refs.~\bcite{Kogut:2000ek,Splittorff:2000mm} for the computations done in the two-color QCD framework.

\subsection{Ab initio methods for QCD at non-zero isospin density}
\label{sec-EB:isospin-abinitio}

Leaving the low-energy effective theory behind, next we discuss ab initio results for the phase diagram. In lattice QCD, the standard approach for studying the phase diagram is to perform simulations in the grand canonical ensemble used already in Sec.~\ref{sec-EB:pionisospin}, where nonzero density is parameterized via the associated chemical potentials. 
As discussed in Sec.~\ref{sec-EB:lattice}, 
QCD at $\muI\neq0$ but vanishing other chemical potential components is free of the complex action problem. However, simulations require the introduction of the aforementioned pion source term~\bcite{Kogut:2002zg}, analogous to the source term used in simulations of two-colour QCD~\bcite{Morrison:1998ud,Kogut:2001na}. Physical results are obtained by removing the regulator in terms of a $\lambda\to0$ extrapolation.

A controlled removal of the pion source is a crucial ingredient for a reliable analysis of the properties of QCD at $\muI\neq0$. Fermionic observables composed of light quark fields typically show a strong dependence on the regulator. Examples for the chiral condensate and the isospin density $n_I$, as well as the pion condensate have been shown in Refs.~\bcite{Brandt:2016zdy,Brandt:2017oyy} and the former is also shown in the left panel of Fig.~\ref{fig-EB:T0-isospin}. The strong dependence is in agreement with the $\lambda$-dependence of observables seen in \chipt, e.g.~\bcite{Splittorff:2002xn} and the comparison in Ref.~\bcite{Endrodi:2014lja}. For the pioneering simulations~\bcite{Kogut:2002tm,Kogut:2002zg}, the resulting steep $\lambda\to0$ extrapolations and the associated large systematic uncertainties have been one of the main obstacles. While for a specific setting with eight flavors of staggered fermions at larger-than-physical quark masses, simulations at $\lambda=0$ are indeed possible~\bcite{deForcrand:2007uz}, simulations with $N_f=2+1$, i.e.\ two physical light and strange quarks at small and intermediate temperatures are not feasible for $\lambda$ much smaller than a tenth of the quark mass. New results with full control over the systematics concerning the $\lambda\to0$ extrapolations have been made possible by the development of an improvement program~\bcite{Brandt:2016zdy,Brandt:2017zck,Brandt:2017oyy}, in particular for light-quark observables. The improvement consists of two steps. First, a valence quark improvement, employing the explicit $\lambda$-dependence of the singular values of the Dirac operator, and second, a leading-order reweighting to the ensemble at $\lambda=0$. The latter improves the extrapolations for all observables. The application of the program to chiral and pion condensates has been discussed in Ref.~\bcite{Brandt:2017oyy}, the application to isospin density $n_I$ and the axial density $\ev{\sigma_A}$ in Ref.~\bcite{Brandt:2018bwq} and the extension to second-order Taylor expansion coefficients, generalized susceptibilities, in Refs.~\bcite{Brandt:2022fij,Brandt:2024dle}. As an example, the impact of the improvement program on the $\lambda\to0$ extrapolation of the chiral condensate is shown in the left panel of Fig.~\ref{fig-EB:T0-isospin}.

An alternative approach to non-zero isospin density is to use the methodology of the canonical ensemble by introducing a fixed isospin number $\I$ on the lattice. This can in principle be realized either in terms of an explicit change of the setting of the simulations, following what is done for simulations at non-zero baryon chemical potential, see e.g.\ Refs.~\bcite{Liu:2003wy,Alexandru:2005ix,deForcrand:2006ec}, or by computing observables which involve the correlation functions of $\I\neq0$ hadrons present in the system~\bcite{Detmold:2012wc}. Assuming all isospin is carried by charged pions, the latter study computed the energy density of the system for a given number of pions at vanishing temperature and, consequently, to the equation of state. The presence of pions forces the system into the BEC phase, which leads to a different vacuum as the one employed in the configuration generation and thus introduces an overlap-type problem, resulting in vast differences in the magnitude of correlators on different field configurations. In the studies using this method, this problem has been kept at bay by using a data analysis based on the assumption of log-normal distribution of the measurements of the associated correlation functions~\bcite{Abbott:2023coj,Abbott:2024vhj}.

\subsection{Pion condensation and equation of state at low temperatures}
\label{sec-EB:isospin-zeroT}

\begin{figure}[t]
\centering
\includegraphics[height=4.2cm]{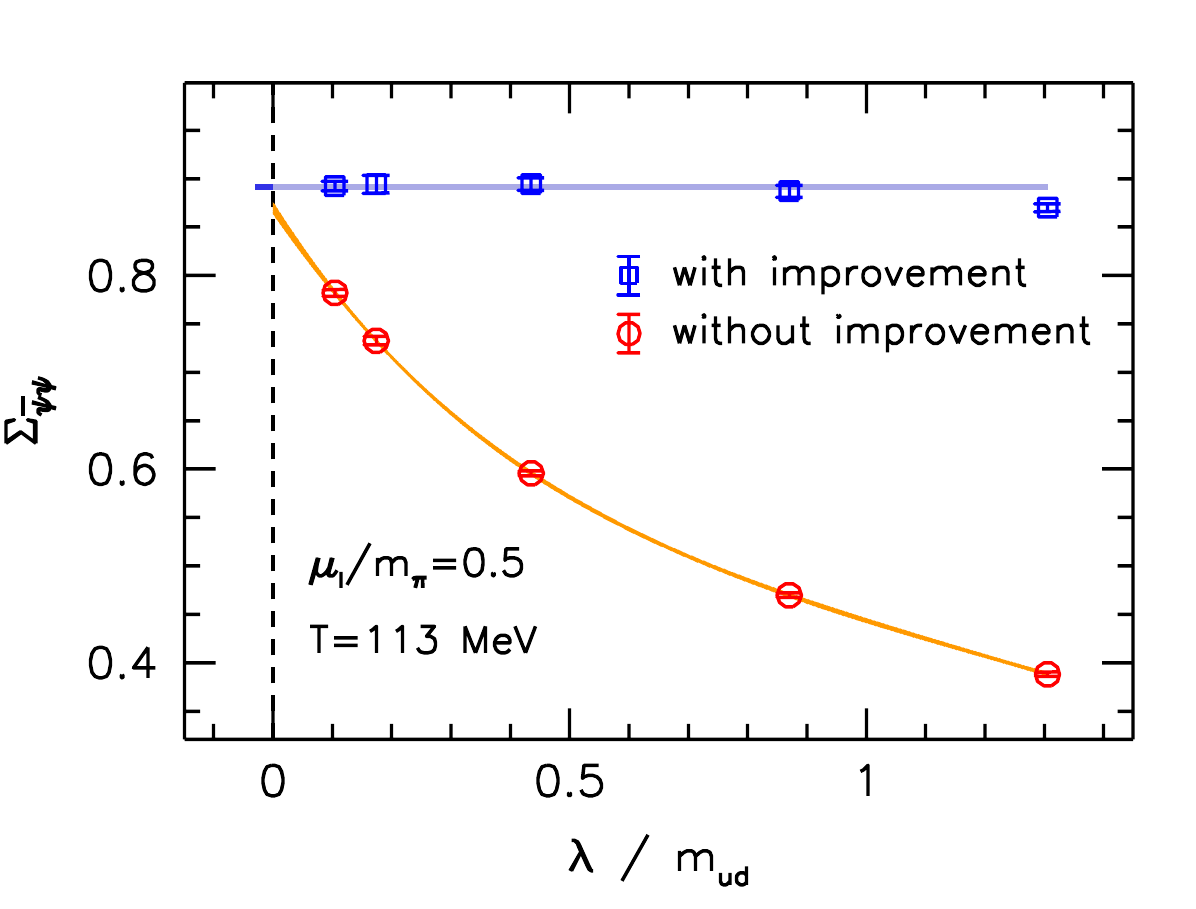}
\quad
\includegraphics[height=4.1cm]{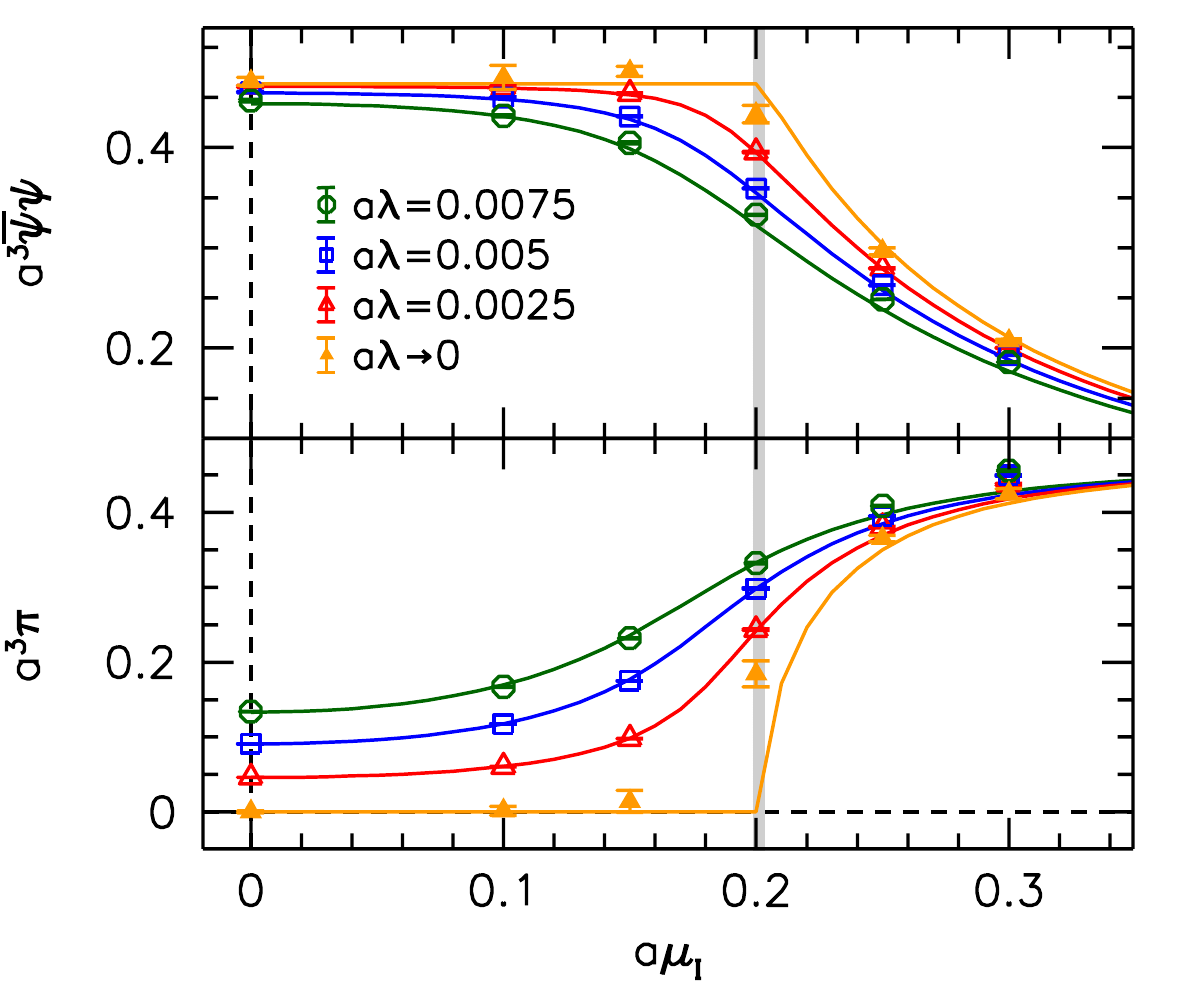}
\quad
\raisebox{.1cm}{\includegraphics[height=4.cm]{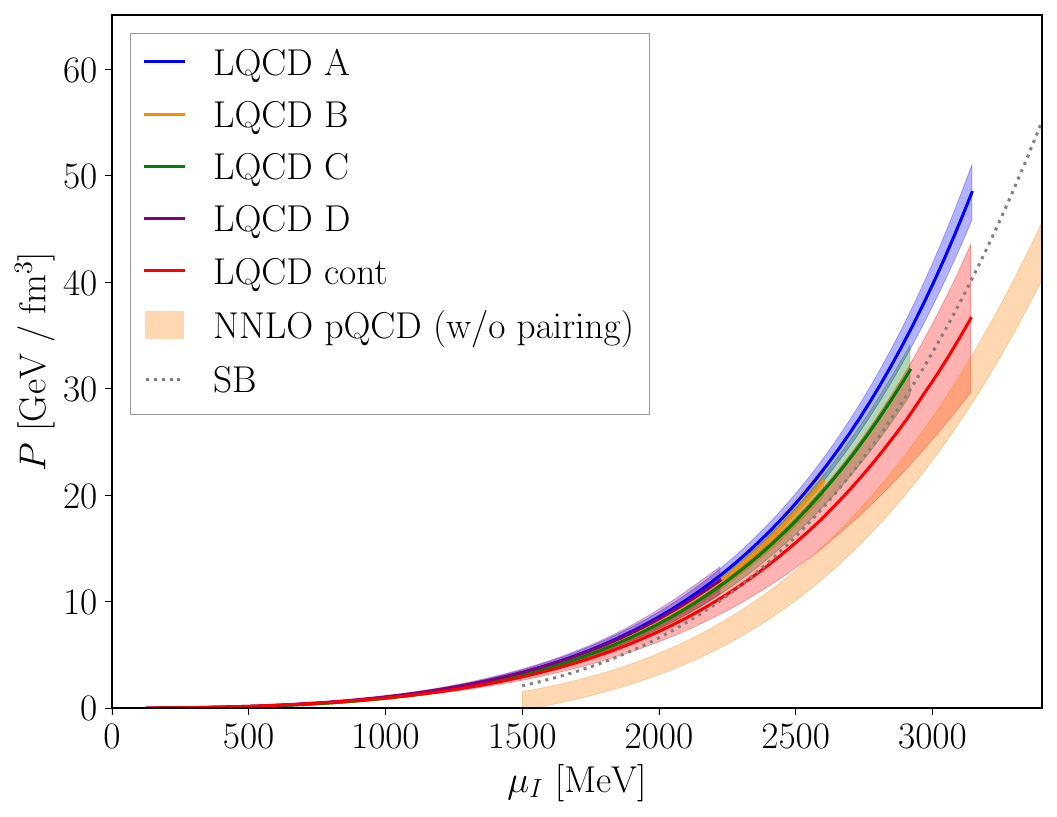}}
\caption{{\it Left panel:} Impact of the improvement program on the $\lambda\to0$ extrapolation of the renormalized chiral condensate $\Sigma_{\bar{\psi}\psi}$, proportional to $\ev{\bar{\psi}\psi}_T-\ev{\bar{\psi}\psi}_{T=0}$, from Ref.~\bcite{Brandt:2017zck}. The red points with the yellow cubic spline represent the unimproved results, while the blue points are obtained using the improvement program together with the remaining $\lambda\to0$ extrapolation (blue band).
{\it Middle panel:} Results for $\ev{\bar{\psi}\psi}$ and $\ev{\pi}$ at $T=0$, obtained with a vacuum pion mass $m_\pi\approx 260$~MeV. Figure adapted from Ref.~\bcite{Endrodi:2014lja}. The solid lines are the results from a fit of the \chipt expressions to the data. The vertical shaded line indicates the critical isospin chemical potential.
{\it Right panel:} Pressure versus isospin chemical potential as obtained in the canonical approach~\bcite{Abbott:2024vhj}. The results correspond to different lattice spacings and a pion mass of $m_\pi\approx 165$~MeV (denoted as LQCD A, B and C) and at one lattice spacing and the physical pion mass (LQCD D). Note that the continuum estimate is based on two lattice spacings.}
\label{fig-EB:T0-isospin}
\end{figure}

The appearance of pion condensation at vanishing temperature has first been investigated in the quenched approximation of QCD~\bcite{Kogut:2002tm} and a short while later in full QCD~\bcite{Kogut:2002zg}, albeit with only two quark flavors and at unphysical pion masses. In both cases the onset of pion condensation has been detected and in the quenched case the critical chemical potential $\muI^c$ was found to lie close to the pion mass used in the simulations, whereas for dynamical quarks only the proportionality $\muI^c\sim m_\pi$ has been established at first. In the two-flavor case and at unphysical pion masses the equivalence $\muI^c=m_\pi$ was later shown in Ref.~\bcite{Endrodi:2014lja} with unimproved staggered fermions. Furthermore, the gradual rotation of $\ev{\bar{\psi}\psi}$ into $\ev{\pi}$, as expected from \chipt -- see Eq.~\eqref{eq-EB:condensates} and the right panel of Fig.~\ref{fig-EB:chipt-pmasses} -- has been observed as well. The middle panel of Fig.~\ref{fig-EB:T0-isospin} shows the results for $\ev{\bar{\psi}\psi}$ and $\ev{\pi}$ at $T=0$ with respect to $a\muI$ for a pion mass $m_\pi\approx 260$~MeV as obtained in~\bcite{Endrodi:2014lja} for different values of $\lambda$ in comparison to a fit with the \chipt expressions (solid lines). The plot clearly indicates the onset of the BEC phase at $\muI^c=m_\pi$ (in fact, this reference used a different normalization for $\muI$ so that the critical chemical potential is at $m_\pi/2$ -- see footnote~\ref{fn:convention}) represented by the vertical shaded line, as well as the rotation of the condensates. At the physical quark masses and including three dynamical quarks, the onset of the BEC phase at $\muI^c=m_\pi$ has been observed at low temperatures in Refs.~\bcite{Brandt:2018bwq,Brandt:2022hwy} in terms of the point where $n_I$ starts to become non-vanishing.

The EoS at $T=0$ with unphysical pion masses has first been determined in Ref.~\bcite{Detmold:2012wc} (first quenched results have been available already in Ref.~\bcite{Kogut:2002tm}) in the canonical setting by calculating the dependence of the energy density $\epsilon$ on the isospin density. In this setting, $n_\I$ is the parameter of the simulations and the chemical potential becomes an observable, $\muI=\partial \epsilon/\partial n_\I$. At vanishing temperature, where the entropy vanishes by definition, this knowledge is sufficient to compute all other relevant thermodynamical quantities. These results have recently been extended to higher isospin densities~\bcite{Abbott:2023coj}, ranging up to chemical potential equivalents of $\muI\lesssim 15m_\pi$, and physical pion masses~\bcite{Abbott:2024vhj}. The pressure versus $\muI$ from Ref.~\bcite{Abbott:2024vhj} is shown in the right panel of Fig.~\ref{fig-EB:T0-isospin}. 
The $p(\muI)$ relation can be obtained equivalently in the grand canonical ensemble as well, where it follows from the dependence of $n_I=\partial p/\partial \muI$ on the isospin chemical potential by integration. Corresponding results at the physical pion mass have been reported in Ref.~\bcite{Brandt:2022hwy}.
Note that due to the Silver-Blaze property, the pressure becomes non-zero only at $\muI=m_\pi$, until which all bulk thermodynamic quantities vanish. Since lattice QCD simulations are necessarily done with a finite space-time volume, there are residual temperature effects that need to be accounted for. In particular, these residual temperature effects impact the Silver Blaze property. This has been corrected in the study of Ref.~\bcite{Brandt:2022hwy} by matching the lattice results for $n_\I(\muI)$ just after the BEC onset to leading-order \chipt.

\begin{figure}[t]
\centering
\raisebox{-.3cm}{\includegraphics[height=4.2cm]{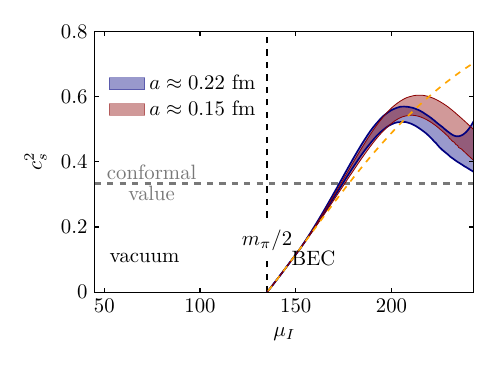}}\quad
\includegraphics[height=3.8cm]{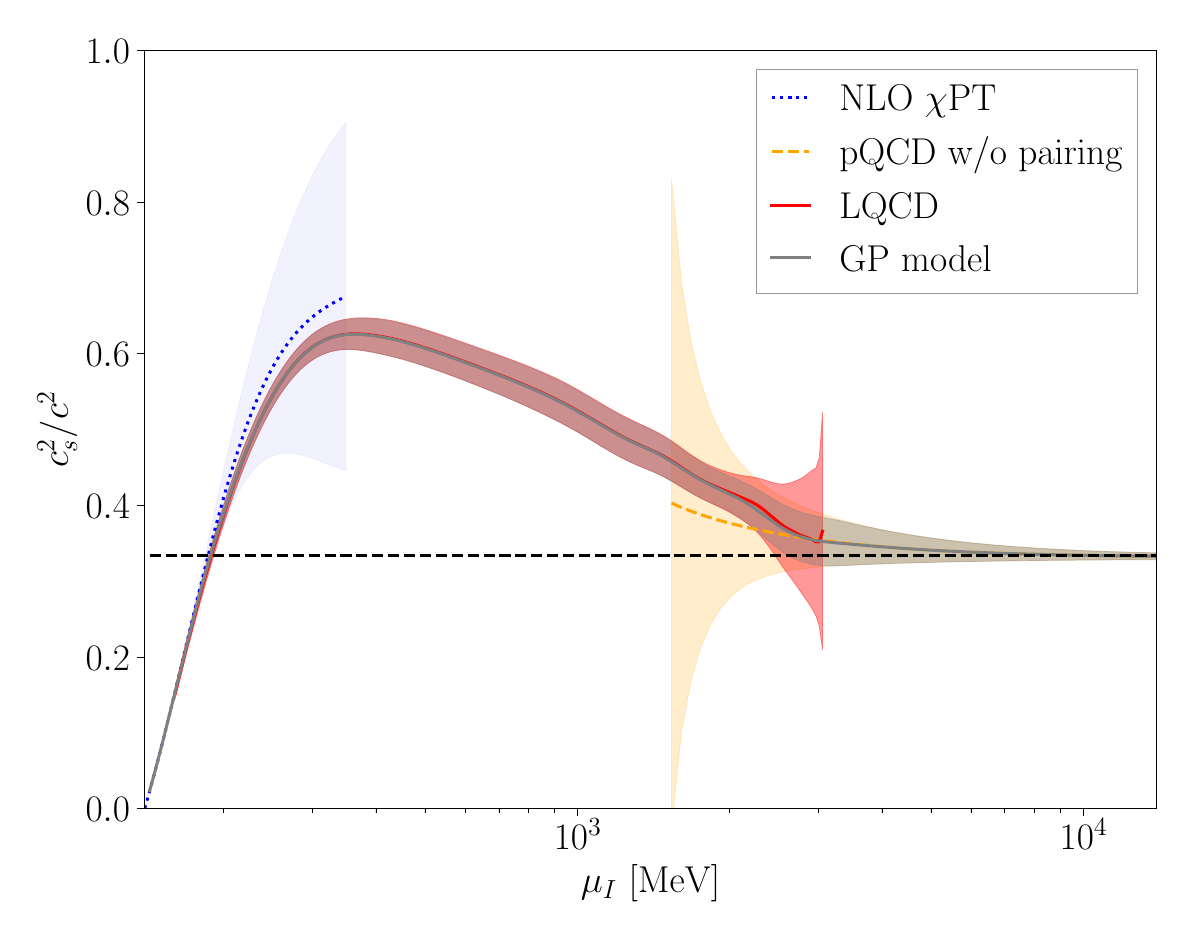}\quad
\raisebox{.1cm}{\includegraphics[height=3.6cm]{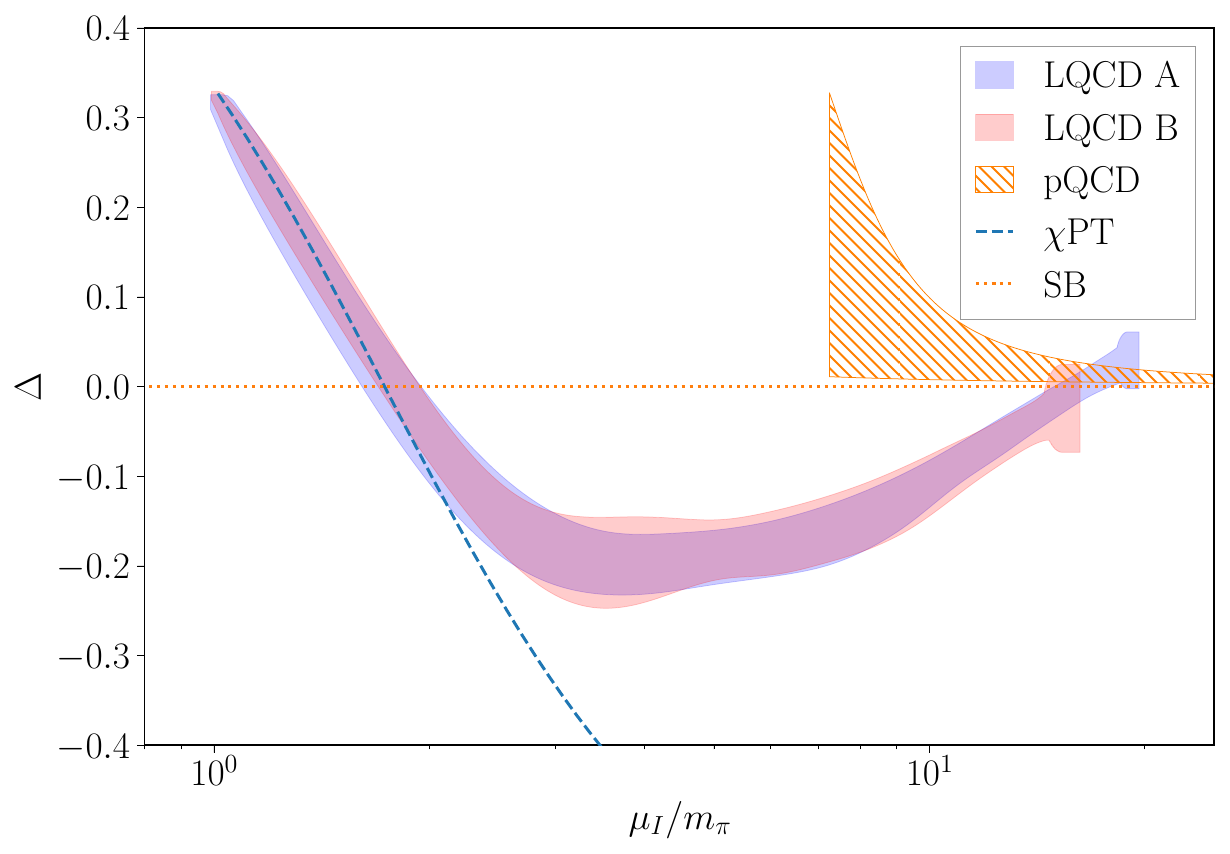}}
\caption{Results for the squared speed of sound $c_s^2$ obtained in the grand canonical ensemble~\bcite{Brandt:2022hwy} (left) and the canonical setting~\bcite{Abbott:2024vhj} (middle), as well as the normalized trace anomaly from the canonical setting~\bcite{Abbott:2023coj} (right). Note that the left panel has been adapted from Ref.~\bcite{Brandt:2022hwy} by converting to the convention used for $\muI$ in this review, cf.\ footnote~\protect\ref{fn:convention}. The dashed horizontal line in the left panel is the conformal value from Ref.~\bcite{Cherman:2009tw} and the dashed yellow curve is \chipt to leading order. In the middle panel ``NLO $\chi$PT'' indicates the \chipt result from Ref.~\bcite{Andersen:2023ofv} and ``pQCD w/o pairing'' the perturbative QCD result from Ref.~\bcite{Fujimoto:2023mvc}. ``GP model'' denotes the curve associated with an interpolation between \chipt, the lattice and the perturbative results using a Gaussian process. In the right panel ``LQCD A'' and ``B'' indicate lattice results at one lattice spacing but two different volumes, ``pQCD'' is the perturbative next-to-leading order result from Ref.~\bcite{Graf:2015tda} and ``$\chi$PT'' the next-to-leading order \chipt prediction~\bcite{Carignano:2016rvs}. 
\label{fig-EB:T0-isospin-eos}}
\end{figure}

A prominent feature of the EoS is the strong increase of the speed of sound $c_s$ in the BEC phase~\bcite{Brandt:2022hwy,Abbott:2023coj,Abbott:2024vhj} -- observed both in simulations at $\muI\neq0$~\bcite{Brandt:2022fij} and in the phenomenologically similar two-color QCD at non-zero baryon chemical potential~\bcite{Iida:2022hyy}. Fig.~\ref{fig-EB:T0-isospin-eos} shows the results from Ref.~\bcite{Brandt:2022hwy} (left panel) and Ref.~\bcite{Abbott:2024vhj} (middle panel) in comparison to next-to-leading order \chipt~\bcite{Andersen:2023ofv}. As visible from the plot, $c_s^2$ initially increases strongly and soon crosses the conformal value~\bcite{Cherman:2009tw} $c_s^2=1/3$, in full agreement with \chipt (see also~\bcite{Adhikari:2019zaj}). It then deviates from \chipt, reaches a maximum at around $\muI\approx 2 m_\pi$, where it starts do decrease towards the Stefan-Boltzmann limit as $\muI\to\infty$. This behavior is also seen in perturbative QCD at next-to-next-to-leading order (NNLO) with the influence of a pairing gap for Cooper pairing of quarks with pionic quantum numbers~\bcite{Fujimoto:2023mvc,Fukushima:2024gmp}, which is also in agreement with the lattice data at large $\muI$. Note that the assumption about the presence of the pairing gap is necessary for the agreement at large $\muI$, as can be seen by comparison to perturbative results without pairing gap~\bcite{Freedman:1976ub,Baluni:1977ms,Kurkela:2009gj}. The observed behavior of the speed of sound is quite remarkable: it is the first time that an equation of state with $c_s^2>1/3$ has been observed in QCD by a first-principles calculation. This lends credibility to similar sound speeds showing up in the model-agnostic sampling of the EoS for neutron stars, see e.g.~\bcite{Annala:2019puf,Altiparmak:2022bke,Marczenko:2022jhl}, which have previously been occasionally considered unlikely. In the same regime where $c_s$ crosses the conformal value, the (normalized) interaction measure $\Delta$ becomes negative~\bcite{Brandt:2022hwy,Abbott:2023coj}, as shown in the right panel of Fig.~\ref{fig-EB:T0-isospin-eos}, passing its conformal value, $\Delta=0$. Similarly to the conformal value as a bound for $c_s^2$, it has been conjectured that $\Delta>0$ for QCD matter, e.g.~\bcite{Fujimoto:2022ohj}, which is found to be violated in this regime.

\subsection{Non-zero temperature phase diagram and equation of state}

\begin{figure}[t]
\centering
\includegraphics[width=.4\textwidth]{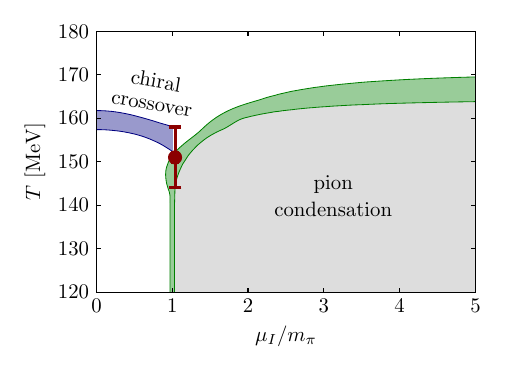}\quad
\includegraphics[width=.4\textwidth]{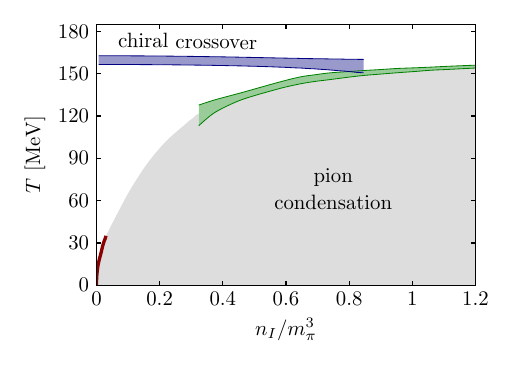}
\caption{Finite temperature phase diagram of isospin-asymmetric QCD with respect to the isospin chemical potential~\bcite{Brandt:2017oyy,Brandt:2018omg} (left) and the isospin density~\bcite{Brandt:2022hwy} (right). Shown is the BEC phase in gray, its phase boundary (green) and the chiral crossover between the hadronic and quark-gluon plasma phases (blue). The red point in the left panel indicates the pseudo-critical point, the meeting point of the chiral crossover line with the BEC phase boundary~\bcite{Brandt:2017oyy}. The red line in the right panel is the prediction from \chipt at next-to-leading order~\bcite{Adhikari:2020kdn}. In the left panel, all results have been extrapolated to the continuum, while in the right panel only the results from $N_t=12$ lattices (the ones with the smallest available lattice spacing) are shown. The continuation of the gray area in the right panel between lattice results and \chipt serves to guide the eye. 
\label{fig-EB:isospin-phasediag}}
\end{figure}

The inclusion of temperature effects within low-energy effective theory has already been discussed in Sec.~\ref{sec-EB:pionisospin} and the resulting phase diagram is shown in the left panel of Fig.~\ref{fig-EB:chipt-pdiag}. The first lattice study of the finite temperature phase diagram was carried out together with the investigation of pion condensation at $T=0$ in Refs.~\bcite{Kogut:2002zg} in two-flavor QCD and was later extended to the three-flavor~\bcite{Kogut:2004qq,Sinclair:2006zm} and eight-flavor cases~\bcite{deForcrand:2007uz}, albeit once more at unphysical pion masses and with unimproved staggered fermions. The continuum phase diagram at physical quark content, three dynamical quark flavors with physical masses and improved staggered quarks -- vital for a reliable continuum extrapolation -- has been obtained first in Ref.~\bcite{Brandt:2017oyy} up to $\muI\lesssim 240$~MeV
and later been extended to $\muI\lesssim 650$~MeV~\bcite{Brandt:2018omg}. The results are shown in the left panel of Fig.~\ref{fig-EB:isospin-phasediag}. 

The phase diagram features the hadronic (low $T$ and $\muI$) and QGP phases (high $T$), as well as the BEC phase (low $T$, high $\muI$), shown by the gray area. These phases are separated by the prolongation of the chiral crossover (hadronic and QGP phases -- blue curve in the plot) into the temperature-chemical potential plane and the phase boundary of the BEC phase (separating it both from the hadronic and the QGP phases -- green curve in the plot). The red point is the estimate for the meeting point of the chiral crossover and the BEC phase boundary~\bcite{Brandt:2017oyy}, dubbed pseudocritical point -- it would be a critical point if the chiral crossover would rather be a true phase transition. 

Remarkable about the phase diagram is the shape of the BEC phase boundary. As dicussed in Sec.~\ref{sec-EB:pionisospin}, from \chipt one would expect a transition temperature which rises monotonically with $\muI$~\bcite{Adhikari:2020kdn}. As can be seen in the figure, the BEC phase boundary remains vertical instead until a temperature of about $T\sim150$~MeV, close to the pseudocritical point. A significant deviation between \chipt and the lattice results shows up at a temperature of about $T\sim30$ to 40 MeV. Close to the pseudocritical point, the phase boundary bends strongly to become almost horizontal for higher $\muI$. Finite size scaling at a temperature of 113~MeV shows consistency with $\text{O}(2)$ scaling~\bcite{Brandt:2017oyy}. The decrease of the chiral crossover temperature with $\muI$ is similar to what happens in the direction of $\muB$, albeit with a different coefficient. The results from the simulations compare well with a functional dependence quadratic in $\muI/T$ all the way up to the BEC phase boundary and a coefficient as obtained in Ref.~\bcite{HotQCD:2018pds} and are also similar in magnitude as the ones obtained from a hadron-resonance gas approximation~\bcite{Toublan:2004ks}.

Since direct simulations at $\muI$ are possible and the phase diagram includes a second-order phase transition to the BEC phase, the simulation results can also be utilized to test the accuracy and efficacy of indirect methods for exploring the EoS and the phase diagram, based on Taylor expansions~\bcite{Allton:2002zi} and analytic continuations from imaginary chemical potentials~\bcite{deForcrand:2003vyj}. Comparisons between the Taylor expansion of the pressure or isospin density and direct simulation results have been discussed in Ref.~\bcite{Brandt:2018omg} and clearly show the expected breakdown of the Taylor expansion at the BEC phase boundary. The comparison to Taylor expansions at different orders (up to twelfth order) has been performed in Refs.~\bcite{Borsanyi:2023tdp,Mitra:2024czm}. This indicates that high-order coefficients are mandatory to reliably detect the breakdown of the Taylor series. From the Taylor series one can also construct estimators for the radius of convergence of the Taylor expansion, which have been compared to the pion condensation phase boundary in Refs.~\bcite{Brandt:2018omg}. Another tool to study the presence of a phase transition is via the Lee-Yang edge singularities~\bcite{Yang:1952be,Stephanov:2006dn}. These have been studied for the isospin chemical potential direction in Ref.~\bcite{Mitra:2024czm} and found to be in reasonable agreement with the transition region for the temperatures where the BEC phase remains present.

\begin{figure}[t]
\centering
\includegraphics[width=.32\textwidth]{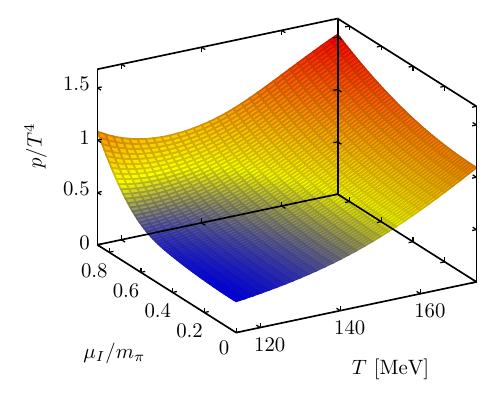}
\includegraphics[width=.32\textwidth]{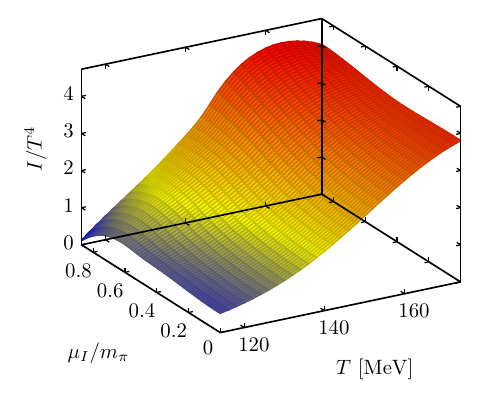}
\includegraphics[width=.34\textwidth]{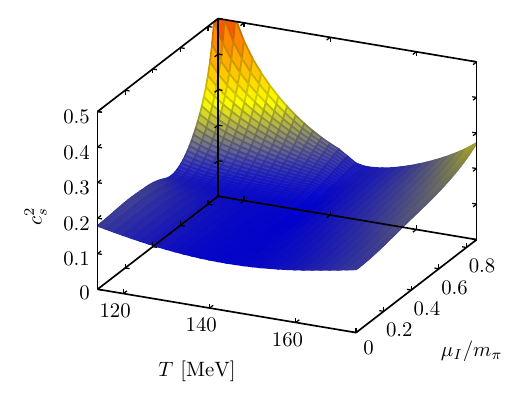}
\caption{Results for the pressure (left) the interaction measure (middle) and $c_s^2$ (right) as obtained in lattice simulations on $N_t=8$ lattices~\bcite{Brandt:2022hwy}. The color coding represents the value of the respective observable.
\label{fig-EB:isospin-eos-finiteT}}
\end{figure}

The EoS at $T\neq0$ can be obtained from the temperature- and chemical potential-dependence of $n_I$ and has been determined in Ref.~\bcite{Brandt:2022hwy}. This also allows for the determination of the phase diagram in the temperature-isospin density plane, shown as obtained on $N_t=12$ lattices in the right panel of Fig.~\ref{fig-EB:isospin-phasediag}. The results for the pressure, the interaction measure and the speed of sound obtained on lattices with $N_t=8$ are shown in Fig.~\ref{fig-EB:isospin-eos-finiteT}. While no continuum limit has been performed in Ref.~\bcite{Brandt:2022hwy}, the approach to the continuum has been investigated using $N_t=8,\,10$ and 12, finding only a mild dependence on the lattice spacing. The interaction measure clearly shows the tendency towards negative values deep in the BEC phase at low temperatures, already discussed in the previous section. With increasing $\muI$ one can also observe a shift in its maximum, indicating a decrease of the thermal transition temperature, in agreement with the phase diagram shown in Fig.~\ref{fig-EB:isospin-phasediag}. The speed of sound also shows the large values in the BEC phase at low temperatures, where it crosses the conformal value, just like we have observed at $T=0$. For the future it would be interesting to test how the results approach hard-thermal loop perturbative QCD~\bcite{Andersen:2015eoa} at even higher temperatures and chemical potentials.

Finally, it is interesting to note that around the transition to the QGP phase, in the region of temperatures around 150 to 160~MeV, the speed of sound does not increase significantly with $\muI$. This is a very similar behaviour to what is expected for QCD in the region of large $\muB$. As discussed above, large values of the speed of sound have been observed for the EoS in neutron stars, corresponding to the region of $T=0$ and large $\muB$. At the same time, the speed of sound has been observed to be almost independent from $\muB$ in the vicinity of the chiral crossover, e.g. Ref.~\bcite{Bollweg:2022fqq}. Thus the results at $\muI\neq0$ not only, once more, lend credibility to the observations at $\muB\neq0$, but also indicate a potential similar behaviour of the bulk thermodynamics of QCD in very different regimes of the parameter space.

\section{Thermodynamic properties of magnetized QCD matter}
\label{sec-EB:magnetic}

We proceed by discussing results for the impact of background electromagnetic fields on the phase diagram and the EoS. Just as for $\muI>0$, the low-energy effective theory can provide useful guidance to this end. Still, most of our knowledge comes from first-principles lattice simulations in this setting as well. In fact, the lattice determination of the phase diagram and of relevant observables proved beneficial also for improving low-energy QCD models and providing a benchmark for functional approaches, too. A comprehensive overview of lattice QCD results at nonzero electromagnetic fields can be found in~\bcite{Endrodi:2024cqn}.

\subsection{Magnetic catalysis and phase structure at low energies}
\label{subsec-EB:Blow}

Much like the case of a non-zero isospin chemical potential, the magnetic field influences the hadron masses in a non-trivial manner. This can be interpreted both directly on the level of hadrons, treating them as charged point-like particles, or indirectly through their fundamental building blocks, the quarks. For weak external magnetic fields, the mass (identified with the energy of the lowest lying state, the so-called lowest Landau-level) of a relativistic free hadron $H$ of charge $q$ changes as
\begin{equation}
    \label{eq-EB:B-hadronmass}
    M_H^2(eB) = M_H^2(0) + \vert qB \vert - g_H s_z qB + \ldots \,,
\end{equation}
where $g_H$ is the gyromagnetic factor of the hadron and $s_z$ its spin projection on the axis of the magnetic field, in our case the $z$-direction. The ellipses represent higher-order terms in $eB$. The non-trivial internal structure of hadrons is manifested through the fact that $g_H$ can strongly deviate from $g=2$.
Following Eq.~\eqref{eq-EB:B-hadronmass}, naively the mass of a point-like $\rho$ meson (with $g_\rho=2$) with a given combination of charge and spin projection vanishes at a critical magnetic field strength, leading to the proposal for the possibility of $\rho$-meson condensation~\bcite{Chernodub:2010qx}. While the presence or absence of $\rho$-meson condensation has been strongly debated in the literature, our current understanding is that no condensation takes place.

In the absence of meson condensation, the main impact of the magnetic field concerns the generic thermodynamic properties of the system in the hadronic and quark-gluon plasma phases, as well as, potentially, the properties of the transition. The external magnetic field has the tendency to strengthen the breaking of chiral symmetry and to enhance the value of the chiral condensate $\ev{\bar{\psi}\psi}$, a phenomenon called magnetic catalysis~\bcite{Gusynin:1994re} (for a detailed review see~\cite{Shovkovy:2012zn}). Essential for this strengthening of the breaking of chiral symmetry is an effective reduction of the dimensionality of the system, $d\to d-2$, by strong magnetic fields, since it forces charged particles (here the quarks) on Landau levels, where motion perpendicular to the magnetic field is suppressed and a state degeneracy $\propto B$ appears. This proliferation of low modes, through the Banks-Casher relation~\bcite{Banks:1979yr}, leads to the enhancement of the condensate. Magnetic catalysis has also been demonstrated to hold within \chipt. Here, the enhancement is proportional to $B$ for weak fields in the case of massless quarks~\bcite{Shushpanov:1997sf,Agasian:1999sx}, whereas it is proportional to $B^2$ at nonzero quark masses~\bcite{Cohen:2007bt}.
In fact, in the limit of weak magnetic fields, magnetic catalysis can be traced back to the positivity of the leading $\beta$-function coefficient in scalar QED~\bcite{Bali:2013txa,Endrodi:2024cqn}, again revealing its universal, robust nature.

Apart from modifying the thermodynamics properties at vanishing chemical potentials, the presence of the external field potentially leads to novel phases in the multi-dimensional parameter space. When the BEC phase at non-zero $\muI$ is subjected to an external magnetic field, it initially remains stable until at a critical $\eb$ the BEC phase undergoes a transition to a type-II superconducting state with an inhomogenous pion condensate~\bcite{Adhikari:2015wva,Adhikari:2018fwm}. One possible form of the inhomogeneity is a lattice of magnetic vortices~\bcite{Adhikari:2018fwm}. However, a study of perturbations with phonon modes raises doubt about the stability of such a lattice structure~\bcite{Adhikari:2022cks}. To scrutinize the exact form of the inhomogeneity, ab initio studies of this system are required in the future. Similarly, in the presence of a non-zero baryon chemical potential, an external magnetic field is expected to trigger the appearance of a new phase with a chiral-soliton lattice~\bcite{Brauner:2016pko}, which might itself potentially trigger charged pion condensation at larger magnetic field.

\subsection{Inverse magnetic catalysis and the phase diagram}
\label{sec-EB:mcimc}

In Sec.~\ref{subsec-EB:Blow} we have seen that at $T=0$ the magnetic catalysis phenomenon, i.e.\ the enhancement of the chiral condensate $\ev{\bar\psi\psi}$ by a background magnetic field is very robust. However, at nonzero temperatures, this picture was observed to change substantially: depending on the quark masses and the precise value of the temperature, the condensate may decrease as $B$ grows, a phenomenon dubbed inverse magnetic catalysis~\bcite{Bruckmann:2013oba}.
In order to understand the origin of inverse magnetic catalysis, we need to investigate the path integral formulation of the expectation value of the condensate. Referring back to~\eqref{eq:partfunc}, for the condensate for flavor $f$, this reads,
\begin{equation}
\ev{\bar\psi_f\psi_f} = 
\frac{1}{\Z}\int \D U_\nu \, e^{-S[U]}\, \prod_{f'} \det \left[ \Ds_{f'}(B)+m_{f'}\right]\, \textmd{Tr} \left[\Ds_f(B)+m_f\right]^{-1}\,.
\label{eq:pbpdef}
\end{equation}
Here we indicated explicitly the dependence of the observable on the magnetic field: on the one hand, through the traced propagator (the valence contribution) and on the other hand, through the determinants (the sea contribution). The robust magnetic catalysis effect mentioned above ensures that for any gluon field configuration $U_\nu$, the magnetic field enhances the valence term. The sea contribution, however, is not fixed by this argument. 

To assess the sea contribution, one needs to investigate the impact of the magnetic field on the typical gluon configurations that dominate the path integral. Around the transition temperature $T_c$, the most important gluonic observable characterizing the configurations is the Polyakov loop $P$, the parallel transporter of a static quark along a closed temporal loop,
\begin{equation}
 P = \frac{1}{V}\int\textmd{d}^3 \bm x\, \textmd{Tr} \,\mathcal{P} \exp\left[ i \int^{1/T}_0 \mathrm{d}x_4 \,\A_4 (\bm x, x_4)\right]
=\frac{1}{V}\sum_{\bm n}\textmd{Tr} \prod_{n_4=0}^{N_t-1} U_4(\bm n,n_4)\,,
 \label{eq:Pdef}
\end{equation}
where we have written down the definition both in terms of continuum and lattice variables and $\mathcal{P}$ denotes the path ordering operator. The Polyakov loop is related to the excess free energy of a static color charge~\bcite{Fukushima:2017csk} and is known to exhibit a strong anti-correlation with the quark condensate in the transition region. It can be shown that the magnetic field prefers larger values of $P$, which, through this anti-correlation, suppresses the condensate. This mechanism was first discovered in~\bcite{Bruckmann:2013oba}, and is explained in detail also in the review~\bcite{Endrodi:2024cqn}. The complete dependence of $\ev{\bar\psi\psi}$ on the magnetic field arises through a competition of the valence and sea contributions. It turns out that for physical quark masses, the sea contribution becomes dominant in a window around $T_c$, leading to the inverse magnetic catalysis of the condensate in the transition region. We note that both~\eqref{eq:pbpdef} and the expectation value of~\eqref{eq:Pdef} are subject to renormalization in the ultraviolet, which, however does not impact the qualitative infrared behavior described above.

\begin{figure}[t]
\centering
\includegraphics[height=5.cm]{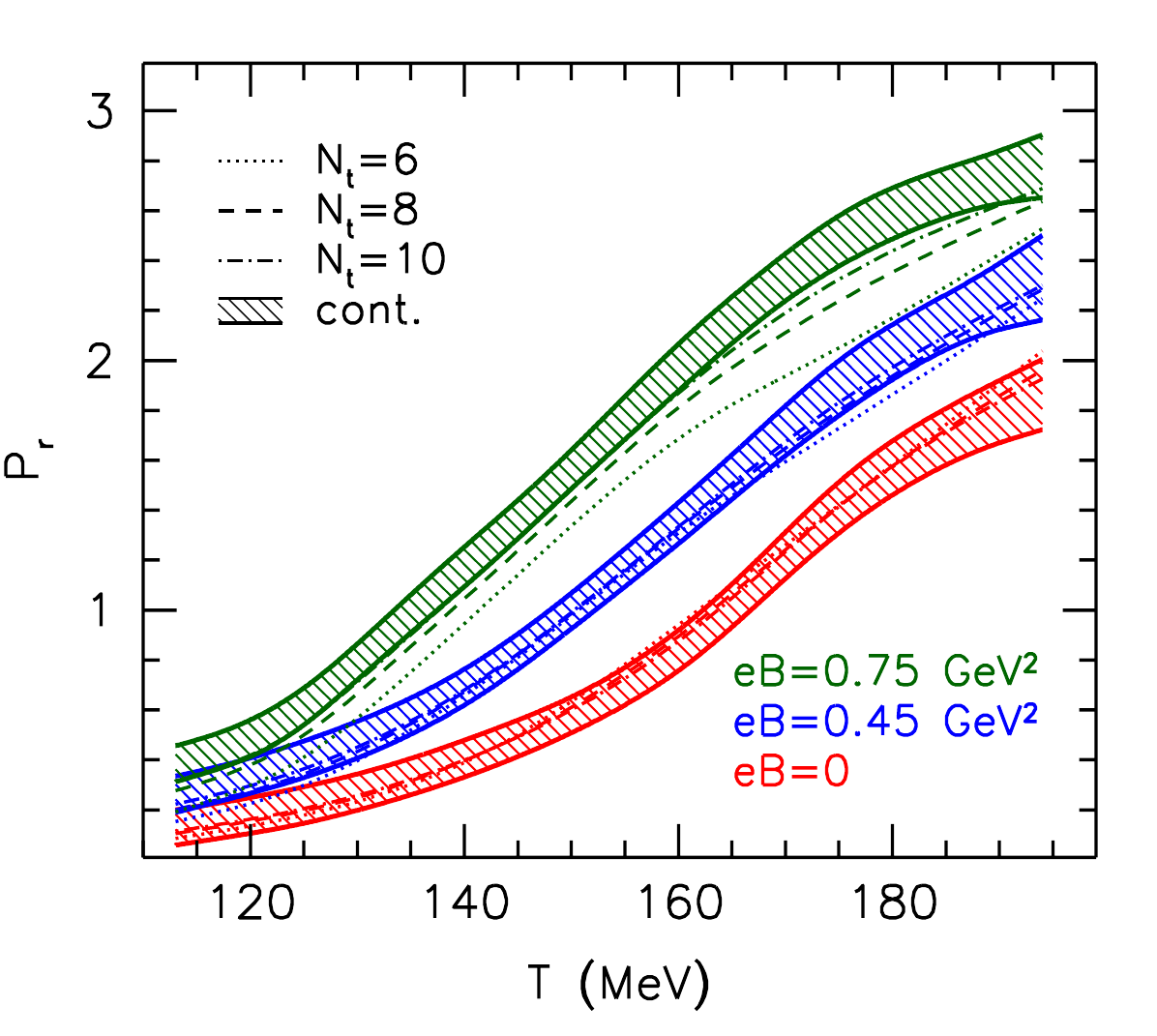}\qquad
\includegraphics[height=5.cm]{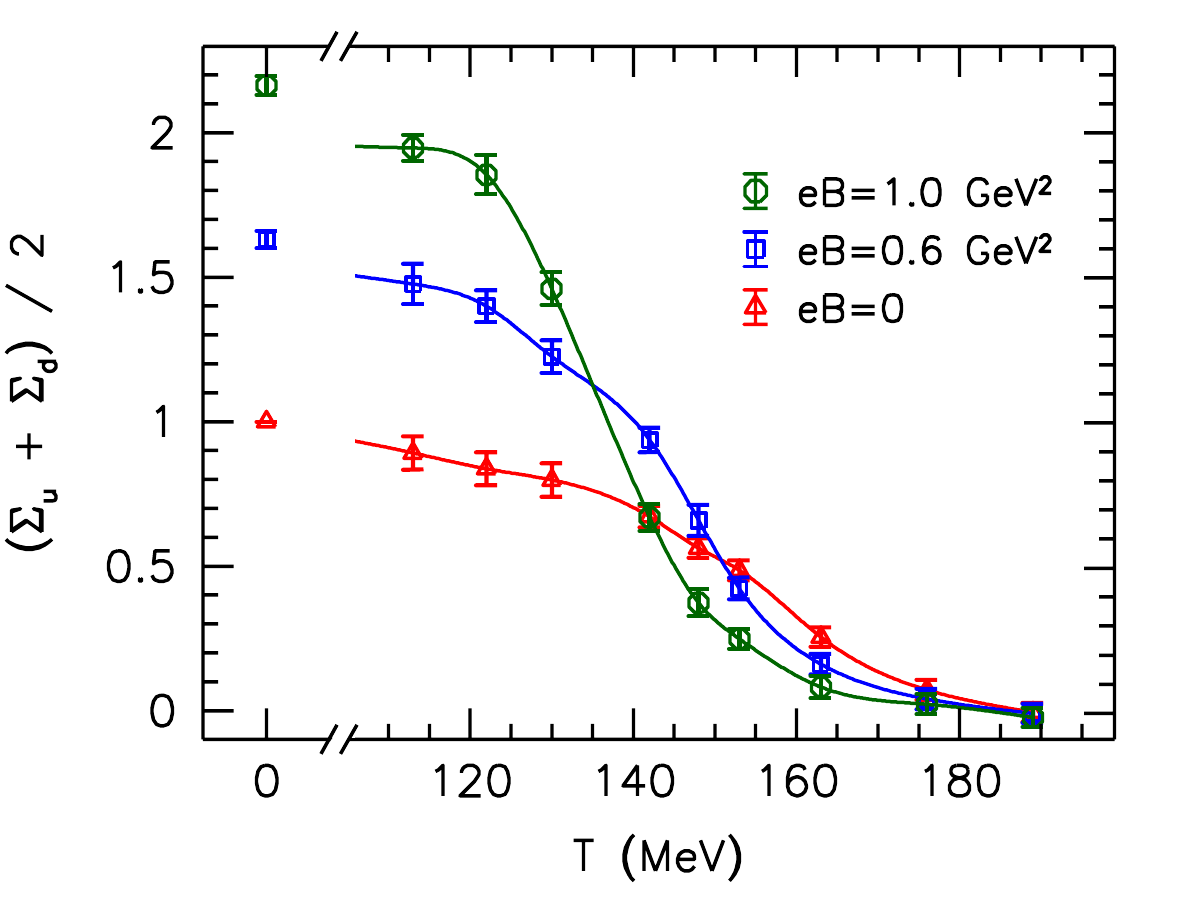}
\caption{{\it Left panel:} Renormalized Polyakov loop as a function of the temperature for different values of the magnetic field~\bcite{Bruckmann:2013oba}. {\it Right panel:} Dependence of the average light quark condensate on the temperature for different values of the magnetic field~\bcite{Bali:2012zg}. Figure from~\bcite{Endrodi:2014vza}.}
\label{chap1:pd1}
\end{figure}

\begin{figure}[t]
\centering
\includegraphics[height=4.8cm]{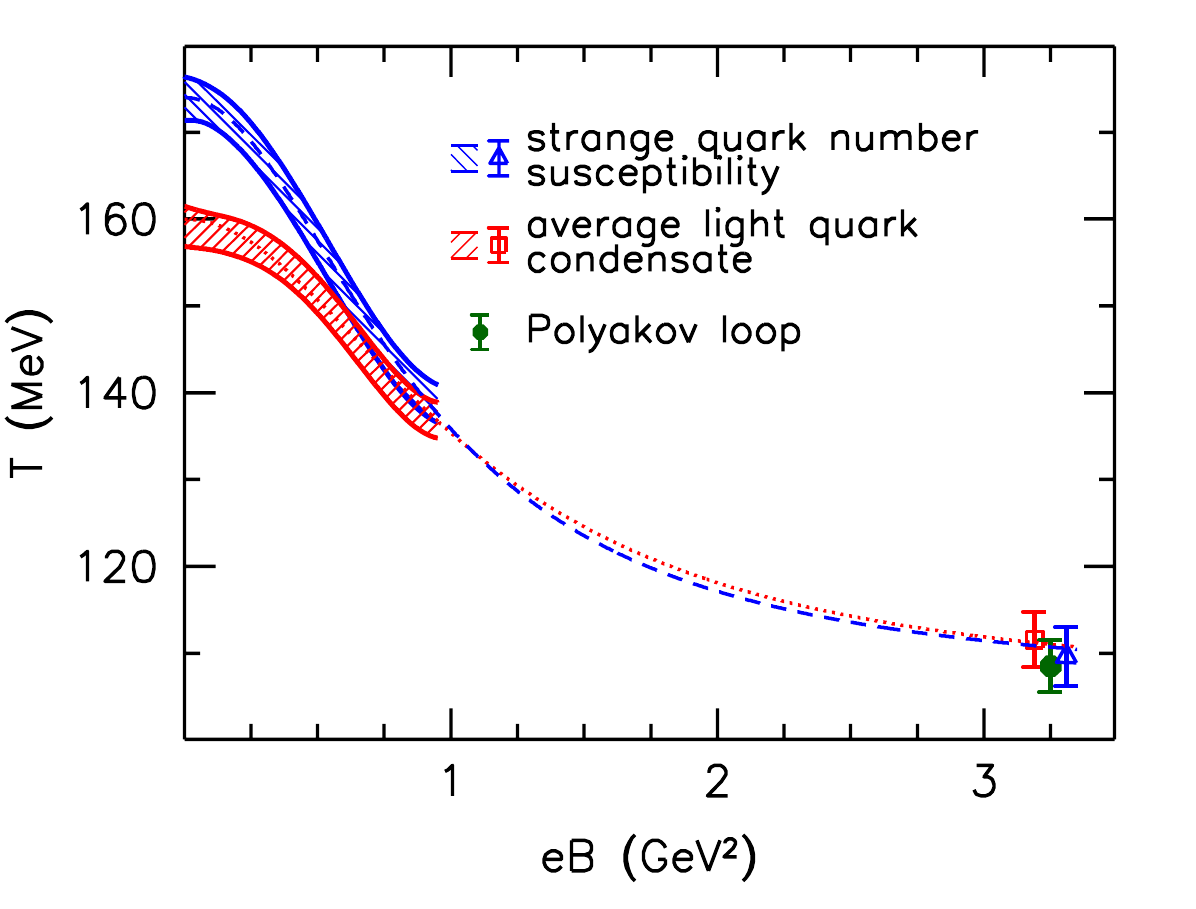}\qquad
\includegraphics[height=5.cm]{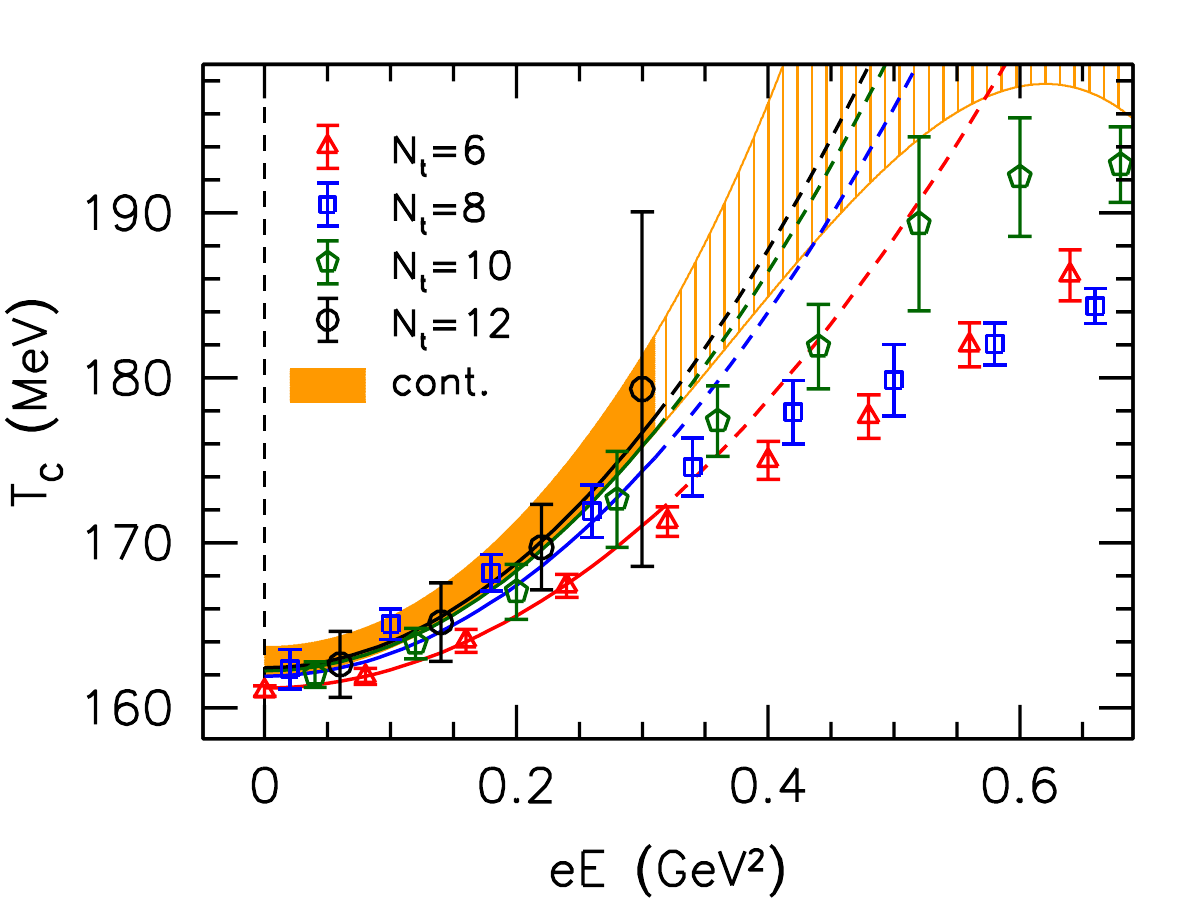}
\caption{{\it Left panel:} Phase diagram of QCD in the magnetic field--temperature plane, indicating the inflection point of the average light quark condensate (red), the strange quark number susceptibility (blue) and the Polyakov loop (green)~\bcite{Endrodi:2015oba}. {\it Right panel:} Phase diagram in the electric field--temperature plane using constant contours of the Polyakov loop at different lattice spacings (points). The orange band indicates the extrapolation to the continuum limit in the region, where the leading-order Taylor expansion is expected to be valid~\bcite{Endrodi:2023wwf}.
}
\label{chap1:pd2}
\end{figure}

This behavior of the Polyakov loop and the average light quark condensates has been observed in lattice simulations employing stout-improved staggered quarks with physical masses~\bcite{Bali:2011qj,Bali:2012zg,Bruckmann:2013oba}. 
The dependence of the Polyakov loop on the temperature and the magnetic field is visualized in the left panel of Fig.~\ref{chap1:pd1}. Analogously, the continuum extrapolated results of $\ev{\bar{\psi}\psi}$ are shown in the right panel of Fig.~\ref{chap1:pd1}, which demonstrates both magnetic catalysis at low and inverse magnetic catalysis at intermediate temperatures. The transition temperature of the crossover may be defined in terms of the inflection point of the chiral condensate. Clearly, $T_c$ is reduced as $B$ grows~\bcite{Bali:2011qj,Bruckmann:2013oba}, which is as also shown by the red curve in the left panel of Fig.~\ref{chap1:pd2}. This plot also includes the inflection point of two different observables, the strange quark number susceptibility (see Eq.~\eqref{eq:qnsusc}) and the Polyakov loop, Eq.~\eqref{eq:Pdef}, which exhibit very similar trends. While the inverse magnetic catalysis of the condensate in the transition region and the reduction of $T_c(B)$ are closely related phenomena, it is important to stress that they are not equivalent. As the mass of the light quarks is increased -- and together with it the vacuum mass $m_\pi$ of the pion -- inverse magnetic catalysis is suppressed and for $m_\pi\gtrsim 500 \textmd{ MeV}$ the condensate is enhanced by $B$ for all temperatures~\bcite{Endrodi:2019zrl}. However, $T_c(B)$ was observed to remain a decreasing function for all pion masses~\bcite{DElia:2018xwo}. The results concerning the behavior at larger-than-physical quark masses were obtained with stout-improved staggered quarks at one lattice spacing.

It has been known that in the asymptotically strong magnetic field limit, QCD approaches an anisotropic pure gauge theory~\bcite{Miransky:2002rp}, which is expected to undergo a first-order deconfining phase transition~\bcite{Cohen:2013zja}. This anisotropic gauge theory has been simulated on the lattice and a first-order transition has indeed been identified~\bcite{Endrodi:2015oba}. This pushed the community to understand the behavior of hot QCD matter at very strong magnetic fields. 
The results for the phase diagram for $eB\le 1\textmd{ GeV}^2$~\bcite{Bali:2011qj} have been extended to $eB\approx 3.25\textmd{ GeV}^2$~\bcite{Endrodi:2015oba} and later up to $eB=9\textmd{ GeV}^2$~\bcite{DElia:2021yvk} using the stout-improved staggered action at one lattice spacing. 
At such extremely strong magnetic fields the crossover transition was observed to become more and more abrupt~\bcite{Endrodi:2015oba} and eventually, indeed turn to a first-order phase transition~\bcite{DElia:2021yvk}.
The existence of a critical endpoint in the QCD phase diagram at strong magnetic fields is particularly interesting as it is expected to induce similar critical behavior in its vicinity as the conjectured critical endpoint at nonzero $\muB$. There are indications that the magnetic critical endpoint is continuously connected to the so-called Roberge-Weiss phase transition at imaginary baryon chemical potential~\bcite{DElia:2025ybj}. We note that the above findings regarding magnetic catalysis and inverse magnetic catalysis have also been recently generalized to inhomogeneous magnetic fields~\bcite{Brandt:2023dir}.

The initial stage of heavy-ion collisions exhibits not only magnetic but also equally strong electric fields $E$. The consistent treatment of background electric fields in studies of QCD thermodynamics is, however, complicated by two independent issues, making these substantially more involved than the case at $B>0$. First, a homogeneous plasma of charged constituents is inevitably driven out of equilibrium by an electric field, and equilibration in the presence of $E$ leads to an inhomogeneous charge profile~\bcite{Endrodi:2022wym,Endrodi:2026kmb}. This inhomogeneous distribution arises at nonzero temperature from the polarization of thermal fluctuations in the medium. This macroscopic charge displacement, quantified by quark number susceptibilities, is desirably separated from the quantum effect of the electric field at microscopic scales~\bcite{Endrodi:2026kmb}. Second, electric fields render the Dirac determinants in~\eqref{eq:partfunc} complex as we pointed out already in Sec.~\ref{sec-EB:lattice}. On the lattice, this necessitates either simulations at imaginary electric fields or a Taylor expansion in $E$. In fact, simulations with imaginary electric fields at nonzero $T$ even face a third problem related to a discontinuous behavior of thermodynamic observables at $E=0$. This stems from the fact that any nonzero electric field expels electric charges from the system, enforcing it to be electrically neutral globally -- unlike $E=0$ situation, where global charge fluctuations are allowed~\bcite{Endrodi:2023wwf}. 
The first lattice determination of the phase diagram was performed in~\bcite{Endrodi:2023wwf} through the Taylor expansion approach, using the stout-improved staggered lattice action and a continuum extrapolation. The results reveal an increase in $T_c(E)$ to leading order in $E$, as shown in the right panel of Fig.~\ref{chap1:pd2}.

\subsection{Equation of state and electromagnetic susceptibilities}
\label{sec-EB:chixi}

Relevant for the physical applications mentioned in Sec.~\ref{sec-EB:intro_pd_eos} is the EoS, which can be derived from the thermodynamic potential $\Omega =-T\log\Z$. For most applications involving background electromagnetic fields $F_{\nu\rho}$, the leading-order behavior of $\Omega$ is the most interesting and, due to parity symmetry, this dependence starts at the quadratic order.
In particular, at zero temperature (Euclidean) Lorentz invariance dictates that $F_{\nu\rho}F_{\nu\rho}/2=E^2+B^2$ is the only possible combination that may appear in $\Omega$. At nonzero temperature or nonzero chemical potentials, Lorentz invariance is partially broken. This can be accounted for by including a Lorentz vector $u_\nu$ that takes the value $u_\nu=\delta_{\nu4}$ in the rest frame of the medium. Now the Lorentz scalar $F_{\nu\rho}F_{\nu\alpha}u_\rho u_\alpha=E^2$ may appear separately in $\Omega$. In practice, this implies different possible coefficients for the leading magnetic and electric responses, which are quantified by the magnetic and electric susceptibilities,
\begin{equation}
\chi = -\frac{1}{V}\left.\frac{\partial^2 \Omega}{\partial (eB)^2}\right|_{B=0}\,,\qquad
\xi = -\frac{1}{V}\left.\frac{\partial^2 \Omega}{\partial (eE)^2}\right|_{E=0}\,.
\label{eq:susc1}
\end{equation}
These observables are subject to additive renormalization, e.g.\ by means of a zero-temperature subtraction so that the renormalized susceptibilities read $\chi^r=\chi(T)-\chi(T=0)$ and $\xi^r=\xi(T)-\xi(T=0)$. The need for multiplicative renormalization is avoided when we take the derivative with respect to the renormalization group invariant combinations $eB$ and $eE$, see~\bcite{Endrodi:2024cqn} for details.

Numerous methods have been developed in the last decade to calculate the susceptibilities on the lattice~\bcite{Levkova:2013qda,Bonati:2013lca,Bonati:2013vba,Bali:2013esa,Bali:2013owa,Bali:2014kia,Bali:2015msa,Bali:2020bcn,Brandt:2024blb}. One possibility is to determine the thermodynamic potential at nonzero values of the background field (in the electric case, one must resort to using imaginary electric fields to avoid the complex action problem) and differentiate numerically subsequently. An alternative is the Taylor expansion of $\Omega$ around $E=B=0$. This is somewhat complicated by the flux quantization condition~\eqref{eq:quantization}, which renders homogeneous magnetic and imaginary electric fields discrete variables. To carry out the derivatives in~\eqref{eq:susc1}, one can instead work with oscillatory fields, e.g.\ $B(x_1)=B \cos(p_1x_1)$ with zero flux and carry out the extrapolation $p_1\to0$ subsequently. In fact, this extrapolation may be implemented directly in the measurements of the susceptibilities.
A summary of lattice techniques to calculate $\chi$ and $\xi$ can be found in~\bcite{Endrodi:2024cqn}. 

\begin{figure}[t]
\centering
\includegraphics[height=5.cm]{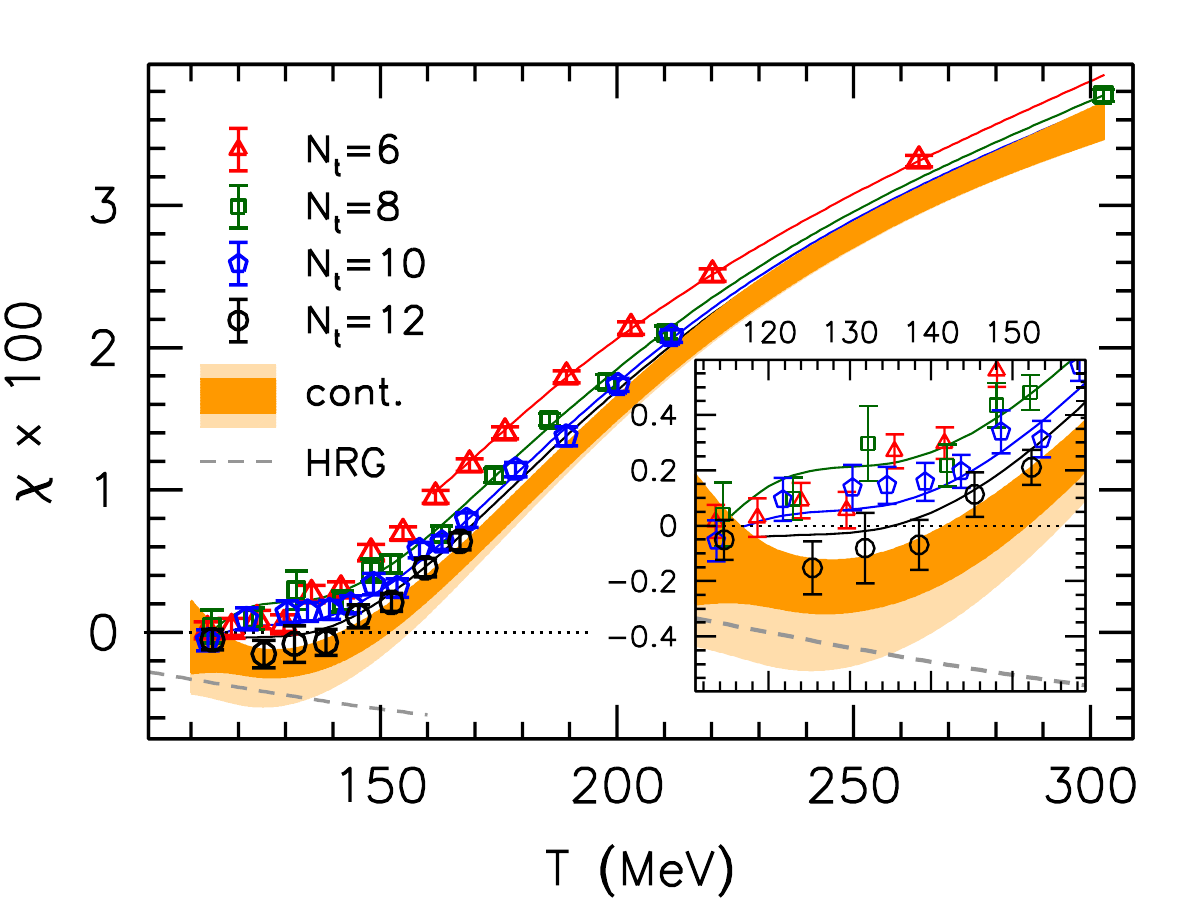}\qquad
\includegraphics[height=5.cm]{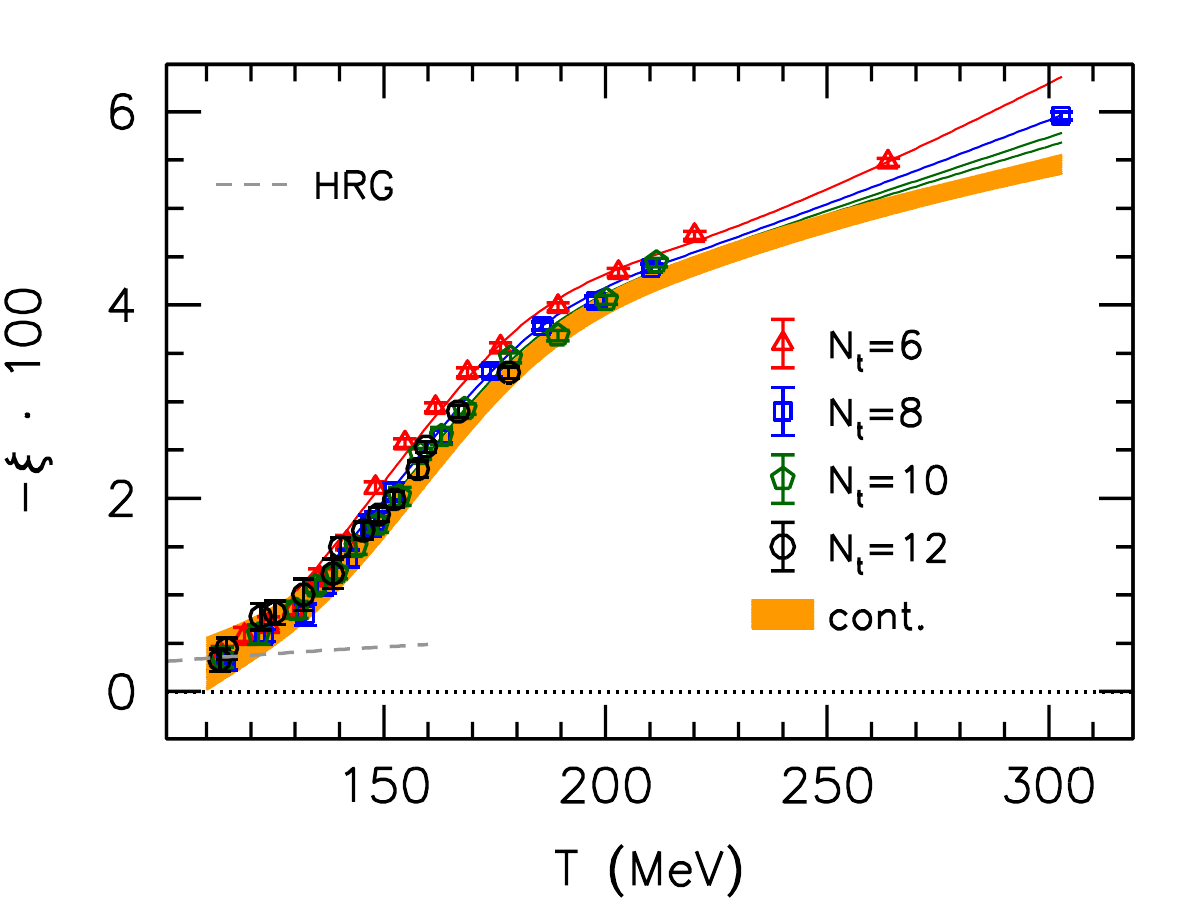}
\caption{{\it Left panel:} Magnetic susceptibility of strongly interacting matter as a function of the temperature obtained using different lattice spacings (colored points) and the continuum extrapolation (orange band), together with the HRG model prediction (dashed gray line)~\bcite{Bali:2020bcn}. {\it Right panel:} The negative of the electric susceptibility on the lattice, extrapolated to the continuum limit~\bcite{Endrodi:2023wwf}, compared to the HRG model prediction~\bcite{Endrodi:2026kmb}, with the same color coding as the left panel. Figure adapted from~\bcite{Endrodi:2023wwf}.}
\label{fig:susc1}
\end{figure}

The latest results for the renormalized magnetic and electric susceptibilities, employing the above explained variant of the Taylor expansion, are shown in Fig.~\ref{fig:susc1}. Here, simulations with the stout-improved staggered lattice action at four different lattice spacings are used to extrapolate to the continuum limit. For the magnetic susceptibility, one observes strong paramagnetism ($\chi^r>0$) at high temperatures, while a weak diamagnetic region ($\chi^r<0$) appears slightly below $T_c$~\bcite{Bali:2020bcn}. This may be understood by considering the magnetic response of the effective degrees of freedom. At low $T$, these are charged pions, which couple to the magnetic field via their angular momentum, a response inherently diamagnetic due to Lenz's law. In turn, at high $T$ the medium consists of quasi-free quarks, which couple to $B$ both via angular momentum and spin and the latter contribution is known to dominate (Pauli paramagnetism). 
The spin term can be directly given in terms of the tensor bilinear observable $\ev{\bar\psi\sigma_{\nu\rho}\psi}$~\bcite{Bali:2012jv,Bali:2020bcn}.
Interestingly, this term turns out to be related to the normalization of the leading-twist photon distribution amplitude, relevant for photon to quark-antiquark dissociation and radiative heavy meson decays~\cite{Rohrwild:2007yt}.
The separation of the susceptibility to spin- and orbital angular momentum-related contributions has been investigated in detail in~\bcite{Bali:2012jv,Bali:2020bcn}.

For the electric susceptibility, the impact of the macroscopic charge displacement is removed, as mentioned at the end of Sec.~\ref{sec-EB:mcimc}.
In contrast to the magnetic susceptibility, $\xi^r$ has the same sign for all temperatures. The negative value indicates that charges originating from thermal fluctuations generate a polarization that opposes the background electric field~\bcite{Endrodi:2026kmb}. The behavior of both $\chi^r$ and $\xi^r$ at low temperatures can be explained quantitatively in terms of hadronic degrees of freedom, using the hadron resonance gas (HRG) model~\bcite{Endrodi:2013cs,Bali:2020bcn,Endrodi:2026kmb}. The HRG prediction compares very well to the continuum extrapolated lattice results, as revealed in Fig.~\ref{fig:susc1}.

The behavior of $\Omega(B)$ and $\Omega(E)$ is expected to be captured well by the susceptibilities for sufficiently weak fields. Quantitatively, it was found that for that magnetic case the $\mathcal{O}(B^2)$ term dominates for $eB\lesssim 0.3\textmd{ GeV}^2$~\bcite{Bali:2020bcn}. Going beyond that requires either a higher-order Taylor expansion or -- in the magnetic case -- explicit calculations of $\Omega(B)$. The latter was carried out in~\bcite{Bali:2014kia} with stout-improved staggered quarks using three lattice spacings. The corresponding continuum estimate for $\Omega(B)$ can be used to calculate all thermodynamic observables like the energy density, the entropy density, the pressure, or the speed of sound. 
Concerning the pressure, magnetic fields may lead to an anisotropy between parallel $p_3$ and perpendicular $p_1=p_2$ pressures. This notion, however, depends on the precise definition of this observable. As was discussed in~\bcite{Bali:2013esa}, one must distinguish  between two situations: when the pressure is defined via a compression of the system with the magnetic field constant (the so-called $B$-scheme), and the compression with the magnetic flux being constant ($\Phi$-scheme). In the $B$-scheme, the pressure is isotropic, $p_1^{(B)}=p_2^{(B)}=p_3^{(B)}$. In contrast, in the $\Phi$-scheme, a compression perpendicular to the magnetic field also pushes the magnetic field lines together, resulting in a pressure anisotropy, $p_1^{(\Phi)}=p_2^{(\Phi)}\neq p_3^{(\Phi)}$. The latter scheme is often employed in neutron star models~\bcite{Ferrer:2010wz}. Further details of the pressure anisotropy and its lattice determination are summarized in~\bcite{Endrodi:2024cqn}.

\subsection{Dense and magnetized QCD matter}
\label{sec:dense-mag-qcd}

Next, we turn to the discussion of phenomena that arise under the combined impact of background magnetic fields and density. This setting is particularly interesting for magnetars as well as for heavy-ion collision phenomenology.

\begin{figure}[t]
\centering
\includegraphics[height=4.6cm]{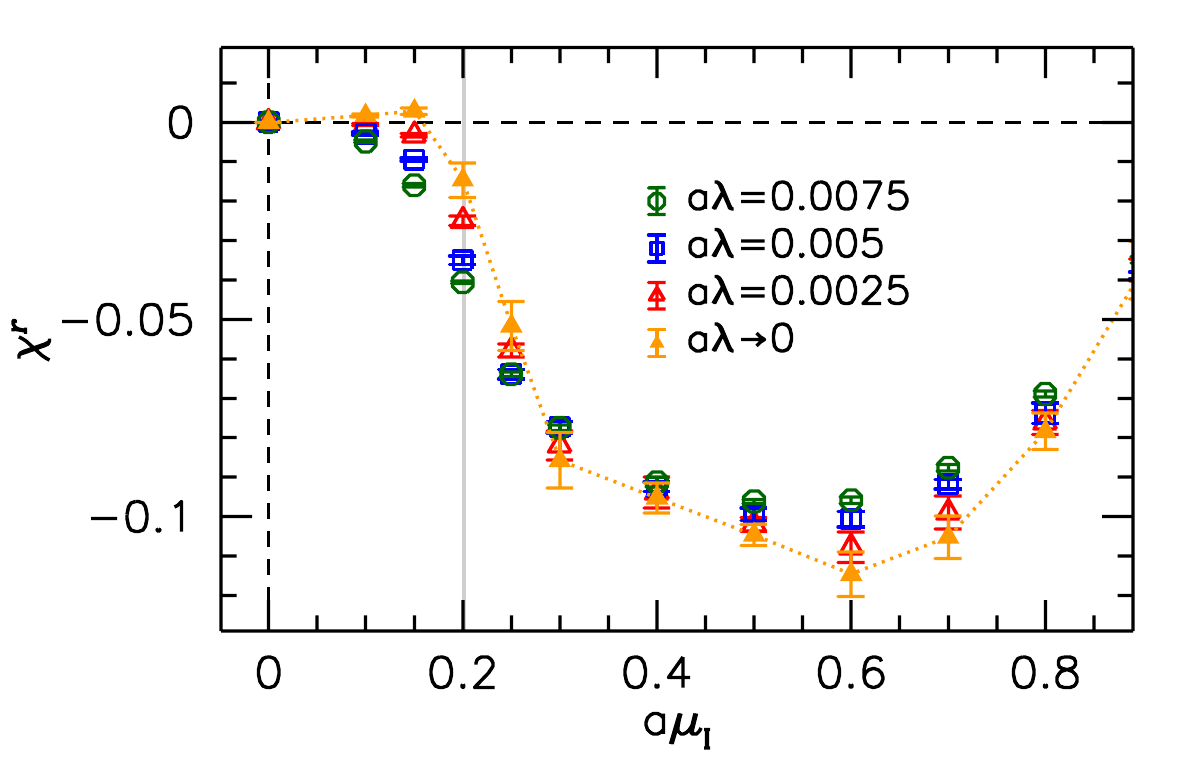} \qquad
\raisebox{.15cm}{\includegraphics[height=4.6cm]{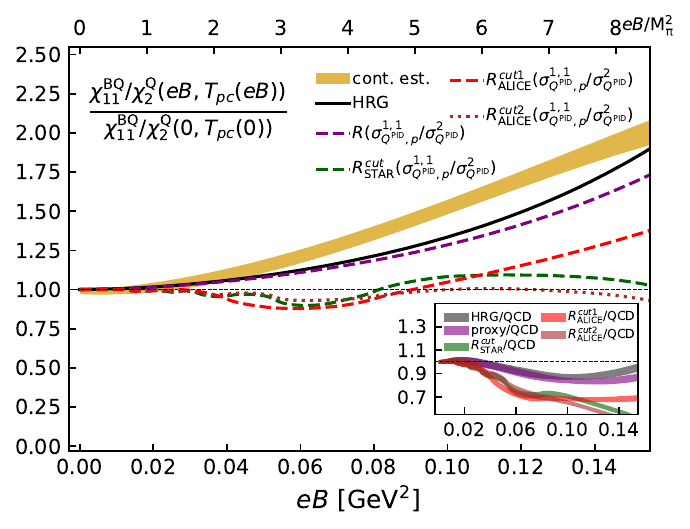}}
\caption{
{\it Left panel:}
Renormalized magnetic susceptibility as a function of the isospin chemical potential for different values of the pion source parameter $\lambda$ (green, blue and red points) and the estimate of the $\lambda\to0$ limit (yellow). The vertical line indicates the onset of pion condensation. Figure taken from~\bcite{Endrodi:2014lja}.
{\it Right panel:}
Double ratio of quark number susceptibilities involving baryon number and electric charge, argued to act as a magnetometer of the strongly interacting medium~\bcite{Ding:2025jfz}.
\label{fig:denseB}}
\end{figure}

At nonzero isospin density or, equivalently nonzero isospin chemical potential, the inclusion of electromagnetic fields renders the action complex as we discussed already in Sec.~\ref{sec-EB:lattice}. To overcome this issue, in~\bcite{Endrodi:2014lja} a Taylor expansion in $B$ was carried out for wide range of $\muI$ values at low temperature, employing field profiles with zero flux (a step function profile with $B>0$ in one half and $B<0$ in the other half of the lattice~\bcite{Levkova:2013qda}). This exploratory work used unimproved staggered quarks at one lattice spacing and three different values of the pion source parameter $\lambda$, together with an extrapolation $\lambda\to0$. The results, shown in the left panel of Fig.~\ref{fig:denseB} reveal the following qualitative picture: the QCD vacuum does not respond to the magnetic field as long as $\muI<m_\pi$ -- yet another manifestation of the Silver-Blaze phenomenon~\bcite{Cohen:2003kd}, which we already encountered in Secs.~\ref{sec-EB:pionisospin} and~\ref{sec-EB:isospin-zeroT}. In turn, as soon as particles appear in the medium, the magnetic field can polarize them. At $\muI\ge m_\pi$ the system is in the pion condensed phase and pions respond to $B$ via their orbital angular momentum as we just saw in Sec.~\ref{sec-EB:chixi}. The diamagnetic nature of this coupling gives rise to the large negative values of $\chi^r$ in this phase. In fact, in the infinite volume limit, where the step function profile can become a true homogeneous magnetic field, $\chi^r$ is expected to approach $-\infty$, physically realizing the Meissner effect: weak magnetic fields are completely expelled from the interior of superconductors. The impact of the force exerted through this magnetic effect on neutron star matter has been estimated in~\bcite{Endrodi:2014lja}.

Turning to heavy-ion collision phenomenology, a key concept is that of a so-called magnetometer: a physical observable that is very sensitive to the background magnetic field and is amenable to detection (through a proxy) in the experiment. Constraining the value of the magnetic field in the initial stages of the collision is important for estimating the strength of further, magnetic field-induced effects like anomalous transport phenomena~\bcite{Kharzeev:2013ffa}, or effects on the anisotropic flow~\bcite{Pang:2016yuh}.

Recent lattice QCD simulations have revealed that promising candidates for the role of the magnetometer can be constructed using quark number susceptibilities,
\begin{equation}
\label{eq:qnsusc}
\chi_{ijk}^{\B\Q\S}=-\frac{1}{V}\left.\frac{\partial^{i+j+k} \Omega/T^4}{\partial (\mu_\B/T)^i\,\partial (\muQ/T)^j\,\partial (\muS/T)^k}\right|_{\muB=\muQ=\muS=0}\,,
\end{equation}
defined as derivatives in the space spanned by the baryon number, electric charge and strangeness chemical potentials, see the discussion around~\eqref{eq:BIS_basis}.
In particular, in~\bcite{Ding:2021cwv} it was argued that the ratio of $\chi_{11}^{\B\Q}\equiv\chi_{110}^{\B\Q\S}$ and $\chi_2^\Q\equiv\chi_{020}^{\B\Q\S}$ is one of the most sensitive combinations. This observable was calculated using the HISQ quark action in~\bcite{Ding:2021cwv} for larger-than-physical quark masses and in~\bcite{Ding:2023bft} for physical quark masses. The results for this ratio, normalized by its $B=0$ value, is shown for weak magnetic fields in the right panel of Fig.~\ref{fig:denseB}. We note that the comparison to experiments is in practice carried out via the HRG model, and the generalization of the model to nonzero $B$~\bcite{Endrodi:2013cs} involves complications that have been discussed recently~\bcite{Vovchenko:2024wbg,Marczenko:2024kko}.

Finally, in dense and magnetized environments, there is a further remarkable effect that arises in QCD matter: the so-called chiral separation effect (CSE). The CSE is the generation of an axial vector current $J_{\nu 5}$ due to the magnetic field $B$ and a baryon chemical potential $\muB$~\bcite{Son:2004tq,Metlitski:2005pr}. It is often discussed analogously to the chiral magnetic effect (CME), the generation of a vector current $J_\nu$ due to the magnetic field and a chiral chemical potential $\mu_5$~\bcite{Fukushima:2008xe},
\begin{equation}
\ev{J_{35}} = C_{\rm CSE} \, eB \, \muB, \qquad
\ev{J_{3}} = C_{\rm CME} \, eB \, \mu_5\,.
\end{equation}
These effects, commonly referred to as anomalous transport phenomena, are most relevant to heavy-ion collision phenomenology, where they are expected to give experimentally observable signatures that connect electric charge separation and chirality of quarks -- or, via the axial anomaly, the topology of gluon fields.

\begin{figure}[t]
\centering
\includegraphics[height=4.6cm]{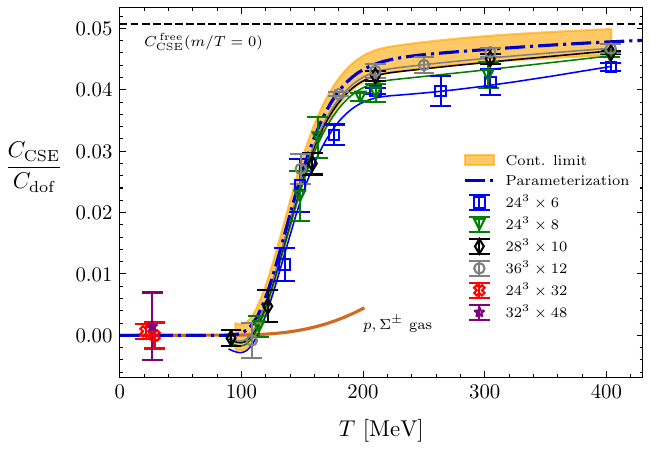} \qquad
\raisebox{-.15cm}{\includegraphics[height=5.1cm]{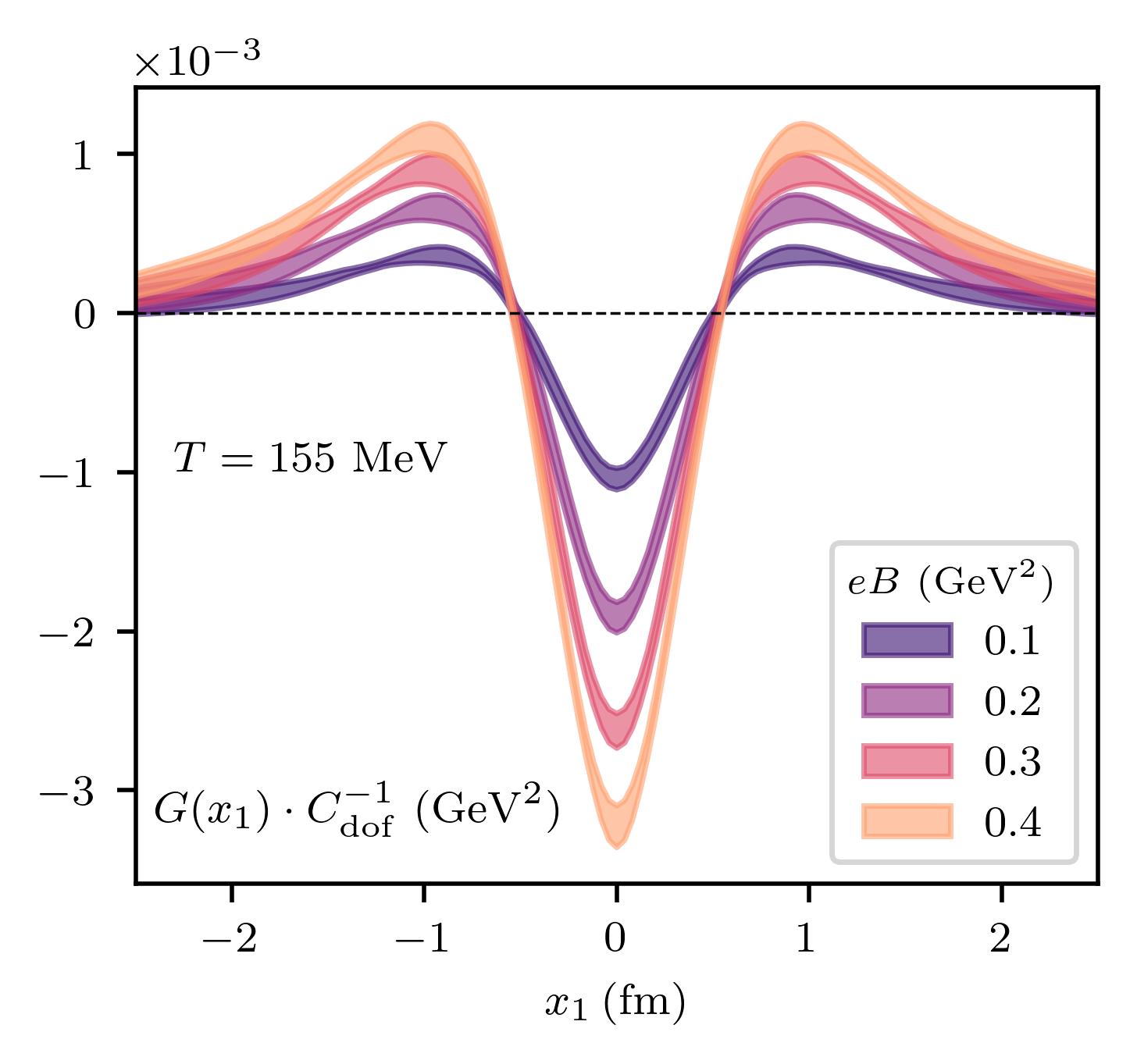}}
\caption{
{\it Left panel:} The CSE conductivity coefficient as a function of the temperature obtained using different lattice spacings (colored points) and a continuum extrapolation (yellow band), compared to the baryon gas model prediction (solid brown line)~\cite{Brandt:2023wgf}. The normalization factor reads $C_{\rm dof}=\sum_f (q_f/e)^2=2/3$.
{\it Right panel:} Localized chiral magnetic currents in equilibrium, generated by an inhomogeneous magnetic field with different magnitudes (colored curves). The normalization factor in this case is $C_{\rm dof}=N_c\sum_f (q_f/e)^2=2$.
\label{fig:cse}}
\end{figure}

Following the initial studies~\bcite{Puhr:2016kzp,Buividovich:2020gnl}, the conductivity coefficient $C_{\rm CSE}$ has been determined using in-equilibrium lattice simulations with continuum extrapolated stout-smeared staggered quarks at physical masses in~\bcite{Brandt:2023wgf}. The results are shown in the left panel of Fig.~\ref{fig:cse} for a range of temperatures, showing the suppression of this coefficient at low $T$, the gradual rise in the transition region and the approach to the high temperature, limit, which can be obtained employing free massless quarks. 

The analogous calculation for the in-equilibrium CME turns out to be very subtle as it requires a careful ultraviolet regularization, compatible with lattice Ward identities. When a discretization with a proper, conserved vector current is used, the conductivity coefficient vanishes, $C_{\rm CME}=0$~\bcite{Brandt:2024wlw} for homogeneous magnetic fields. Studies with using a non-conserved vector current erroneously give a nonzero result~\bcite{Yamamoto:2011gk,Yamamoto:2011ks}. For inhomogeneous magnetic fields, the in-equilibrium response $\ev{J_3(x_1)}$ develops a non-trivial spatial structure, as visualized in the right panel of Fig.~\ref{fig:cse}. Here, the magnetic field is assumed to have the profile $\bm B(x_1)=B\, \bm e_3 / \cosh^2(x_1/\epsilon)$, with $\epsilon\approx0.6\textmd{ fm}$ a typical width relevant for off-central heavy-ion collisions. As visible from the figure, the generated current changes sign in the transverse direction, with the total integral remaining zero. These results may guide experimental searches for signatures of the chiral magnetic effect in heavy-ion collisions.

\section{Summary, applications and open questions}
\label{sec-EB:applications}

Scrutinizing the properties of strongly interacting matter under extreme conditions has been one of the major endeavours in experimental and theoretical high-energy nuclear and particle physics in the past decades.
In this review we have focused on the phenomenologically interesting regions of the parameter space, where direct non-perturbative results can be obtained from lattice QCD, supported by predictions from chiral perturbation theory at low energies. These include QCD at pure isospin chemical potential ($\muI\neq0$ but $\muB=\muS=0$) as well as QCD in external magnetic fields. For the former, we have discussed the results concerning the onset of a phase with a Bose-Einstein condensate of charged pions, as well as the EoS in this phase. The latter involves remarkable features such as an excess of the speed of sound and a strong suppression of the interaction measure, both crossing through the associated conformal values, previously often considered as bounds for these quantities. In the case of external magnetic fields, we have discussed magnetic catalysis at low temperatures, as well as inverse magnetic catalysis in the vicinity of the transition temperature. A related feature is the decrease in the transition temperature with the magnetic field and the strengthening of the transition, reaching a critical endpoint at very strong field strength values. Furthermore, we discussed the EoS and the associated electromagnetic susceptibilities, as well as how nonzero isospin or baryon densities affect magnetized QCD matter.

Apart from these features, there are many aspects of magnetized and isospin-asymmetric QCD that still demand further investigations in future studies:

{\bf\boldmath BCS phase and pairing gaps at large $\muI$:}
As already mentioned in Sec.~\ref{sec-EB:intro_pd_eos}, at large $\muI$ perturbation theory predicts the presence of a BCS-type superconducting phase~\bcite{Son:2000by} with pseudoscalar $\bar{u}d$ Cooper pairs, separated from the BEC phase by a continuous crossover. 
While a number of different signatures for the identification of the BCS phase have been proposed, including conformal points~\bcite{Carignano:2016rvs}, the specific behavior of the constituent-quark mass~\bcite{Adhikari:2018cea}, or a characteristic scaling of the density, as in two-color QCD~\bcite{Cotter:2012mb}, one of the main difficulties is to develop observables which can unambiguously differentiate between the BEC and BCS superconducting phases and are directly accessible to lattice QCD.
Some first signs for the presence of a pairing gap of the size predicted in perturbative QCD~\bcite{Fujimoto:2023mvc}, which is enhanced compared to the gap for color superconductivity, have been obtained in Ref.~\bcite{Abbott:2024vhj} by comparing the EoS to the perturbative prediction including the gap. However, direct attempts to see the presence of a gap at smaller $\muI$~\bcite{Brandt:2019hel}, using a Banks-Casher type relation~\bcite{Kanazawa:2013crb}, so far remain inconclusive.

{\bf\boldmath BEC phase boundary at large $\muI$:}
A related question concerns the behavior of the phase boundary of the BEC (or BCS) phase at large $\muI$ and nonzero $T$. A line of first order deconfinement transitions in the BEC phase at asymptotically large $\muI$, ending in a critical endpoint, has been proposed already in Ref.~\bcite{Son:2000xc,Son:2000by}. This picture is consistent with the findings in an effective theory for this regime~\bcite{Cohen:2015soa}. Whether the picture of two different transitions for deconfinement and the melting of the condensate is accurate has so far not been tested in first-principles simulations. While simulations with unimproved staggered quarks, unphysical quark masses and on coarse lattices have seen signs for the onset of a first-order transition on the BEC phase boundary~\bcite{Kogut:2004zg}, a systematic analysis of the order of the transition is desirable to draw final conclusions.

{\bf\boldmath Phase diagram at large $\muI$ and small $\muB$ or $\muS$:}
A particularly interesting question about the multi-dimensional phase diagram concerns the prolongation of the BEC phase boundary towards non-zero $\muB$ and/or $\muS$. At vanishing temperature, the Silver-Blaze property and the expected onset of pion condensation at 
$\muI=m_\pi$ is suggestive of 
a phase boundary perpendicular to the isospin axis in the $\muB$ and $\muS$ directions
until the corresponding excitation energy is reached.
On the lattice, the phase boundary is only accessible indirectly, for instance through Taylor expansion~\bcite{Brandt:2024dle}. As observed in this work, a major problem for the extraction of the expansion coefficients is the increase of uncertainties in the $\lambda\to0$ extrapolations within the BEC phase, rendering the study of this phase boundary challenging. 

{\bf Location and features of the magnetic critical endpoint:}
The critical endpoint in the $B-T$ plane is expected to have similar features as the conjectured endpoint in the $\muB-T$ plane. This gives lattice practitioners the opportunity to investigate the critical behavior in this, sign problem-free, setting. From the theory point of view, an interesting open question concerns chiral symmetry restoration and the behavior of Dirac eigenvalues near the critical point. A further intriguing question is whether the magnetic critical point is continuously connected to the Roberge-Weiss endpoint at imaginary chemical potentials.

{\bf Out-of-equilibrium conductivities:}
The calculation of static susceptibilities, like $\chi$ and $\xi$ discussed in this chapter, should be extended to dynamic conductivities relevant for out-of-equilibrium processes. For the electric conductivity, the impact of the magnetic fields has already been assessed on the lattice~\bcite{Astrakhantsev:2019zkr}.
Future lattice QCD simulations should use similar spectral reconstruction techniques to calculate the chiral magnetic conductivity as well, see~\bcite{Brandt:2025now} for a first result. The complete understanding of this effect will require input from chiral hydrodynamics involving conserved vector as well as anomalous axial currents.

Apart from enhancing our understanding of the properties of strongly interacting matter under extreme conditions and the phases present in the multi-dimensional parameter space, the results presented in this review are also of direct relevance for a number of physical systems, which we will now briefly summarize. Furthermore, we will comment on possible further applications which might allow progress in currently inaccessible regions of the parameter space.

{\bf The early Universe at large lepton flavor asymmetries:}
As mentioned in the introduction, the early Universe might have featured a lepton flavor asymmetry, considerably larger than the matter/antimatter asymmetry~\bcite{Schwarz:2009ii}. Moreover, the observed bound on the lepton asymmetry constrains the sum of lepton flavor asymmetries $l_\alpha$~\bcite{Oldengott:2017tzj}, $\vert l_e + l_\mu + l_\tau \vert < 0.012$, so that the individual $l_\alpha$ might be even larger with a cancellation taking place~\bcite{Wygas:2018otj,Middeldorf-Wygas:2020glx}. As shown in Refs.~\bcite{Wygas:2018otj,Middeldorf-Wygas:2020glx,Vovchenko:2020crk},
these lepton flavor asymmetries lead to large $\muQ$, or equivalently $\muI$, so that the early Universe might have passed through the BEC phase. The EoS at $\muI\neq0$ is an essential input to studies of this kind, since it allows to trace the path of the Universe within the BEC phase, where indirect methods like Taylor expansion break down.

{\bf Gravitationally stable self-bound pion condensates -- pion stars:}
It is interesting to note that the EoS at large $\muI$, when used as input for the Tolman-Oppenheimer-Volkoff (TOV) equations, allows for gravitationally self-bound pion condensates, as first noted in Ref.~\bcite{Carignano:2016lxe}. The properties of the resulting boson stars, dubbed pion stars, have been further worked out in Ref.~\bcite{Brandt:2018bwq} (see also~\bcite{Andersen:2018nzq,Andersen:2022aig,Stashko:2023gnn,Chen:2024cxh}), where the decay rate of the pion condensate due to the weak interactions has also been calculated.

{\bf Bridging the gap between theory and experiment for magnetic field-induced phenomena in heavy-ion collisions:}
To connect first-principles lattice QCD results at nonzero magnetic fields to heavy-ion collision phenomenology, it is necessary to benchmark the initial magnetic field generated in the early stages of the collisions. The magnetometer observables discussed in Sec.~\ref{sec:dense-mag-qcd}, together with their matching through HRG model calculations, should be understood in more detail. In particular, with such a matching in place, lattice QCD-based predictions for observables related to the chiral magnetic effect should be implemented in hydrodynamic simulations in order to guide the search for CME signatures in the experiment.

{\bf Phase-quenched QCD and constraints on the nuclear EoS:}
For two mass-degenerate quark flavors, the partition function at $\muI\neq0$ provides an upper bound for the same partition function with a pure baryon chemical potential of $\muB=3\muI/2$. 
This is because the former system corresponds to the system at $\muB\neq0$ after neglecting the complex phase of the determinants, often referred to as the phase-quenched (PQ) theory~\bcite{Cohen:2003ut}.
Using the standard relation between the partition function and the pressure, one therefore arrives at the inequality $p_{\rm QCD}\leq p_{\rm PQ}$. 
Furthermore, in perturbation theory the difference between the PQ theory and QCD is given by a diagram of $\mathcal{O}(\alpha_s^3)$~\bcite{Moore:2023glb} with a  small prefactor~\bcite{Navarrete:2024zgz}, suggesting that the pressure difference is suppressed at large $\muB$.
One particular case where exact and strict bounds on the EoS may be very useful is in the model-agnostic sampling of the neutron star EoS. In this case, where $\mu_u<\mu_d$ to ensure charge neutrality, additional inequalities may also be derived~\bcite{Lee:2004hc,Fujimoto:2023unl},
and the PQ setup can be generalized to different chemical potentials for each quark flavor~\bcite{Moore:2023glb}. The resulting bounds on the neutron star EoS using results at $T=0$ and $\muI\neq0$ have been investigated in Ref.~\bcite{Abbott:2024vhj} and their potential impact on sampling in~\bcite{Fujimoto:2024pcd}. 
In addition, it has recently been argued that the same bounds might also be useful in the $T\neq0$ case~\bcite{Gorda:2025cwu} and alternative types of bounds related to imaginary chemical potentials have recently been suggested~\bcite{Cohen:2025ahp}. The efficacy and utility of these bounds need to be studied in detail.

{\bf Benchmarking and parameter tuning for functional methods and models:}
Apart from their direct relevance for physical systems, both the phase diagram 
and the EoS can be used to benchmark and tune low-energy models of QCD. In turn, the latter can be used to investigate the theory in parameter spaces, where the complex action problem hinders direct simulations. 
In fact, the lattice results on the $T-B$ phase diagram have for long served as benchmarks for a range of such effective models, see e.g.~\bcite{Andersen:2014xxa} for a review.
The situation is similar in the $\muI>0$ case: while a review of model results is beyond the scope of this chapter, 
one particular example where the tuning with lattice QCD data has been very beneficial is in a renormalization group invariant formulation of the quark-meson model~\bcite{Adhikari:2018cea,Folkestad:2018psc,Brandt:2025tkg}, which leads to very good agreement with the phase diagram and EoS for all values of the isospin chemical potentials~\bcite{Brandt:2025tkg}.

\begin{ack}[Acknowledgments]

This work was supported by the Deutsche Forschungsgemeinschaft (DFG, German Research
Foundation) through CRC-TR 211 - project number 315477589. 
GE acknowledges funding by the Hungarian National Research, Development and Innovation Office - NKFIH (Research Grant Hungary 150241) and the European Research Council (Consolidator Grant 101125637 CoStaMM).
Views and opinions expressed are however those of the authors only and do not necessarily reflect those of the European Union or the European Research Council. Neither the European Union nor the granting authority can be held responsible for them.
BB also received support by MKW NRW under the funding code NW21-024-A.
\end{ack}

\begin{thebibliography*}{100}
\providecommand{\bibtype}[1]{}
\providecommand{\url}[1]{{\tt #1}}
\providecommand{\urlprefix}{URL }
\expandafter\ifx\csname urlstyle\endcsname\relax
  \providecommand{\doi}[1]{doi:\discretionary{}{}{}#1}\else
  \providecommand{\doi}{doi:\discretionary{}{}{}\begingroup
  \urlstyle{rm}\Url}\fi
\providecommand{\bibinfo}[2]{#2}
\providecommand{\eprint}[2][]{\url{#2}}
\makeatletter\def\@biblabel#1{\bibinfo{label}{[#1]}}\makeatother

\bibtype{Article}%
\bibitem{Oldengott:2017tzj}
\bibinfo{author}{Isabel~M. Oldengott}, \bibinfo{author}{Dominik~J. Schwarz},
  \bibinfo{title}{{Improved constraints on lepton asymmetry from the cosmic
  microwave background}}, \bibinfo{journal}{EPL} \bibinfo{volume}{119}
  (\bibinfo{number}{2}) (\bibinfo{year}{2017}) \bibinfo{pages}{29001},
  \bibinfo{doi}{\doi{10.1209/0295-5075/119/29001}}, \eprint{1706.01705}.

\bibtype{Article}%
\bibitem{Duncan:1992hi}
\bibinfo{author}{Robert~C. Duncan}, \bibinfo{author}{Christopher Thompson},
  \bibinfo{title}{{Formation of very strongly magnetized neutron stars -
  implications for gamma-ray bursts}}, \bibinfo{journal}{Astrophys. J. Lett.}
  \bibinfo{volume}{392} (\bibinfo{year}{1992}) \bibinfo{pages}{L9},
  \bibinfo{doi}{\doi{10.1086/186413}}.

\bibtype{Article}%
\bibitem{Skokov:2009qp}
\bibinfo{author}{V. Skokov}, \bibinfo{author}{A.~Yu. Illarionov},
  \bibinfo{author}{V. Toneev}, \bibinfo{title}{{Estimate of the magnetic field
  strength in heavy-ion collisions}}, \bibinfo{journal}{Int. J. Mod. Phys. A}
  \bibinfo{volume}{24} (\bibinfo{year}{2009}) \bibinfo{pages}{5925--5932},
  \bibinfo{doi}{\doi{10.1142/S0217751X09047570}}, \eprint{0907.1396}.

\bibtype{Article}%
\bibitem{Grasso:2000wj}
\bibinfo{author}{Dario Grasso}, \bibinfo{author}{Hector~R. Rubinstein},
  \bibinfo{title}{{Magnetic fields in the early universe}},
  \bibinfo{journal}{Phys. Rept.} \bibinfo{volume}{348} (\bibinfo{year}{2001})
  \bibinfo{pages}{163--266},
  \bibinfo{doi}{\doi{10.1016/S0370-1573(00)00110-1}},
  \eprint{astro-ph/0009061}.

\bibtype{Article}%
\bibitem{Fischer:2026uni}
\bibinfo{author}{Christian~S. Fischer}, \bibinfo{author}{Jan~M. Pawlowski},
  \bibinfo{title}{{Phase structure and observables at high densities from first
  principles QCD}}  (\bibinfo{year}{2026}), \eprint{2603.11135}.

\bibtype{Article}%
\bibitem{Allton:2002zi}
\bibinfo{author}{C.~R. Allton}, \bibinfo{author}{S. Ejiri},
  \bibinfo{author}{S.~J. Hands}, \bibinfo{author}{O. Kaczmarek},
  \bibinfo{author}{F. Karsch}, \bibinfo{author}{E. Laermann},
  \bibinfo{author}{C. Schmidt}, \bibinfo{author}{L. Scorzato},
  \bibinfo{title}{{The QCD thermal phase transition in the presence of a small
  chemical potential}}, \bibinfo{journal}{Phys. Rev. D} \bibinfo{volume}{66}
  (\bibinfo{year}{2002}) \bibinfo{pages}{074507},
  \bibinfo{doi}{\doi{10.1103/PhysRevD.66.074507}}, \eprint{hep-lat/0204010}.

\bibtype{Article}%
\bibitem{Wygas:2018otj}
\bibinfo{author}{Mandy~M. Wygas}, \bibinfo{author}{Isabel~M. Oldengott},
  \bibinfo{author}{Dietrich B{\"o}deker}, \bibinfo{author}{Dominik~J. Schwarz},
  \bibinfo{title}{{Cosmic QCD Epoch at Nonvanishing Lepton Asymmetry}},
  \bibinfo{journal}{Phys. Rev. Lett.} \bibinfo{volume}{121}
  (\bibinfo{number}{20}) (\bibinfo{year}{2018}) \bibinfo{pages}{201302},
  \bibinfo{doi}{\doi{10.1103/PhysRevLett.121.201302}}, \eprint{1807.10815}.

\bibtype{Article}%
\bibitem{Aoki:2006we}
\bibinfo{author}{Y. Aoki}, \bibinfo{author}{G. Endr\H{o}di},
  \bibinfo{author}{Z. Fodor}, \bibinfo{author}{S.~D. Katz},
  \bibinfo{author}{K.~K. Szab\'o}, \bibinfo{title}{{The Order of the quantum
  chromodynamics transition predicted by the standard model of particle
  physics}}, \bibinfo{journal}{Nature} \bibinfo{volume}{443}
  (\bibinfo{year}{2006}) \bibinfo{pages}{675--678},
  \bibinfo{doi}{\doi{10.1038/nature05120}}, \eprint{hep-lat/0611014}.

\bibtype{Article}%
\bibitem{Bhattacharya:2014ara}
\bibinfo{author}{Tanmoy Bhattacharya}, et al., \bibinfo{title}{{QCD Phase
  Transition with Chiral Quarks and Physical Quark Masses}},
  \bibinfo{journal}{Phys. Rev. Lett.} \bibinfo{volume}{113}
  (\bibinfo{number}{8}) (\bibinfo{year}{2014}) \bibinfo{pages}{082001},
  \bibinfo{doi}{\doi{10.1103/PhysRevLett.113.082001}}, \eprint{1402.5175}.

\bibtype{Article}%
\bibitem{Aarts:2023vsf}
\bibinfo{author}{Gert Aarts}, et al., \bibinfo{title}{{Phase Transitions in
  Particle Physics}: {Results and Perspectives from Lattice Quantum
  Chromo-Dynamics}}, \bibinfo{journal}{Prog. Part. Nucl. Phys.}
  \bibinfo{volume}{133} (\bibinfo{year}{2023}) \bibinfo{pages}{104070},
  \bibinfo{doi}{\doi{10.1016/j.ppnp.2023.104070}}, \eprint{2301.04382}.

\bibtype{Article}%
\bibitem{Endrodi:2015oba}
\bibinfo{author}{Gergely Endr\H{o}di}, \bibinfo{title}{{Critical point in the
  QCD phase diagram for extremely strong background magnetic fields}},
  \bibinfo{journal}{JHEP} \bibinfo{volume}{07} (\bibinfo{year}{2015})
  \bibinfo{pages}{173}, \bibinfo{doi}{\doi{10.1007/JHEP07(2015)173}},
  \eprint{1504.08280}.

\bibtype{Article}%
\bibitem{DElia:2021yvk}
\bibinfo{author}{Massimo D'Elia}, \bibinfo{author}{Lorenzo Maio},
  \bibinfo{author}{Francesco Sanfilippo}, \bibinfo{author}{Alfredo Stanzione},
  \bibinfo{title}{{Phase diagram of QCD in a magnetic background}},
  \bibinfo{journal}{Phys. Rev. D} \bibinfo{volume}{105} (\bibinfo{number}{3})
  (\bibinfo{year}{2022}) \bibinfo{pages}{034511},
  \bibinfo{doi}{\doi{10.1103/PhysRevD.105.034511}}, \eprint{2111.11237}.

\bibtype{Article}%
\bibitem{Brandt:2017oyy}
\bibinfo{author}{B.~B. Brandt}, \bibinfo{author}{G. Endr\H{o}di},
  \bibinfo{author}{S. Schmalzbauer}, \bibinfo{title}{{QCD phase diagram for
  nonzero isospin-asymmetry}}, \bibinfo{journal}{Phys. Rev. D}
  \bibinfo{volume}{97} (\bibinfo{number}{5}) (\bibinfo{year}{2018})
  \bibinfo{pages}{054514}, \bibinfo{doi}{\doi{10.1103/PhysRevD.97.054514}},
  \eprint{1712.08190}.

\bibtype{Article}%
\bibitem{Kaiser:2026msy}
\bibinfo{author}{Norbert Kaiser}, \bibinfo{author}{Wolfram Weise},
  \bibinfo{title}{{Liquid-gas phase transition of nuclear matter}}
  (\bibinfo{year}{2026}), \eprint{2602.09916}.

\bibtype{Article}%
\bibitem{Alford:2007xm}
\bibinfo{author}{Mark~G. Alford}, \bibinfo{author}{Andreas Schmitt},
  \bibinfo{author}{Krishna Rajagopal}, \bibinfo{author}{Thomas Sch{\"a}fer},
  \bibinfo{title}{{Color superconductivity in dense quark matter}},
  \bibinfo{journal}{Rev. Mod. Phys.} \bibinfo{volume}{80}
  (\bibinfo{year}{2008}) \bibinfo{pages}{1455--1515},
  \bibinfo{doi}{\doi{10.1103/RevModPhys.80.1455}}, \eprint{0709.4635}.

\bibtype{Article}%
\bibitem{Buballa:2014tba}
\bibinfo{author}{Michael Buballa}, \bibinfo{author}{Stefano Carignano},
  \bibinfo{title}{{Inhomogeneous chiral condensates}}, \bibinfo{journal}{Prog.
  Part. Nucl. Phys.} \bibinfo{volume}{81} (\bibinfo{year}{2015})
  \bibinfo{pages}{39--96}, \bibinfo{doi}{\doi{10.1016/j.ppnp.2014.11.001}},
  \eprint{1406.1367}.

\bibtype{Article}%
\bibitem{Fu:2019hdw}
\bibinfo{author}{Wei-jie Fu}, \bibinfo{author}{Jan~M. Pawlowski},
  \bibinfo{author}{Fabian Rennecke}, \bibinfo{title}{{QCD phase structure at
  finite temperature and density}}, \bibinfo{journal}{Phys. Rev. D}
  \bibinfo{volume}{101} (\bibinfo{number}{5}) (\bibinfo{year}{2020})
  \bibinfo{pages}{054032}, \bibinfo{doi}{\doi{10.1103/PhysRevD.101.054032}},
  \eprint{1909.02991}.

\bibtype{Article}%
\bibitem{McLerran:2007qj}
\bibinfo{author}{Larry McLerran}, \bibinfo{author}{Robert~D. Pisarski},
  \bibinfo{title}{{Phases of cold, dense quarks at large N(c)}},
  \bibinfo{journal}{Nucl. Phys. A} \bibinfo{volume}{796} (\bibinfo{year}{2007})
  \bibinfo{pages}{83--100},
  \bibinfo{doi}{\doi{10.1016/j.nuclphysa.2007.08.013}}, \eprint{0706.2191}.

\bibtype{Article}%
\bibitem{Lattimer:2000nx}
\bibinfo{author}{J.~M. Lattimer}, \bibinfo{author}{M. Prakash},
  \bibinfo{title}{{Neutron star structure and the equation of state}},
  \bibinfo{journal}{Astrophys. J.} \bibinfo{volume}{550} (\bibinfo{year}{2001})
  \bibinfo{pages}{426}, \bibinfo{doi}{\doi{10.1086/319702}},
  \eprint{astro-ph/0002232}.

\bibtype{Article}%
\bibitem{Boyanovsky:2006bf}
\bibinfo{author}{D. Boyanovsky}, \bibinfo{author}{H.~J. de Vega},
  \bibinfo{author}{D.~J. Schwarz}, \bibinfo{title}{{Phase transitions in the
  early and the present universe}}, \bibinfo{journal}{Ann. Rev. Nucl. Part.
  Sci.} \bibinfo{volume}{56} (\bibinfo{year}{2006}) \bibinfo{pages}{441--500},
  \bibinfo{doi}{\doi{10.1146/annurev.nucl.56.080805.140539}},
  \eprint{hep-ph/0602002}.

\bibtype{Article}%
\bibitem{Teaney:2001av}
\bibinfo{author}{D. Teaney}, \bibinfo{author}{J. Lauret},
  \bibinfo{author}{E.~V. Shuryak}, \bibinfo{title}{{A Hydrodynamic Description
  of Heavy Ion Collisions at the SPS and RHIC}}  (\bibinfo{year}{2001}),
  \eprint{nucl-th/0110037}.

\bibtype{Article}%
\bibitem{Kolb:2003dz}
\bibinfo{author}{Peter~F. Kolb}, \bibinfo{author}{Ulrich~W. Heinz},
  \bibinfo{title}{{Hydrodynamic description of ultrarelativistic heavy ion
  collisions}}  (\bibinfo{year}{2003}) \bibinfo{pages}{634--714},
  \eprint{nucl-th/0305084}.

\bibtype{Article}%
\bibitem{Brandt:2018bwq}
\bibinfo{author}{Bastian~B. Brandt}, \bibinfo{author}{Gergely Endr\H{o}di},
  \bibinfo{author}{Eduardo~S. Fraga}, \bibinfo{author}{Mauricio Hippert},
  \bibinfo{author}{Jurgen Schaffner-Bielich}, \bibinfo{author}{Sebastian
  Schmalzbauer}, \bibinfo{title}{{New class of compact stars: Pion stars}},
  \bibinfo{journal}{Phys. Rev. D} \bibinfo{volume}{98} (\bibinfo{number}{9})
  (\bibinfo{year}{2018}) \bibinfo{pages}{094510},
  \bibinfo{doi}{\doi{10.1103/PhysRevD.98.094510}}, \eprint{1802.06685}.

\bibtype{Article}%
\bibitem{Abbott:2023coj}
\bibinfo{author}{Ryan Abbott}, \bibinfo{author}{William Detmold},
  \bibinfo{author}{Fernando Romero-L{\'o}pez}, \bibinfo{author}{Zohreh
  Davoudi}, \bibinfo{author}{Marc Illa}, \bibinfo{author}{Assumpta
  Parre{\~n}o}, \bibinfo{author}{Robert~J. Perry}, \bibinfo{author}{Phiala~E.
  Shanahan}, \bibinfo{author}{Michael~L. Wagman}
  (\bibinfo{collaboration}{NPLQCD}), \bibinfo{title}{{Lattice quantum
  chromodynamics at large isospin density}}, \bibinfo{journal}{Phys. Rev. D}
  \bibinfo{volume}{108} (\bibinfo{number}{11}) (\bibinfo{year}{2023})
  \bibinfo{pages}{114506}, \bibinfo{doi}{\doi{10.1103/PhysRevD.108.114506}},
  \eprint{2307.15014}.

\bibtype{Article}%
\bibitem{Abbott:2024vhj}
\bibinfo{author}{Ryan Abbott}, \bibinfo{author}{William Detmold},
  \bibinfo{author}{Marc Illa}, \bibinfo{author}{Assumpta Parre{\~n}o},
  \bibinfo{author}{Robert~J. Perry}, \bibinfo{author}{Fernando
  Romero-L{\'o}pez}, \bibinfo{author}{Phiala~E. Shanahan},
  \bibinfo{author}{Michael~L. Wagman} (\bibinfo{collaboration}{NPLQCD}),
  \bibinfo{title}{{QCD Constraints on Isospin-Dense Matter and the Nuclear
  Equation of State}}, \bibinfo{journal}{Phys. Rev. Lett.}
  \bibinfo{volume}{134} (\bibinfo{number}{1}) (\bibinfo{year}{2025})
  \bibinfo{pages}{011903}, \bibinfo{doi}{\doi{10.1103/PhysRevLett.134.011903}},
  \eprint{2406.09273}.

\bibtype{Article}%
\bibitem{Vovchenko:2020crk}
\bibinfo{author}{Volodymyr Vovchenko}, \bibinfo{author}{Bastian~B. Brandt},
  \bibinfo{author}{Francesca Cuteri}, \bibinfo{author}{Gergely Endr{\H{o}}di},
  \bibinfo{author}{Fazlollah Hajkarim}, \bibinfo{author}{J{\"u}rgen
  Schaffner-Bielich}, \bibinfo{title}{{Pion Condensation in the Early Universe
  at Nonvanishing Lepton Flavor Asymmetry and Its Gravitational Wave
  Signatures}}, \bibinfo{journal}{Phys. Rev. Lett.} \bibinfo{volume}{126}
  (\bibinfo{number}{1}) (\bibinfo{year}{2021}) \bibinfo{pages}{012701},
  \bibinfo{doi}{\doi{10.1103/PhysRevLett.126.012701}}, \eprint{2009.02309}.

\bibtype{Article}%
\bibitem{Brandt:2022hwy}
\bibinfo{author}{Bastian~B. Brandt}, \bibinfo{author}{Francesca Cuteri},
  \bibinfo{author}{Gergely Endr\H{o}di}, \bibinfo{title}{{Equation of state and
  speed of sound of isospin-asymmetric QCD on the lattice}},
  \bibinfo{journal}{JHEP} \bibinfo{volume}{07} (\bibinfo{year}{2023})
  \bibinfo{pages}{055}, \bibinfo{doi}{\doi{10.1007/JHEP07(2023)055}},
  \eprint{2212.14016}.

\bibtype{Article}%
\bibitem{Endrodi:2024cqn}
\bibinfo{author}{Gergely Endr\H{o}di}, \bibinfo{title}{{QCD with background
  electromagnetic fields on the lattice: A review}}, \bibinfo{journal}{Prog.
  Part. Nucl. Phys.} \bibinfo{volume}{141} (\bibinfo{year}{2025})
  \bibinfo{pages}{104153}, \bibinfo{doi}{\doi{10.1016/j.ppnp.2024.104153}},
  \eprint{2406.19780}.

\bibtype{Article}%
\bibitem{Kamikado:2012bt}
\bibinfo{author}{Kazuhiko Kamikado}, \bibinfo{author}{Nils Strodthoff},
  \bibinfo{author}{Lorenz von Smekal}, \bibinfo{author}{Jochen Wambach},
  \bibinfo{title}{{Fluctuations in the quark-meson model for QCD with isospin
  chemical potential}}, \bibinfo{journal}{Phys. Lett. B} \bibinfo{volume}{718}
  (\bibinfo{year}{2013}) \bibinfo{pages}{1044--1053},
  \bibinfo{doi}{\doi{10.1016/j.physletb.2012.11.055}}, \eprint{1207.0400}.

\bibtype{Article}%
\bibitem{Fujimoto:2023mvc}
\bibinfo{author}{Yuki Fujimoto}, \bibinfo{title}{{Enhanced contribution of the
  pairing gap to the QCD equation of state at large isospin chemical
  potential}}, \bibinfo{journal}{Phys. Rev. D} \bibinfo{volume}{109}
  (\bibinfo{number}{5}) (\bibinfo{year}{2024}) \bibinfo{pages}{054035},
  \bibinfo{doi}{\doi{10.1103/PhysRevD.109.054035}}, \eprint{2312.11443}.

\bibtype{Book}%
\bibitem{Gattringer:2010zz}
\bibinfo{author}{Christof Gattringer}, \bibinfo{author}{Christian~B. Lang},
  \bibinfo{title}{{Quantum chromodynamics on the lattice}},
  \bibinfo{comment}{vol.} \bibinfo{volume}{788}, \bibinfo{publisher}{Springer},
  \bibinfo{address}{Berlin} \bibinfo{year}{2010}, ISBN
  \bibinfo{isbn}{978-3-642-01849-7, 978-3-642-01850-3},
  \bibinfo{doi}{\doi{10.1007/978-3-642-01850-3}}.

\bibtype{Article}%
\bibitem{tHooft:1979rtg}
\bibinfo{author}{Gerard 't Hooft}, \bibinfo{title}{{A Property of Electric and
  Magnetic Flux in Nonabelian Gauge Theories}}, \bibinfo{journal}{Nucl. Phys.
  B} \bibinfo{volume}{153} (\bibinfo{year}{1979}) \bibinfo{pages}{141--160},
  \bibinfo{doi}{\doi{10.1016/0550-3213(79)90595-9}}.

\bibtype{Article}%
\bibitem{deForcrand:2003vyj}
\bibinfo{author}{Philippe de Forcrand}, \bibinfo{author}{Owe Philipsen},
  \bibinfo{title}{{The QCD phase diagram for three degenerate flavors and small
  baryon density}}, \bibinfo{journal}{Nucl. Phys. B} \bibinfo{volume}{673}
  (\bibinfo{year}{2003}) \bibinfo{pages}{170--186},
  \bibinfo{doi}{\doi{10.1016/j.nuclphysb.2003.09.005}},
  \eprint{hep-lat/0307020}.

\bibtype{Article}%
\bibitem{Alford:1998sd}
\bibinfo{author}{Mark~G. Alford}, \bibinfo{author}{Anton Kapustin},
  \bibinfo{author}{Frank Wilczek}, \bibinfo{title}{{Imaginary chemical
  potential and finite fermion density on the lattice}},
  \bibinfo{journal}{Phys. Rev. D} \bibinfo{volume}{59} (\bibinfo{year}{1999})
  \bibinfo{pages}{054502}, \bibinfo{doi}{\doi{10.1103/PhysRevD.59.054502}},
  \eprint{hep-lat/9807039}.

\bibtype{Article}%
\bibitem{Kogut:2002zg}
\bibinfo{author}{J.~B. Kogut}, \bibinfo{author}{D.~K. Sinclair},
  \bibinfo{title}{{Lattice QCD at finite isospin density at zero and finite
  temperature}}, \bibinfo{journal}{Phys. Rev. D} \bibinfo{volume}{66}
  (\bibinfo{year}{2002}) \bibinfo{pages}{034505},
  \bibinfo{doi}{\doi{10.1103/PhysRevD.66.034505}}, \eprint{hep-lat/0202028}.

\bibtype{Article}%
\bibitem{Gasser:1983yg}
\bibinfo{author}{J. Gasser}, \bibinfo{author}{H. Leutwyler},
  \bibinfo{title}{{Chiral Perturbation Theory to One Loop}},
  \bibinfo{journal}{Annals Phys.} \bibinfo{volume}{158} (\bibinfo{year}{1984})
  \bibinfo{pages}{142}, \bibinfo{doi}{\doi{10.1016/0003-4916(84)90242-2}}.

\bibtype{Article}%
\bibitem{Scherer:2002tk}
\bibinfo{author}{Stefan Scherer}, \bibinfo{title}{{Introduction to chiral
  perturbation theory}}, \bibinfo{journal}{Adv. Nucl. Phys.}
  \bibinfo{volume}{27} (\bibinfo{year}{2003}) \bibinfo{pages}{277},
  \eprint{hep-ph/0210398}.

\bibtype{Article}%
\bibitem{Witten:1979vv}
\bibinfo{author}{Edward Witten}, \bibinfo{title}{{Current Algebra Theorems for
  the U(1) Goldstone Boson}}, \bibinfo{journal}{Nucl. Phys. B}
  \bibinfo{volume}{156} (\bibinfo{year}{1979}) \bibinfo{pages}{269--283},
  \bibinfo{doi}{\doi{10.1016/0550-3213(79)90031-2}}.

\bibtype{Article}%
\bibitem{Veneziano:1979ec}
\bibinfo{author}{G. Veneziano}, \bibinfo{title}{{U(1) Without Instantons}},
  \bibinfo{journal}{Nucl. Phys. B} \bibinfo{volume}{159} (\bibinfo{year}{1979})
  \bibinfo{pages}{213--224}, \bibinfo{doi}{\doi{10.1016/0550-3213(79)90332-8}}.

\bibtype{Article}%
\bibitem{Gavai:2002fi}
\bibinfo{author}{Rajiv~V. Gavai}, \bibinfo{author}{Sourendu Gupta},
  \bibinfo{title}{{The Phase transition in QCD with broken SU(2) flavor
  symmetry}}, \bibinfo{journal}{Phys. Rev. D} \bibinfo{volume}{66}
  (\bibinfo{year}{2002}) \bibinfo{pages}{094510},
  \bibinfo{doi}{\doi{10.1103/PhysRevD.66.094510}}, \eprint{hep-lat/0208019}.

\bibtype{Article}%
\bibitem{Dagotto:1986gw}
\bibinfo{author}{E. Dagotto}, \bibinfo{author}{F. Karsch}, \bibinfo{author}{A.
  Moreo}, \bibinfo{title}{{The Strong Coupling Limit of SU(2) {QCD} at Finite
  Baryon Density}}, \bibinfo{journal}{Phys. Lett. B} \bibinfo{volume}{169}
  (\bibinfo{year}{1986}) \bibinfo{pages}{421--427},
  \bibinfo{doi}{\doi{10.1016/0370-2693(86)90383-7}}.

\bibtype{Article}%
\bibitem{Dagotto:1986ms}
\bibinfo{author}{Elbio Dagotto}, \bibinfo{author}{Adriana Moreo},
  \bibinfo{author}{Ulli Wolff}, \bibinfo{title}{{Study of Lattice SU($N$) {QCD}
  at Finite Baryon Density}}, \bibinfo{journal}{Phys. Rev. Lett.}
  \bibinfo{volume}{57} (\bibinfo{year}{1986}) \bibinfo{pages}{1292},
  \bibinfo{doi}{\doi{10.1103/PhysRevLett.57.1292}}.

\bibtype{Article}%
\bibitem{Dagotto:1986xt}
\bibinfo{author}{Elbio Dagotto}, \bibinfo{author}{Adriana Moreo},
  \bibinfo{author}{Ulli Wolff}, \bibinfo{title}{{Lattice SU($N$) {QCD} at
  Finite Temperature and Density in the Strong Coupling Limit}},
  \bibinfo{journal}{Phys. Lett. B} \bibinfo{volume}{186} (\bibinfo{year}{1987})
  \bibinfo{pages}{395--400}, \bibinfo{doi}{\doi{10.1016/0370-2693(87)90315-7}}.

\bibtype{Article}%
\bibitem{Hands:1999md}
\bibinfo{author}{Simon Hands}, \bibinfo{author}{John~B. Kogut},
  \bibinfo{author}{Maria-Paola Lombardo}, \bibinfo{author}{Susan~E. Morrison},
  \bibinfo{title}{{Symmetries and spectrum of SU(2) lattice gauge theory at
  finite chemical potential}}, \bibinfo{journal}{Nucl. Phys. B}
  \bibinfo{volume}{558} (\bibinfo{year}{1999}) \bibinfo{pages}{327--346},
  \bibinfo{doi}{\doi{10.1016/S0550-3213(99)00364-8}}, \eprint{hep-lat/9902034}.

\bibtype{Article}%
\bibitem{Kogut:1999iv}
\bibinfo{author}{J.~B. Kogut}, \bibinfo{author}{Misha~A. Stephanov},
  \bibinfo{author}{D. Toublan}, \bibinfo{title}{{On two color QCD with baryon
  chemical potential}}, \bibinfo{journal}{Phys. Lett. B} \bibinfo{volume}{464}
  (\bibinfo{year}{1999}) \bibinfo{pages}{183--191},
  \bibinfo{doi}{\doi{10.1016/S0370-2693(99)00971-5}}, \eprint{hep-ph/9906346}.

\bibtype{Article}%
\bibitem{Kogut:2000ek}
\bibinfo{author}{J.~B. Kogut}, \bibinfo{author}{Misha~A. Stephanov},
  \bibinfo{author}{D. Toublan}, \bibinfo{author}{J.~J.~M. Verbaarschot},
  \bibinfo{author}{A. Zhitnitsky}, \bibinfo{title}{{QCD - like theories at
  finite baryon density}}, \bibinfo{journal}{Nucl. Phys. B}
  \bibinfo{volume}{582} (\bibinfo{year}{2000}) \bibinfo{pages}{477--513},
  \bibinfo{doi}{\doi{10.1016/S0550-3213(00)00242-X}}, \eprint{hep-ph/0001171}.

\bibtype{Article}%
\bibitem{Splittorff:2000mm}
\bibinfo{author}{K. Splittorff}, \bibinfo{author}{D.~T. Son},
  \bibinfo{author}{Misha~A. Stephanov}, \bibinfo{title}{{QCD - like theories at
  finite baryon and isospin density}}, \bibinfo{journal}{Phys. Rev. D}
  \bibinfo{volume}{64} (\bibinfo{year}{2001}) \bibinfo{pages}{016003},
  \bibinfo{doi}{\doi{10.1103/PhysRevD.64.016003}}, \eprint{hep-ph/0012274}.

\bibtype{Article}%
\bibitem{Begun:2022bxj}
\bibinfo{author}{A. Begun}, \bibinfo{author}{V.~G. Bornyakov},
  \bibinfo{author}{V.~A. Goy}, \bibinfo{author}{A. Nakamura},
  \bibinfo{author}{R.~N. Rogalyov}, \bibinfo{title}{{Study of two color QCD on
  large lattices}}, \bibinfo{journal}{Phys. Rev. D} \bibinfo{volume}{105}
  (\bibinfo{number}{11}) (\bibinfo{year}{2022}) \bibinfo{pages}{114505},
  \bibinfo{doi}{\doi{10.1103/PhysRevD.105.114505}}, \eprint{2203.04909}.

\bibtype{Article}%
\bibitem{Iida:2022hyy}
\bibinfo{author}{Kei Iida}, \bibinfo{author}{Etsuko Itou},
  \bibinfo{title}{{Velocity of sound beyond the high-density relativistic limit
  from lattice simulation of dense two-color QCD}}, \bibinfo{journal}{PTEP}
  \bibinfo{volume}{2022} (\bibinfo{number}{11}) (\bibinfo{year}{2022})
  \bibinfo{pages}{111B01}, \bibinfo{doi}{\doi{10.1093/ptep/ptac137}},
  \eprint{2207.01253}.

\bibtype{Article}%
\bibitem{Hands:2024nkx}
\bibinfo{author}{Simon Hands}, \bibinfo{author}{Seyong Kim},
  \bibinfo{author}{Dale Lawlor}, \bibinfo{author}{Andrew Lee-Mitchell},
  \bibinfo{author}{Jon-Ivar Skullerud}, \bibinfo{title}{{Dense QC$_2$D: What's
  up with that?!?}}, \bibinfo{journal}{PoS} \bibinfo{volume}{LATTICE2024}
  (\bibinfo{year}{2025}) \bibinfo{pages}{165},
  \bibinfo{doi}{\doi{10.22323/1.466.0165}}, \eprint{2412.15872}.

\bibtype{Article}%
\bibitem{Son:2000by}
\bibinfo{author}{D.~T. Son}, \bibinfo{author}{Misha~A. Stephanov},
  \bibinfo{title}{{QCD at finite isospin density: From pion to quark -
  anti-quark condensation}}, \bibinfo{journal}{Phys. Atom. Nucl.}
  \bibinfo{volume}{64} (\bibinfo{year}{2001}) \bibinfo{pages}{834--842},
  \bibinfo{doi}{\doi{10.1134/1.1378872}}, \eprint{hep-ph/0011365}.

\bibtype{Article}%
\bibitem{Cohen:2003kd}
\bibinfo{author}{Thomas D~. Cohen}, \bibinfo{title}{{Functional integrals for
  QCD at nonzero chemical potential and zero density}}, \bibinfo{journal}{Phys.
  Rev. Lett.} \bibinfo{volume}{91} (\bibinfo{year}{2003})
  \bibinfo{pages}{222001}, \bibinfo{doi}{\doi{10.1103/PhysRevLett.91.222001}},
  \eprint{hep-ph/0307089}.

\bibtype{Article}%
\bibitem{Son:2000xc}
\bibinfo{author}{D.~T. Son}, \bibinfo{author}{Misha~A. Stephanov},
  \bibinfo{title}{{QCD at finite isospin density}}, \bibinfo{journal}{Phys.
  Rev. Lett.} \bibinfo{volume}{86} (\bibinfo{year}{2001})
  \bibinfo{pages}{592--595}, \bibinfo{doi}{\doi{10.1103/PhysRevLett.86.592}},
  \eprint{hep-ph/0005225}.

\bibtype{Article}%
\bibitem{Kogut:2001id}
\bibinfo{author}{J.~B. Kogut}, \bibinfo{author}{D. Toublan},
  \bibinfo{title}{{QCD at small nonzero quark chemical potentials}},
  \bibinfo{journal}{Phys. Rev. D} \bibinfo{volume}{64} (\bibinfo{year}{2001})
  \bibinfo{pages}{034007}, \bibinfo{doi}{\doi{10.1103/PhysRevD.64.034007}},
  \eprint{hep-ph/0103271}.

\bibtype{Article}%
\bibitem{Mammarella:2015pxa}
\bibinfo{author}{Andrea Mammarella}, \bibinfo{author}{Massimo Mannarelli},
  \bibinfo{title}{{Intriguing aspects of meson condensation}},
  \bibinfo{journal}{Phys. Rev. D} \bibinfo{volume}{92} (\bibinfo{number}{8})
  (\bibinfo{year}{2015}) \bibinfo{pages}{085025},
  \bibinfo{doi}{\doi{10.1103/PhysRevD.92.085025}}, \eprint{1507.02934}.

\bibtype{Article}%
\bibitem{Adhikari:2019mdk}
\bibinfo{author}{Prabal Adhikari}, \bibinfo{author}{Jens~O. Andersen},
  \bibinfo{author}{Patrick Kneschke}, \bibinfo{title}{{Two-flavor chiral
  perturbation theory at nonzero isospin: Pion condensation at zero
  temperature}}, \bibinfo{journal}{Eur. Phys. J. C} \bibinfo{volume}{79}
  (\bibinfo{number}{10}) (\bibinfo{year}{2019}) \bibinfo{pages}{874},
  \bibinfo{doi}{\doi{10.1140/epjc/s10052-019-7381-4}}, \eprint{1904.03887}.

\bibtype{Article}%
\bibitem{Carignano:2016lxe}
\bibinfo{author}{Stefano Carignano}, \bibinfo{author}{Luca Lepori},
  \bibinfo{author}{Andrea Mammarella}, \bibinfo{author}{Massimo Mannarelli},
  \bibinfo{author}{Giulia Pagliaroli}, \bibinfo{title}{{Scrutinizing the pion
  condensed phase}}, \bibinfo{journal}{Eur. Phys. J. A} \bibinfo{volume}{53}
  (\bibinfo{number}{2}) (\bibinfo{year}{2017}) \bibinfo{pages}{35},
  \bibinfo{doi}{\doi{10.1140/epja/i2017-12221-x}}, \eprint{1610.06097}.

\bibtype{Article}%
\bibitem{Adhikari:2020ufo}
\bibinfo{author}{Prabal Adhikari}, \bibinfo{author}{Jens~O. Andersen},
  \bibinfo{title}{{Quark and pion condensates at finite isospin density in
  chiral perturbation theory}}, \bibinfo{journal}{Eur. Phys. J. C}
  \bibinfo{volume}{80} (\bibinfo{number}{11}) (\bibinfo{year}{2020})
  \bibinfo{pages}{1028}, \bibinfo{doi}{\doi{10.1140/epjc/s10052-020-08574-8}},
  \eprint{2003.12567}.

\bibtype{Article}%
\bibitem{Adhikari:2020qda}
\bibinfo{author}{Prabal Adhikari}, \bibinfo{author}{Jens~O. Andersen},
  \bibinfo{author}{Martin~A. Mojahed}, \bibinfo{title}{{Quark, pion and axial
  condensates in three-flavor finite isospin chiral perturbation theory}},
  \bibinfo{journal}{Eur. Phys. J. C} \bibinfo{volume}{81} (\bibinfo{number}{5})
  (\bibinfo{year}{2021}) \bibinfo{pages}{449},
  \bibinfo{doi}{\doi{10.1140/epjc/s10052-021-09212-7}}, \eprint{2012.04339}.

\bibtype{Article}%
\bibitem{Brauner:2016lkh}
\bibinfo{author}{Tomas Brauner}, \bibinfo{author}{Xu-Guang Huang},
  \bibinfo{title}{{Vector meson condensation in a pion superfluid}},
  \bibinfo{journal}{Phys. Rev. D} \bibinfo{volume}{94} (\bibinfo{number}{9})
  (\bibinfo{year}{2016}) \bibinfo{pages}{094003},
  \bibinfo{doi}{\doi{10.1103/PhysRevD.94.094003}}, \eprint{1610.00426}.

\bibtype{Article}%
\bibitem{Loewe:2002tw}
\bibinfo{author}{Marcelo Loewe}, \bibinfo{author}{Cristian Villavicencio},
  \bibinfo{title}{{Thermal pions at finite isospin chemical potential}},
  \bibinfo{journal}{Phys. Rev. D} \bibinfo{volume}{67} (\bibinfo{year}{2003})
  \bibinfo{pages}{074034}, \bibinfo{doi}{\doi{10.1103/PhysRevD.67.074034}},
  \eprint{hep-ph/0212275}.

\bibtype{Article}%
\bibitem{Loewe:2004mu}
\bibinfo{author}{M. Loewe}, \bibinfo{author}{C. Villavicencio},
  \bibinfo{title}{{Thermal pion masses in the second phase: |mu(I)|
  {\ensuremath{>}} m(pi)}}, \bibinfo{journal}{Phys. Rev. D}
  \bibinfo{volume}{70} (\bibinfo{year}{2004}) \bibinfo{pages}{074005},
  \bibinfo{doi}{\doi{10.1103/PhysRevD.70.074005}}, \eprint{hep-ph/0404232}.

\bibtype{Article}%
\bibitem{Adhikari:2020kdn}
\bibinfo{author}{Prabal Adhikari}, \bibinfo{author}{Jens~O. Andersen},
  \bibinfo{author}{Martin~A. Mojahed}, \bibinfo{title}{{Condensates and
  pressure of two-flavor chiral perturbation theory at nonzero isospin and
  temperature}}, \bibinfo{journal}{Eur. Phys. J. C} \bibinfo{volume}{81}
  (\bibinfo{number}{2}) (\bibinfo{year}{2021}) \bibinfo{pages}{173},
  \bibinfo{doi}{\doi{10.1140/epjc/s10052-021-08948-6}}, \eprint{2010.13655}.

\bibtype{Article}%
\bibitem{Splittorff:2002xn}
\bibinfo{author}{K. Splittorff}, \bibinfo{author}{D. Toublan},
  \bibinfo{author}{J.~J.~M. Verbaarschot}, \bibinfo{title}{{Thermodynamics of
  chiral symmetry at low densities}}, \bibinfo{journal}{Nucl. Phys. B}
  \bibinfo{volume}{639} (\bibinfo{year}{2002}) \bibinfo{pages}{524--548},
  \bibinfo{doi}{\doi{10.1016/S0550-3213(02)00440-6}}, \eprint{hep-ph/0204076}.

\bibtype{Article}%
\bibitem{Brandt:2017zck}
\bibinfo{author}{Bastian~B. Brandt}, \bibinfo{author}{Gergely Endr\H{o}di},
  \bibinfo{author}{Sebastian Schmalzbauer}, \bibinfo{title}{{QCD at finite
  isospin chemical potential}}, \bibinfo{journal}{EPJ Web Conf.}
  \bibinfo{volume}{175} (\bibinfo{year}{2018}) \bibinfo{pages}{07020},
  \bibinfo{doi}{\doi{10.1051/epjconf/201817507020}}, \eprint{1709.10487}.

\bibtype{Article}%
\bibitem{Adhikari:2019zaj}
\bibinfo{author}{Prabal Adhikari}, \bibinfo{author}{Jens~O. Andersen},
  \bibinfo{title}{{QCD at finite isospin density: chiral perturbation theory
  confronts lattice data}}, \bibinfo{journal}{Phys. Lett. B}
  \bibinfo{volume}{804} (\bibinfo{year}{2020}) \bibinfo{pages}{135352},
  \bibinfo{doi}{\doi{10.1016/j.physletb.2020.135352}}, \eprint{1909.01131}.

\bibtype{Article}%
\bibitem{Andersen:2023ivj}
\bibinfo{author}{Jens~O. Andersen}, \bibinfo{author}{Qing Yu},
  \bibinfo{author}{Hua Zhou}, \bibinfo{title}{{Pion condensation in QCD at
  finite isospin density, the dilute Bose gas, and speedy Goldstone bosons}},
  \bibinfo{journal}{Phys. Rev. D} \bibinfo{volume}{109} (\bibinfo{number}{3})
  (\bibinfo{year}{2024}) \bibinfo{pages}{034022},
  \bibinfo{doi}{\doi{10.1103/PhysRevD.109.034022}}, \eprint{2306.14472}.

\bibtype{Article}%
\bibitem{Carignano:2016rvs}
\bibinfo{author}{Stefano Carignano}, \bibinfo{author}{Andrea Mammarella},
  \bibinfo{author}{Massimo Mannarelli}, \bibinfo{title}{{Equation of state of
  imbalanced cold matter from chiral perturbation theory}},
  \bibinfo{journal}{Phys. Rev. D} \bibinfo{volume}{93} (\bibinfo{number}{5})
  (\bibinfo{year}{2016}) \bibinfo{pages}{051503},
  \bibinfo{doi}{\doi{10.1103/PhysRevD.93.051503}}, \eprint{1602.01317}.

\bibtype{Article}%
\bibitem{Andersen:2023ofv}
\bibinfo{author}{Jens~O. Andersen}, \bibinfo{author}{Martin~Kj{\o}llesdal
  Johnsrud}, \bibinfo{author}{Qing Yu}, \bibinfo{author}{Hua Zhou},
  \bibinfo{title}{{Chiral perturbation theory and Bose-Einstein condensation in
  QCD}}, \bibinfo{journal}{Phys. Rev. D} \bibinfo{volume}{111}
  (\bibinfo{number}{3}) (\bibinfo{year}{2025}) \bibinfo{pages}{034017},
  \bibinfo{doi}{\doi{10.1103/PhysRevD.111.034017}}, \eprint{2312.13092}.

\bibtype{Article}%
\bibitem{Cherman:2009tw}
\bibinfo{author}{Aleksey Cherman}, \bibinfo{author}{Thomas~D. Cohen},
  \bibinfo{author}{Abhinav Nellore}, \bibinfo{title}{{A Bound on the speed of
  sound from holography}}, \bibinfo{journal}{Phys. Rev. D} \bibinfo{volume}{80}
  (\bibinfo{year}{2009}) \bibinfo{pages}{066003},
  \bibinfo{doi}{\doi{10.1103/PhysRevD.80.066003}}, \eprint{0905.0903}.

\bibtype{Article}%
\bibitem{Adhikari:2019mlf}
\bibinfo{author}{Prabal Adhikari}, \bibinfo{author}{Jens~O. Andersen},
  \bibinfo{title}{{Pion and kaon condensation at zero temperature in
  three-flavor $\chi$PPT at nonzero isospin and strange chemical potentials at
  next-to-leading order}}, \bibinfo{journal}{JHEP} \bibinfo{volume}{06}
  (\bibinfo{year}{2020}) \bibinfo{pages}{170},
  \bibinfo{doi}{\doi{10.1007/JHEP06(2020)170}}, \eprint{1909.10575}.

\bibtype{Article}%
\bibitem{Mannarelli:2019hgn}
\bibinfo{author}{Massimo Mannarelli}, \bibinfo{title}{{Meson condensation}},
  \bibinfo{journal}{Particles} \bibinfo{volume}{2} (\bibinfo{number}{3})
  (\bibinfo{year}{2019}) \bibinfo{pages}{411--443},
  \bibinfo{doi}{\doi{10.3390/particles2030025}}, \eprint{1908.02042}.

\bibtype{Inproceedings}%
\bibitem{Morrison:1998ud}
\bibinfo{author}{Susan Morrison}, \bibinfo{author}{Simon Hands},
  \bibinfo{title}{{Two colors QCD at nonzero chemical potential}}, in:
  \bibinfo{booktitle}{{3rd International Conference on Strong and Electroweak
  Matter}} \bibinfo{year}{1998}, pp. \bibinfo{pages}{364--368},
  \eprint{hep-lat/9902012}.

\bibtype{Article}%
\bibitem{Kogut:2001na}
\bibinfo{author}{J.~B. Kogut}, \bibinfo{author}{D.~K. Sinclair},
  \bibinfo{author}{S.~J. Hands}, \bibinfo{author}{S.~E. Morrison},
  \bibinfo{title}{{Two color QCD at nonzero quark number density}},
  \bibinfo{journal}{Phys. Rev. D} \bibinfo{volume}{64} (\bibinfo{year}{2001})
  \bibinfo{pages}{094505}, \bibinfo{doi}{\doi{10.1103/PhysRevD.64.094505}},
  \eprint{hep-lat/0105026}.

\bibtype{Article}%
\bibitem{Brandt:2016zdy}
\bibinfo{author}{Bastian~B. Brandt}, \bibinfo{author}{Gergely Endr\H{o}di},
  \bibinfo{title}{{QCD phase diagram with isospin chemical potential}},
  \bibinfo{journal}{PoS} \bibinfo{volume}{LATTICE2016} (\bibinfo{year}{2016})
  \bibinfo{pages}{039}, \bibinfo{doi}{\doi{10.22323/1.256.0039}},
  \eprint{1611.06758}.

\bibtype{Article}%
\bibitem{Endrodi:2014lja}
\bibinfo{author}{G. Endr\H{o}di}, \bibinfo{title}{{Magnetic structure of
  isospin-asymmetric QCD matter in neutron stars}}, \bibinfo{journal}{Phys.
  Rev. D} \bibinfo{volume}{90} (\bibinfo{number}{9}) (\bibinfo{year}{2014})
  \bibinfo{pages}{094501}, \bibinfo{doi}{\doi{10.1103/PhysRevD.90.094501}},
  \eprint{1407.1216}.

\bibtype{Article}%
\bibitem{Kogut:2002tm}
\bibinfo{author}{J.~B. Kogut}, \bibinfo{author}{D.~K. Sinclair},
  \bibinfo{title}{{Quenched lattice QCD at finite isospin density and related
  theories}}, \bibinfo{journal}{Phys. Rev. D} \bibinfo{volume}{66}
  (\bibinfo{year}{2002}) \bibinfo{pages}{014508},
  \bibinfo{doi}{\doi{10.1103/PhysRevD.66.014508}}, \eprint{hep-lat/0201017}.

\bibtype{Article}%
\bibitem{deForcrand:2007uz}
\bibinfo{author}{Philippe de Forcrand}, \bibinfo{author}{Mikhail~A. Stephanov},
  \bibinfo{author}{Urs Wenger}, \bibinfo{title}{{On the phase diagram of QCD at
  finite isospin density}}, \bibinfo{journal}{PoS}
  \bibinfo{volume}{LATTICE2007} (\bibinfo{year}{2007}) \bibinfo{pages}{237},
  \eprint{0711.0023}.

\bibtype{Article}%
\bibitem{Brandt:2022fij}
\bibinfo{author}{Bastian~B. Brandt}, \bibinfo{author}{Francesca Cuteri},
  \bibinfo{author}{Gergely Endr{\"o}di}, \bibinfo{title}{{Equation of state and
  Taylor expansions at nonzero isospin chemical potential}},
  \bibinfo{journal}{PoS} \bibinfo{volume}{LATTICE2022} (\bibinfo{year}{2023})
  \bibinfo{pages}{144}, \bibinfo{doi}{\doi{10.22323/1.430.0144}},
  \eprint{2212.01431}.

\bibtype{Article}%
\bibitem{Brandt:2024dle}
\bibinfo{author}{Bastian~B. Brandt}, \bibinfo{author}{Gergely Endr{\H{o}}di},
  \bibinfo{author}{Gergely Mark{\'o}}, \bibinfo{title}{{Equation of state of
  isospin asymmetric QCD with small baryon chemical potentials}},
  \bibinfo{journal}{PoS} \bibinfo{volume}{LATTICE2024} (\bibinfo{year}{2025})
  \bibinfo{pages}{176}, \bibinfo{doi}{\doi{10.22323/1.466.0176}},
  \eprint{2411.12918}.

\bibtype{Inproceedings}%
\bibitem{Liu:2003wy}
\bibinfo{author}{Keh-Fei Liu}, \bibinfo{title}{{A Finite baryon density
  algorithm}}, in: \bibinfo{booktitle}{{3rd International Workshop on Numerical
  Analysis and Lattice QCD}} \bibinfo{year}{2003}, pp.
  \bibinfo{pages}{101--111}, \eprint{hep-lat/0312027}.

\bibtype{Article}%
\bibitem{Alexandru:2005ix}
\bibinfo{author}{Andrei Alexandru}, \bibinfo{author}{Manfried Faber},
  \bibinfo{author}{Ivan Horvath}, \bibinfo{author}{Keh-Fei Liu},
  \bibinfo{title}{{Lattice QCD at finite density via a new canonical
  approach}}, \bibinfo{journal}{Phys. Rev. D} \bibinfo{volume}{72}
  (\bibinfo{year}{2005}) \bibinfo{pages}{114513},
  \bibinfo{doi}{\doi{10.1103/PhysRevD.72.114513}}, \eprint{hep-lat/0507020}.

\bibtype{Article}%
\bibitem{deForcrand:2006ec}
\bibinfo{author}{Philippe de Forcrand}, \bibinfo{author}{Slavo Kratochvila},
  \bibinfo{title}{{Finite density QCD with a canonical approach}},
  \bibinfo{journal}{Nucl. Phys. B Proc. Suppl.} \bibinfo{volume}{153}
  (\bibinfo{year}{2006}) \bibinfo{pages}{62--67},
  \bibinfo{doi}{\doi{10.1016/j.nuclphysbps.2006.01.007}},
  \eprint{hep-lat/0602024}.

\bibtype{Article}%
\bibitem{Detmold:2012wc}
\bibinfo{author}{William Detmold}, \bibinfo{author}{Kostas Orginos},
  \bibinfo{author}{Zhifeng Shi}, \bibinfo{title}{{Lattice QCD at non-zero
  isospin chemical potential}}, \bibinfo{journal}{Phys. Rev. D}
  \bibinfo{volume}{86} (\bibinfo{year}{2012}) \bibinfo{pages}{054507},
  \bibinfo{doi}{\doi{10.1103/PhysRevD.86.054507}}, \eprint{1205.4224}.

\bibtype{Article}%
\bibitem{Fukushima:2024gmp}
\bibinfo{author}{Kenji Fukushima}, \bibinfo{author}{Shuhei Minato},
  \bibinfo{title}{{Speed of sound and trace anomaly in a unified treatment of
  the two-color diquark superfluid, the pion-condensed high-isospin matter, and
  the 2SC quark matter}}, \bibinfo{journal}{Phys. Rev. D} \bibinfo{volume}{111}
  (\bibinfo{number}{9}) (\bibinfo{year}{2025}) \bibinfo{pages}{094006},
  \bibinfo{doi}{\doi{10.1103/PhysRevD.111.094006}}, \eprint{2411.03781}.

\bibtype{Article}%
\bibitem{Freedman:1976ub}
\bibinfo{author}{Barry~A. Freedman}, \bibinfo{author}{Larry~D. McLerran},
  \bibinfo{title}{{Fermions and Gauge Vector Mesons at Finite Temperature and
  Density. 3. The Ground State Energy of a Relativistic Quark Gas}},
  \bibinfo{journal}{Phys. Rev. D} \bibinfo{volume}{16} (\bibinfo{year}{1977})
  \bibinfo{pages}{1169}, \bibinfo{doi}{\doi{10.1103/PhysRevD.16.1169}}.

\bibtype{Article}%
\bibitem{Baluni:1977ms}
\bibinfo{author}{Varouzhan Baluni}, \bibinfo{title}{{Nonabelian Gauge Theories
  of Fermi Systems: Chromotheory of Highly Condensed Matter}},
  \bibinfo{journal}{Phys. Rev. D} \bibinfo{volume}{17} (\bibinfo{year}{1978})
  \bibinfo{pages}{2092}, \bibinfo{doi}{\doi{10.1103/PhysRevD.17.2092}}.

\bibtype{Article}%
\bibitem{Kurkela:2009gj}
\bibinfo{author}{Aleksi Kurkela}, \bibinfo{author}{Paul Romatschke},
  \bibinfo{author}{Aleksi Vuorinen}, \bibinfo{title}{{Cold Quark Matter}},
  \bibinfo{journal}{Phys. Rev. D} \bibinfo{volume}{81} (\bibinfo{year}{2010})
  \bibinfo{pages}{105021}, \bibinfo{doi}{\doi{10.1103/PhysRevD.81.105021}},
  \eprint{0912.1856}.

\bibtype{Article}%
\bibitem{Annala:2019puf}
\bibinfo{author}{Eemeli Annala}, \bibinfo{author}{Tyler Gorda},
  \bibinfo{author}{Aleksi Kurkela}, \bibinfo{author}{Joonas N{\"a}ttil{\"a}},
  \bibinfo{author}{Aleksi Vuorinen}, \bibinfo{title}{{Evidence for quark-matter
  cores in massive neutron stars}}, \bibinfo{journal}{Nature Phys.}
  \bibinfo{volume}{16} (\bibinfo{number}{9}) (\bibinfo{year}{2020})
  \bibinfo{pages}{907--910}, \bibinfo{doi}{\doi{10.1038/s41567-020-0914-9}},
  \eprint{1903.09121}.

\bibtype{Article}%
\bibitem{Altiparmak:2022bke}
\bibinfo{author}{Sinan Altiparmak}, \bibinfo{author}{Christian Ecker},
  \bibinfo{author}{Luciano Rezzolla}, \bibinfo{title}{{On the Sound Speed in
  Neutron Stars}}, \bibinfo{journal}{Astrophys. J. Lett.} \bibinfo{volume}{939}
  (\bibinfo{number}{2}) (\bibinfo{year}{2022}) \bibinfo{pages}{L34},
  \bibinfo{doi}{\doi{10.3847/2041-8213/ac9b2a}}, \eprint{2203.14974}.

\bibtype{Article}%
\bibitem{Marczenko:2022jhl}
\bibinfo{author}{Micha{\l} Marczenko}, \bibinfo{author}{Larry McLerran},
  \bibinfo{author}{Krzysztof Redlich}, \bibinfo{author}{Chihiro Sasaki},
  \bibinfo{title}{{Reaching percolation and conformal limits in neutron
  stars}}, \bibinfo{journal}{Phys. Rev. C} \bibinfo{volume}{107}
  (\bibinfo{number}{2}) (\bibinfo{year}{2023}) \bibinfo{pages}{025802},
  \bibinfo{doi}{\doi{10.1103/PhysRevC.107.025802}}, \eprint{2207.13059}.

\bibtype{Article}%
\bibitem{Fujimoto:2022ohj}
\bibinfo{author}{Yuki Fujimoto}, \bibinfo{author}{Kenji Fukushima},
  \bibinfo{author}{Larry~D. McLerran}, \bibinfo{author}{Michal Praszalowicz},
  \bibinfo{title}{{Trace Anomaly as Signature of Conformality in Neutron
  Stars}}, \bibinfo{journal}{Phys. Rev. Lett.} \bibinfo{volume}{129}
  (\bibinfo{number}{25}) (\bibinfo{year}{2022}) \bibinfo{pages}{252702},
  \bibinfo{doi}{\doi{10.1103/PhysRevLett.129.252702}}, \eprint{2207.06753}.

\bibtype{Article}%
\bibitem{Graf:2015tda}
\bibinfo{author}{Thorben Graf}, \bibinfo{author}{Juergen Schaffner-Bielich},
  \bibinfo{author}{Eduardo~S. Fraga}, \bibinfo{title}{{The impact of quark
  masses on pQCD thermodynamics}}, \bibinfo{journal}{Eur. Phys. J. A}
  \bibinfo{volume}{52} (\bibinfo{number}{7}) (\bibinfo{year}{2016})
  \bibinfo{pages}{208}, \bibinfo{doi}{\doi{10.1140/epja/i2016-16208-9}},
  \eprint{1507.08941}.

\bibtype{Article}%
\bibitem{Kogut:2004qq}
\bibinfo{author}{J.~B. Kogut}, \bibinfo{author}{D.~K. Sinclair},
  \bibinfo{title}{{The Finite temperature transition for 3-flavor lattice QCD
  at finite isospin density}}, \bibinfo{journal}{Nucl. Phys. B Proc. Suppl.}
  \bibinfo{volume}{140} (\bibinfo{year}{2005}) \bibinfo{pages}{526--528},
  \bibinfo{doi}{\doi{10.1016/j.nuclphysbps.2004.11.156}},
  \eprint{hep-lat/0407041}.

\bibtype{Article}%
\bibitem{Sinclair:2006zm}
\bibinfo{author}{D.~K. Sinclair}, \bibinfo{author}{J.~B. Kogut},
  \bibinfo{title}{{Searching for the elusive critical endpoint at finite
  temperature and isospin density}}, \bibinfo{journal}{PoS}
  \bibinfo{volume}{LAT2006} (\bibinfo{year}{2006}) \bibinfo{pages}{147},
  \bibinfo{doi}{\doi{10.22323/1.032.0147}}, \eprint{hep-lat/0609041}.

\bibtype{Article}%
\bibitem{Brandt:2018omg}
\bibinfo{author}{Bastian~B. Brandt}, \bibinfo{author}{Gergely Endr\H{o}di},
  \bibinfo{title}{{Reliability of Taylor expansions in QCD}},
  \bibinfo{journal}{Phys. Rev. D} \bibinfo{volume}{99} (\bibinfo{number}{1})
  (\bibinfo{year}{2019}) \bibinfo{pages}{014518},
  \bibinfo{doi}{\doi{10.1103/PhysRevD.99.014518}}, \eprint{1810.11045}.

\bibtype{Article}%
\bibitem{HotQCD:2018pds}
\bibinfo{author}{A. Bazavov}, et al. (\bibinfo{collaboration}{HotQCD}),
  \bibinfo{title}{{Chiral crossover in QCD at zero and non-zero chemical
  potentials}}, \bibinfo{journal}{Phys. Lett. B} \bibinfo{volume}{795}
  (\bibinfo{year}{2019}) \bibinfo{pages}{15--21},
  \bibinfo{doi}{\doi{10.1016/j.physletb.2019.05.013}}, \eprint{1812.08235}.

\bibtype{Article}%
\bibitem{Toublan:2004ks}
\bibinfo{author}{D. Toublan}, \bibinfo{author}{John~B. Kogut},
  \bibinfo{title}{{The QCD phase diagram at nonzero baryon, isospin and
  strangeness chemical potentials: Results from a hadron resonance gas model}},
  \bibinfo{journal}{Phys. Lett. B} \bibinfo{volume}{605} (\bibinfo{year}{2005})
  \bibinfo{pages}{129--136},
  \bibinfo{doi}{\doi{10.1016/j.physletb.2004.11.018}}, \eprint{hep-ph/0409310}.

\bibtype{Article}%
\bibitem{Borsanyi:2023tdp}
\bibinfo{author}{Szabolcs Borsanyi}, \bibinfo{author}{Zoltan Fodor},
  \bibinfo{author}{Matteo Giordano}, \bibinfo{author}{Jana~N. Guenther},
  \bibinfo{author}{Sandor~D. Katz}, \bibinfo{author}{Attila Pasztor},
  \bibinfo{author}{Chik~Him Wong}, \bibinfo{title}{{Can rooted staggered
  fermions describe nonzero baryon density at low temperatures?}},
  \bibinfo{journal}{Phys. Rev. D} \bibinfo{volume}{109} (\bibinfo{number}{5})
  (\bibinfo{year}{2024}) \bibinfo{pages}{054509},
  \bibinfo{doi}{\doi{10.1103/PhysRevD.109.054509}}, \eprint{2308.06105}.

\bibtype{Article}%
\bibitem{Mitra:2024czm}
\bibinfo{author}{Sabarnya Mitra}, \bibinfo{title}{{Estimates of Lee-Yang zeros
  and a possible critical point on the pion condensate boundary in the QCD
  isospin phase diagram using an unbiased exponential resummation on the
  lattice}}, \bibinfo{journal}{Phys. Rev. D} \bibinfo{volume}{112}
  (\bibinfo{number}{1}) (\bibinfo{year}{2025}) \bibinfo{pages}{014511},
  \bibinfo{doi}{\doi{10.1103/h24l-2h8j}}, \eprint{2401.14299}.

\bibtype{Article}%
\bibitem{Yang:1952be}
\bibinfo{author}{Chen-Ning Yang}, \bibinfo{author}{T.~D. Lee},
  \bibinfo{title}{{Statistical theory of equations of state and phase
  transitions. 1. Theory of condensation}}, \bibinfo{journal}{Phys. Rev.}
  \bibinfo{volume}{87} (\bibinfo{year}{1952}) \bibinfo{pages}{404--409},
  \bibinfo{doi}{\doi{10.1103/PhysRev.87.404}}.

\bibtype{Article}%
\bibitem{Stephanov:2006dn}
\bibinfo{author}{M.~A. Stephanov}, \bibinfo{title}{{QCD critical point and
  complex chemical potential singularities}}, \bibinfo{journal}{Phys. Rev. D}
  \bibinfo{volume}{73} (\bibinfo{year}{2006}) \bibinfo{pages}{094508},
  \bibinfo{doi}{\doi{10.1103/PhysRevD.73.094508}}, \eprint{hep-lat/0603014}.

\bibtype{Article}%
\bibitem{Andersen:2015eoa}
\bibinfo{author}{Jens~O. Andersen}, \bibinfo{author}{Najmul Haque},
  \bibinfo{author}{Munshi~G. Mustafa}, \bibinfo{author}{Michael Strickland},
  \bibinfo{title}{{Three-loop hard-thermal-loop perturbation theory
  thermodynamics at finite temperature and finite baryonic and isospin chemical
  potential}}, \bibinfo{journal}{Phys. Rev. D} \bibinfo{volume}{93}
  (\bibinfo{number}{5}) (\bibinfo{year}{2016}) \bibinfo{pages}{054045},
  \bibinfo{doi}{\doi{10.1103/PhysRevD.93.054045}}, \eprint{1511.04660}.

\bibtype{Article}%
\bibitem{Bollweg:2022fqq}
\bibinfo{author}{D. Bollweg}, \bibinfo{author}{D.~A. Clarke},
  \bibinfo{author}{J. Goswami}, \bibinfo{author}{O. Kaczmarek},
  \bibinfo{author}{F. Karsch}, \bibinfo{author}{Swagato Mukherjee},
  \bibinfo{author}{P. Petreczky}, \bibinfo{author}{C. Schmidt},
  \bibinfo{author}{Sipaz Sharma} (\bibinfo{collaboration}{HotQCD}),
  \bibinfo{title}{{Equation of state and speed of sound of (2+1)-flavor QCD in
  strangeness-neutral matter at nonvanishing net baryon-number density}},
  \bibinfo{journal}{Phys. Rev. D} \bibinfo{volume}{108} (\bibinfo{number}{1})
  (\bibinfo{year}{2023}) \bibinfo{pages}{014510},
  \bibinfo{doi}{\doi{10.1103/PhysRevD.108.014510}}, \eprint{2212.09043}.

\bibtype{Article}%
\bibitem{Chernodub:2010qx}
\bibinfo{author}{M.~N. Chernodub}, \bibinfo{title}{{Superconductivity of QCD
  vacuum in strong magnetic field}}, \bibinfo{journal}{Phys. Rev. D}
  \bibinfo{volume}{82} (\bibinfo{year}{2010}) \bibinfo{pages}{085011},
  \bibinfo{doi}{\doi{10.1103/PhysRevD.82.085011}}, \eprint{1008.1055}.

\bibtype{Article}%
\bibitem{Gusynin:1994re}
\bibinfo{author}{V.~P. Gusynin}, \bibinfo{author}{V.~A. Miransky},
  \bibinfo{author}{I.~A. Shovkovy}, \bibinfo{title}{{Catalysis of dynamical
  flavor symmetry breaking by a magnetic field in (2+1)-dimensions}},
  \bibinfo{journal}{Phys. Rev. Lett.} \bibinfo{volume}{73}
  (\bibinfo{year}{1994}) \bibinfo{pages}{3499--3502},
  \bibinfo{doi}{\doi{10.1103/PhysRevLett.73.3499}}, \bibinfo{note}{[Erratum:
  Phys.Rev.Lett. 76, 1005 (1996)]}, \eprint{hep-ph/9405262}.

\bibtype{Article}%
\bibitem{Shovkovy:2012zn}
\bibinfo{author}{Igor~A. Shovkovy}, \bibinfo{title}{{Magnetic Catalysis: A
  Review}}, \bibinfo{journal}{Lect. Notes Phys.} \bibinfo{volume}{871}
  (\bibinfo{year}{2013}) \bibinfo{pages}{13--49},
  \bibinfo{doi}{\doi{10.1007/978-3-642-37305-3_2}}, \eprint{1207.5081}.

\bibtype{Article}%
\bibitem{Banks:1979yr}
\bibinfo{author}{Tom Banks}, \bibinfo{author}{A. Casher},
  \bibinfo{title}{{Chiral Symmetry Breaking in Confining Theories}},
  \bibinfo{journal}{Nucl. Phys. B} \bibinfo{volume}{169} (\bibinfo{year}{1980})
  \bibinfo{pages}{103--125}, \bibinfo{doi}{\doi{10.1016/0550-3213(80)90255-2}}.

\bibtype{Article}%
\bibitem{Shushpanov:1997sf}
\bibinfo{author}{I.~A. Shushpanov}, \bibinfo{author}{Andrei~V. Smilga},
  \bibinfo{title}{{Quark condensate in a magnetic field}},
  \bibinfo{journal}{Phys. Lett. B} \bibinfo{volume}{402} (\bibinfo{year}{1997})
  \bibinfo{pages}{351--358},
  \bibinfo{doi}{\doi{10.1016/S0370-2693(97)00441-3}}, \eprint{hep-ph/9703201}.

\bibtype{Article}%
\bibitem{Agasian:1999sx}
\bibinfo{author}{Nikita~O. Agasian}, \bibinfo{author}{I.~A. Shushpanov},
  \bibinfo{title}{{The Quark and gluon condensates and low-energy QCD theorems
  in a magnetic field}}, \bibinfo{journal}{Phys. Lett. B} \bibinfo{volume}{472}
  (\bibinfo{year}{2000}) \bibinfo{pages}{143--149},
  \bibinfo{doi}{\doi{10.1016/S0370-2693(99)01414-8}}, \eprint{hep-ph/9911254}.

\bibtype{Article}%
\bibitem{Cohen:2007bt}
\bibinfo{author}{Thomas~D. Cohen}, \bibinfo{author}{David~A. McGady},
  \bibinfo{author}{Elizabeth~S. Werbos}, \bibinfo{title}{{The Chiral condensate
  in a constant electromagnetic field}}, \bibinfo{journal}{Phys. Rev. C}
  \bibinfo{volume}{76} (\bibinfo{year}{2007}) \bibinfo{pages}{055201},
  \bibinfo{doi}{\doi{10.1103/PhysRevC.76.055201}}, \eprint{0706.3208}.

\bibtype{Article}%
\bibitem{Bali:2013txa}
\bibinfo{author}{G.~S. Bali}, \bibinfo{author}{F. Bruckmann},
  \bibinfo{author}{G. Endr\H{o}di}, \bibinfo{author}{A. Sch\"afer},
  \bibinfo{title}{{Magnetization and pressures at nonzero magnetic fields in
  QCD}}, \bibinfo{journal}{PoS} \bibinfo{volume}{LATTICE2013}
  (\bibinfo{year}{2014}) \bibinfo{pages}{182},
  \bibinfo{doi}{\doi{10.22323/1.187.0182}}, \eprint{1310.8145}.

\bibtype{Article}%
\bibitem{Adhikari:2015wva}
\bibinfo{author}{Prabal Adhikari}, \bibinfo{author}{Thomas~D. Cohen},
  \bibinfo{author}{Julia Sakowitz}, \bibinfo{title}{{Finite Isospin Chiral
  Perturbation Theory in a Magnetic Field}}, \bibinfo{journal}{Phys. Rev. C}
  \bibinfo{volume}{91} (\bibinfo{number}{4}) (\bibinfo{year}{2015})
  \bibinfo{pages}{045202}, \bibinfo{doi}{\doi{10.1103/PhysRevC.91.045202}},
  \eprint{1501.02737}.

\bibtype{Article}%
\bibitem{Adhikari:2018fwm}
\bibinfo{author}{Prabal Adhikari}, \bibinfo{title}{{Magnetic Vortex Lattices in
  Finite Isospin Chiral Perturbation Theory}}, \bibinfo{journal}{Phys. Lett. B}
  \bibinfo{volume}{790} (\bibinfo{year}{2019}) \bibinfo{pages}{211--217},
  \bibinfo{doi}{\doi{10.1016/j.physletb.2019.01.027}}, \eprint{1810.03663}.

\bibtype{Article}%
\bibitem{Adhikari:2022cks}
\bibinfo{author}{Prabal Adhikari}, \bibinfo{author}{Elizabeth Leeser},
  \bibinfo{author}{Jake Markowski}, \bibinfo{title}{{Phonon modes of magnetic
  vortex lattices in finite isospin chiral perturbation theory}},
  \bibinfo{journal}{Mod. Phys. Lett. A} \bibinfo{volume}{38}
  (\bibinfo{number}{14n15}) (\bibinfo{year}{2023}) \bibinfo{pages}{2350078},
  \bibinfo{doi}{\doi{10.1142/S0217732323500785}}, \eprint{2205.13369}.

\bibtype{Article}%
\bibitem{Brauner:2016pko}
\bibinfo{author}{Tomas Brauner}, \bibinfo{author}{Naoki Yamamoto},
  \bibinfo{title}{{Chiral Soliton Lattice and Charged Pion Condensation in
  Strong Magnetic Fields}}, \bibinfo{journal}{JHEP} \bibinfo{volume}{04}
  (\bibinfo{year}{2017}) \bibinfo{pages}{132},
  \bibinfo{doi}{\doi{10.1007/JHEP04(2017)132}}, \eprint{1609.05213}.

\bibtype{Article}%
\bibitem{Bruckmann:2013oba}
\bibinfo{author}{Falk Bruckmann}, \bibinfo{author}{Gergely Endr\H{o}di},
  \bibinfo{author}{Tamas~G. Kov\'acs}, \bibinfo{title}{{Inverse magnetic
  catalysis and the Polyakov loop}}, \bibinfo{journal}{JHEP}
  \bibinfo{volume}{04} (\bibinfo{year}{2013}) \bibinfo{pages}{112},
  \bibinfo{doi}{\doi{10.1007/JHEP04(2013)112}}, \eprint{1303.3972}.

\bibtype{Article}%
\bibitem{Fukushima:2017csk}
\bibinfo{author}{Kenji Fukushima}, \bibinfo{author}{Vladimir Skokov},
  \bibinfo{title}{{Polyakov loop modeling for hot QCD}},
  \bibinfo{journal}{Prog. Part. Nucl. Phys.} \bibinfo{volume}{96}
  (\bibinfo{year}{2017}) \bibinfo{pages}{154--199},
  \bibinfo{doi}{\doi{10.1016/j.ppnp.2017.05.002}}, \eprint{1705.00718}.

\bibtype{Article}%
\bibitem{Bali:2011qj}
\bibinfo{author}{G.~S. Bali}, \bibinfo{author}{F. Bruckmann},
  \bibinfo{author}{G. Endr\H{o}di}, \bibinfo{author}{Z. Fodor},
  \bibinfo{author}{S.~D. Katz}, \bibinfo{author}{S. Krieg}, \bibinfo{author}{A.
  Sch{\"a}fer}, \bibinfo{author}{K.~K. Szab\'o}, \bibinfo{title}{{The QCD phase
  diagram for external magnetic fields}}, \bibinfo{journal}{JHEP}
  \bibinfo{volume}{02} (\bibinfo{year}{2012}) \bibinfo{pages}{044},
  \bibinfo{doi}{\doi{10.1007/JHEP02(2012)044}}, \eprint{1111.4956}.

\bibtype{Article}%
\bibitem{Bali:2012zg}
\bibinfo{author}{G.~S. Bali}, \bibinfo{author}{F. Bruckmann},
  \bibinfo{author}{G. Endr\H{o}di}, \bibinfo{author}{Z. Fodor},
  \bibinfo{author}{S.~D. Katz}, \bibinfo{author}{A. Sch{\"a}fer},
  \bibinfo{title}{{QCD quark condensate in external magnetic fields}},
  \bibinfo{journal}{Phys. Rev. D} \bibinfo{volume}{86} (\bibinfo{year}{2012})
  \bibinfo{pages}{071502}, \bibinfo{doi}{\doi{10.1103/PhysRevD.86.071502}},
  \eprint{1206.4205}.

\bibtype{Article}%
\bibitem{Endrodi:2019zrl}
\bibinfo{author}{Gergely Endr\H{o}di}, \bibinfo{author}{Matteo Giordano},
  \bibinfo{author}{Sandor~D. Katz}, \bibinfo{author}{T.~G. Kov\'acs},
  \bibinfo{author}{Ferenc Pittler}, \bibinfo{title}{{Magnetic catalysis and
  inverse catalysis for heavy pions}}, \bibinfo{journal}{JHEP}
  \bibinfo{volume}{07} (\bibinfo{year}{2019}) \bibinfo{pages}{007},
  \bibinfo{doi}{\doi{10.1007/JHEP07(2019)007}}, \eprint{1904.10296}.

\bibtype{Article}%
\bibitem{DElia:2018xwo}
\bibinfo{author}{Massimo D'Elia}, \bibinfo{author}{Floriano Manigrasso},
  \bibinfo{author}{Francesco Negro}, \bibinfo{author}{Francesco Sanfilippo},
  \bibinfo{title}{{QCD phase diagram in a magnetic background for different
  values of the pion mass}}, \bibinfo{journal}{Phys. Rev. D}
  \bibinfo{volume}{98} (\bibinfo{number}{5}) (\bibinfo{year}{2018})
  \bibinfo{pages}{054509}, \bibinfo{doi}{\doi{10.1103/PhysRevD.98.054509}},
  \eprint{1808.07008}.

\bibtype{Article}%
\bibitem{Endrodi:2014vza}
\bibinfo{author}{Gergely Endr\H{o}di}, \bibinfo{title}{{QCD in magnetic fields:
  from Hofstadter's butterfly to the phase diagram}}, \bibinfo{journal}{PoS}
  \bibinfo{volume}{LATTICE2014} (\bibinfo{year}{2014}) \bibinfo{pages}{018},
  \bibinfo{doi}{\doi{10.22323/1.214.0018}}, \eprint{1410.8028}.

\bibtype{Article}%
\bibitem{Miransky:2002rp}
\bibinfo{author}{V.~A. Miransky}, \bibinfo{author}{I.~A. Shovkovy},
  \bibinfo{title}{{Magnetic catalysis and anisotropic confinement in QCD}},
  \bibinfo{journal}{Phys. Rev. D} \bibinfo{volume}{66} (\bibinfo{year}{2002})
  \bibinfo{pages}{045006}, \bibinfo{doi}{\doi{10.1103/PhysRevD.66.045006}},
  \eprint{hep-ph/0205348}.

\bibtype{Article}%
\bibitem{Cohen:2013zja}
\bibinfo{author}{Thomas~D. Cohen}, \bibinfo{author}{Naoki Yamamoto},
  \bibinfo{title}{{New critical point for QCD in a magnetic field}},
  \bibinfo{journal}{Phys. Rev. D} \bibinfo{volume}{89} (\bibinfo{number}{5})
  (\bibinfo{year}{2014}) \bibinfo{pages}{054029},
  \bibinfo{doi}{\doi{10.1103/PhysRevD.89.054029}}, \eprint{1310.2234}.

\bibtype{Article}%
\bibitem{DElia:2025ybj}
\bibinfo{author}{Massimo D'Elia}, \bibinfo{author}{Lorenzo Maio},
  \bibinfo{author}{Kevin Zambello}, \bibinfo{author}{Giuseppe Zanichelli},
  \bibinfo{title}{{Roberge-Weiss transition for QCD in a magnetic background}},
  \bibinfo{journal}{Phys. Rev. D} \bibinfo{volume}{111} (\bibinfo{number}{9})
  (\bibinfo{year}{2025}) \bibinfo{pages}{094509},
  \bibinfo{doi}{\doi{10.1103/PhysRevD.111.094509}}, \eprint{2502.19294}.

\bibtype{Article}%
\bibitem{Brandt:2023dir}
\bibinfo{author}{B.~B. Brandt}, \bibinfo{author}{F. Cuteri},
  \bibinfo{author}{G. Endr{\H{o}}di}, \bibinfo{author}{G. Mark{\'o}},
  \bibinfo{author}{L. Sandbote}, \bibinfo{author}{A.~D.~M. Valois},
  \bibinfo{title}{{Thermal QCD in a non-uniform magnetic background}},
  \bibinfo{journal}{JHEP} \bibinfo{volume}{11} (\bibinfo{year}{2023})
  \bibinfo{pages}{229}, \bibinfo{doi}{\doi{10.1007/JHEP11(2023)229}},
  \eprint{2305.19029}.

\bibtype{Article}%
\bibitem{Endrodi:2022wym}
\bibinfo{author}{Gergely Endr\H{o}di}, \bibinfo{author}{Gergely Mark\'o},
  \bibinfo{title}{{On electric fields in hot QCD: perturbation theory}},
  \bibinfo{journal}{JHEP} \bibinfo{volume}{12} (\bibinfo{year}{2022})
  \bibinfo{pages}{015}, \bibinfo{doi}{\doi{10.1007/JHEP12(2022)015}},
  \eprint{2208.14306}.

\bibtype{Article}%
\bibitem{Endrodi:2026kmb}
\bibinfo{author}{Gergely Endr{\H{o}}di}, \bibinfo{author}{Gergely Mark{\'o}},
  \bibinfo{author}{Leon Sandbote}, \bibinfo{title}{{On electric fields in hot
  QCD: infrared regularization dependence}}, \bibinfo{journal}{JHEP}
  \bibinfo{volume}{04} (\bibinfo{year}{2026}) \bibinfo{pages}{062},
  \bibinfo{doi}{\doi{10.1007/JHEP04(2026)062}}, \eprint{2601.01478}.

\bibtype{Article}%
\bibitem{Endrodi:2023wwf}
\bibinfo{author}{Gergely Endr\H{o}di}, \bibinfo{author}{Gergely Mark\'o},
  \bibinfo{title}{{QCD phase diagram and equation of state in background
  electric fields}}, \bibinfo{journal}{Phys. Rev. D} \bibinfo{volume}{109}
  (\bibinfo{number}{3}) (\bibinfo{year}{2024}) \bibinfo{pages}{034506},
  \bibinfo{doi}{\doi{10.1103/PhysRevD.109.034506}}, \eprint{2309.07058}.

\bibtype{Article}%
\bibitem{Levkova:2013qda}
\bibinfo{author}{L. Levkova}, \bibinfo{author}{C. DeTar},
  \bibinfo{title}{{Quark-gluon plasma in an external magnetic field}},
  \bibinfo{journal}{Phys. Rev. Lett.} \bibinfo{volume}{112}
  (\bibinfo{number}{1}) (\bibinfo{year}{2014}) \bibinfo{pages}{012002},
  \bibinfo{doi}{\doi{10.1103/PhysRevLett.112.012002}}, \eprint{1309.1142}.

\bibtype{Article}%
\bibitem{Bonati:2013lca}
\bibinfo{author}{Claudio Bonati}, \bibinfo{author}{Massimo D'Elia},
  \bibinfo{author}{Marco Mariti}, \bibinfo{author}{Francesco Negro},
  \bibinfo{author}{Francesco Sanfilippo}, \bibinfo{title}{{Magnetic
  Susceptibility of Strongly Interacting Matter across the Deconfinement
  Transition}}, \bibinfo{journal}{Phys. Rev. Lett.} \bibinfo{volume}{111}
  (\bibinfo{year}{2013}) \bibinfo{pages}{182001},
  \bibinfo{doi}{\doi{10.1103/PhysRevLett.111.182001}}, \eprint{1307.8063}.

\bibtype{Article}%
\bibitem{Bonati:2013vba}
\bibinfo{author}{Claudio Bonati}, \bibinfo{author}{Massimo D'Elia},
  \bibinfo{author}{Marco Mariti}, \bibinfo{author}{Francesco Negro},
  \bibinfo{author}{Francesco Sanfilippo}, \bibinfo{title}{{Magnetic
  susceptibility and equation of state of $N_f=2+1$ QCD with physical quark
  masses}}, \bibinfo{journal}{Phys. Rev. D} \bibinfo{volume}{89}
  (\bibinfo{number}{5}) (\bibinfo{year}{2014}) \bibinfo{pages}{054506},
  \bibinfo{doi}{\doi{10.1103/PhysRevD.89.054506}}, \eprint{1310.8656}.

\bibtype{Article}%
\bibitem{Bali:2013esa}
\bibinfo{author}{G.~S. Bali}, \bibinfo{author}{F. Bruckmann},
  \bibinfo{author}{G. Endr\H{o}di}, \bibinfo{author}{F. Gruber},
  \bibinfo{author}{A. Sch{\"a}fer}, \bibinfo{title}{{Magnetic field-induced
  gluonic (inverse) catalysis and pressure (an)isotropy in QCD}},
  \bibinfo{journal}{JHEP} \bibinfo{volume}{04} (\bibinfo{year}{2013})
  \bibinfo{pages}{130}, \bibinfo{doi}{\doi{10.1007/JHEP04(2013)130}},
  \eprint{1303.1328}.

\bibtype{Article}%
\bibitem{Bali:2013owa}
\bibinfo{author}{G.~S. Bali}, \bibinfo{author}{F. Bruckmann},
  \bibinfo{author}{G. Endr\H{o}di}, \bibinfo{author}{A. Sch{\"a}fer},
  \bibinfo{title}{{Paramagnetic squeezing of QCD matter}},
  \bibinfo{journal}{Phys. Rev. Lett.} \bibinfo{volume}{112}
  (\bibinfo{year}{2014}) \bibinfo{pages}{042301},
  \bibinfo{doi}{\doi{10.1103/PhysRevLett.112.042301}}, \eprint{1311.2559}.

\bibtype{Article}%
\bibitem{Bali:2014kia}
\bibinfo{author}{G.~S. Bali}, \bibinfo{author}{F. Bruckmann},
  \bibinfo{author}{G. Endr\H{o}di}, \bibinfo{author}{S.~D. Katz},
  \bibinfo{author}{A. Sch\"afer}, \bibinfo{title}{{The QCD equation of state in
  background magnetic fields}}, \bibinfo{journal}{JHEP} \bibinfo{volume}{08}
  (\bibinfo{year}{2014}) \bibinfo{pages}{177},
  \bibinfo{doi}{\doi{10.1007/JHEP08(2014)177}}, \eprint{1406.0269}.

\bibtype{Article}%
\bibitem{Bali:2015msa}
\bibinfo{author}{Gunnar Bali}, \bibinfo{author}{Gergely Endr\H{o}di},
  \bibinfo{title}{{Hadronic vacuum polarization and muon g\ensuremath{-}2 from
  magnetic susceptibilities on the lattice}}, \bibinfo{journal}{Phys. Rev. D}
  \bibinfo{volume}{92} (\bibinfo{number}{5}) (\bibinfo{year}{2015})
  \bibinfo{pages}{054506}, \bibinfo{doi}{\doi{10.1103/PhysRevD.92.054506}},
  \eprint{1506.08638}.

\bibtype{Article}%
\bibitem{Bali:2020bcn}
\bibinfo{author}{Gunnar~S. Bali}, \bibinfo{author}{Gergely Endr\H{o}di},
  \bibinfo{author}{Stefano Piemonte}, \bibinfo{title}{{Magnetic susceptibility
  of QCD matter and its decomposition from the lattice}},
  \bibinfo{journal}{JHEP} \bibinfo{volume}{07} (\bibinfo{year}{2020})
  \bibinfo{pages}{183}, \bibinfo{doi}{\doi{10.1007/JHEP07(2020)183}},
  \eprint{2004.08778}.

\bibtype{Article}%
\bibitem{Brandt:2024blb}
\bibinfo{author}{B.~B. Brandt}, \bibinfo{author}{G. Endr\H{o}di},
  \bibinfo{author}{G. Mark\'o}, \bibinfo{author}{A.~D.~M. Valois},
  \bibinfo{title}{{Steady electric currents in magnetized QCD and their use for
  the equation of state}}, \bibinfo{journal}{JHEP} \bibinfo{volume}{07}
  (\bibinfo{year}{2024}) \bibinfo{pages}{027},
  \bibinfo{doi}{\doi{10.1007/JHEP07(2024)027}}, \eprint{2405.06557}.

\bibtype{Article}%
\bibitem{Bali:2012jv}
\bibinfo{author}{G.~S. Bali}, \bibinfo{author}{F. Bruckmann},
  \bibinfo{author}{M. Constantinou}, \bibinfo{author}{M. Costa},
  \bibinfo{author}{G. Endr\H{o}di}, \bibinfo{author}{S.~D. Katz},
  \bibinfo{author}{H. Panagopoulos}, \bibinfo{author}{A. Sch{\"a}fer},
  \bibinfo{title}{{Magnetic susceptibility of QCD at zero and at finite
  temperature from the lattice}}, \bibinfo{journal}{Phys. Rev. D}
  \bibinfo{volume}{86} (\bibinfo{year}{2012}) \bibinfo{pages}{094512},
  \bibinfo{doi}{\doi{10.1103/PhysRevD.86.094512}}, \eprint{1209.6015}.

\bibtype{Article}%
\bibitem{Rohrwild:2007yt}
\bibinfo{author}{J. Rohrwild}, \bibinfo{title}{{Determination of the magnetic
  susceptibility of the quark condensate using radiative heavy meson decays}},
  \bibinfo{journal}{JHEP} \bibinfo{volume}{09} (\bibinfo{year}{2007})
  \bibinfo{pages}{073}, \bibinfo{doi}{\doi{10.1088/1126-6708/2007/09/073}},
  \eprint{0708.1405}.

\bibtype{Article}%
\bibitem{Endrodi:2013cs}
\bibinfo{author}{G. Endr\H{o}di}, \bibinfo{title}{{QCD equation of state at
  nonzero magnetic fields in the Hadron Resonance Gas model}},
  \bibinfo{journal}{JHEP} \bibinfo{volume}{04} (\bibinfo{year}{2013})
  \bibinfo{pages}{023}, \bibinfo{doi}{\doi{10.1007/JHEP04(2013)023}},
  \eprint{1301.1307}.

\bibtype{Article}%
\bibitem{Ferrer:2010wz}
\bibinfo{author}{Efrain~J. Ferrer}, \bibinfo{author}{Vivian de~la Incera},
  \bibinfo{author}{Jason~P. Keith}, \bibinfo{author}{Israel Portillo},
  \bibinfo{author}{Paul~L. Springsteen}, \bibinfo{title}{{Equation of State of
  a Dense and Magnetized Fermion System}}, \bibinfo{journal}{Phys. Rev. C}
  \bibinfo{volume}{82} (\bibinfo{year}{2010}) \bibinfo{pages}{065802},
  \bibinfo{doi}{\doi{10.1103/PhysRevC.82.065802}}, \eprint{1009.3521}.

\bibtype{Article}%
\bibitem{Ding:2025jfz}
\bibinfo{author}{Heng-Tong Ding}, \bibinfo{author}{Jin-Biao Gu},
  \bibinfo{author}{Arpith Kumar}, \bibinfo{author}{Sheng-Tai Li},
  \bibinfo{title}{{Second order fluctuations of conserved charges in external
  magnetic fields}}, \bibinfo{journal}{Phys. Rev. D} \bibinfo{volume}{111}
  (\bibinfo{number}{11}) (\bibinfo{year}{2025}) \bibinfo{pages}{114522},
  \bibinfo{doi}{\doi{10.1103/tgm5-jvyf}}, \eprint{2503.18467}.

\bibtype{Article}%
\bibitem{Kharzeev:2013ffa}
\bibinfo{author}{Dmitri~E. Kharzeev}, \bibinfo{title}{{The Chiral Magnetic
  Effect and Anomaly-Induced Transport}}, \bibinfo{journal}{Prog. Part. Nucl.
  Phys.} \bibinfo{volume}{75} (\bibinfo{year}{2014}) \bibinfo{pages}{133--151},
  \bibinfo{doi}{\doi{10.1016/j.ppnp.2014.01.002}}, \eprint{1312.3348}.

\bibtype{Article}%
\bibitem{Pang:2016yuh}
\bibinfo{author}{Long-Gang Pang}, \bibinfo{author}{Gergely Endr\H{o}di},
  \bibinfo{author}{Hannah Petersen}, \bibinfo{title}{{Magnetic-field-induced
  squeezing effect at energies available at the BNL Relativistic Heavy Ion
  Collider and at the CERN Large Hadron Collider}}, \bibinfo{journal}{Phys.
  Rev. C} \bibinfo{volume}{93} (\bibinfo{number}{4}) (\bibinfo{year}{2016})
  \bibinfo{pages}{044919}, \bibinfo{doi}{\doi{10.1103/PhysRevC.93.044919}},
  \eprint{1602.06176}.

\bibtype{Article}%
\bibitem{Ding:2021cwv}
\bibinfo{author}{H.~T. Ding}, \bibinfo{author}{S.~T. Li}, \bibinfo{author}{Q.
  Shi}, \bibinfo{author}{X.~D. Wang}, \bibinfo{title}{{Fluctuations and
  correlations of net baryon number, electric charge and strangeness in a
  background magnetic field}}, \bibinfo{journal}{Eur. Phys. J. A}
  \bibinfo{volume}{57} (\bibinfo{number}{6}) (\bibinfo{year}{2021})
  \bibinfo{pages}{202}, \bibinfo{doi}{\doi{10.1140/epja/s10050-021-00519-3}},
  \eprint{2104.06843}.

\bibtype{Article}%
\bibitem{Ding:2023bft}
\bibinfo{author}{Heng-Tong Ding}, \bibinfo{author}{Jin-Biao Gu},
  \bibinfo{author}{Arpith Kumar}, \bibinfo{author}{Sheng-Tai Li},
  \bibinfo{author}{Jun-Hong Liu}, \bibinfo{title}{{Baryon Electric Charge
  Correlation as a Magnetometer of QCD}}, \bibinfo{journal}{Phys. Rev. Lett.}
  \bibinfo{volume}{132} (\bibinfo{number}{20}) (\bibinfo{year}{2024})
  \bibinfo{pages}{201903}, \bibinfo{doi}{\doi{10.1103/PhysRevLett.132.201903}},
  \eprint{2312.08860}.

\bibtype{Article}%
\bibitem{Vovchenko:2024wbg}
\bibinfo{author}{Volodymyr Vovchenko}, \bibinfo{title}{{Magnetic field effect
  on hadron yield ratios and fluctuations in a hadron resonance gas}},
  \bibinfo{journal}{Phys. Rev. C} \bibinfo{volume}{110} (\bibinfo{number}{3})
  (\bibinfo{year}{2024}) \bibinfo{pages}{034914},
  \bibinfo{doi}{\doi{10.1103/PhysRevC.110.034914}}, \eprint{2405.16306}.

\bibtype{Article}%
\bibitem{Marczenko:2024kko}
\bibinfo{author}{Micha{\l} Marczenko}, \bibinfo{author}{Micha{\l}
  Szyma{\'n}ski}, \bibinfo{author}{Pok~Man Lo}, \bibinfo{author}{Bithika
  Karmakar}, \bibinfo{author}{Pasi Huovinen}, \bibinfo{author}{Chihiro Sasaki},
  \bibinfo{author}{Krzysztof Redlich}, \bibinfo{title}{{Magnetic effects in the
  hadron resonance gas}}, \bibinfo{journal}{Phys. Rev. C} \bibinfo{volume}{110}
  (\bibinfo{number}{6}) (\bibinfo{year}{2024}) \bibinfo{pages}{065203},
  \bibinfo{doi}{\doi{10.1103/PhysRevC.110.065203}}, \eprint{2405.15745}.

\bibtype{Article}%
\bibitem{Son:2004tq}
\bibinfo{author}{D.~T. Son}, \bibinfo{author}{Ariel~R. Zhitnitsky},
  \bibinfo{title}{{Quantum anomalies in dense matter}}, \bibinfo{journal}{Phys.
  Rev. D} \bibinfo{volume}{70} (\bibinfo{year}{2004}) \bibinfo{pages}{074018},
  \bibinfo{doi}{\doi{10.1103/PhysRevD.70.074018}}, \eprint{hep-ph/0405216}.

\bibtype{Article}%
\bibitem{Metlitski:2005pr}
\bibinfo{author}{Max~A. Metlitski}, \bibinfo{author}{Ariel~R. Zhitnitsky},
  \bibinfo{title}{{Anomalous axion interactions and topological currents in
  dense matter}}, \bibinfo{journal}{Phys. Rev. D} \bibinfo{volume}{72}
  (\bibinfo{year}{2005}) \bibinfo{pages}{045011},
  \bibinfo{doi}{\doi{10.1103/PhysRevD.72.045011}}, \eprint{hep-ph/0505072}.

\bibtype{Article}%
\bibitem{Fukushima:2008xe}
\bibinfo{author}{Kenji Fukushima}, \bibinfo{author}{Dmitri~E. Kharzeev},
  \bibinfo{author}{Harmen~J. Warringa}, \bibinfo{title}{{The Chiral Magnetic
  Effect}}, \bibinfo{journal}{Phys. Rev. D} \bibinfo{volume}{78}
  (\bibinfo{year}{2008}) \bibinfo{pages}{074033},
  \bibinfo{doi}{\doi{10.1103/PhysRevD.78.074033}}, \eprint{0808.3382}.

\bibtype{Article}%
\bibitem{Brandt:2023wgf}
\bibinfo{author}{Bastian~B. Brandt}, \bibinfo{author}{Gergely Endr\H{o}di},
  \bibinfo{author}{Eduardo Garnacho-Velasco}, \bibinfo{author}{Gergely
  Mark\'o}, \bibinfo{title}{{The chiral separation effect from lattice QCD at
  the physical point}}, \bibinfo{journal}{JHEP} \bibinfo{volume}{02}
  (\bibinfo{year}{2024}) \bibinfo{pages}{142},
  \bibinfo{doi}{\doi{10.1007/JHEP02(2024)142}}, \eprint{2312.02945}.

\bibtype{Article}%
\bibitem{Puhr:2016kzp}
\bibinfo{author}{M. Puhr}, \bibinfo{author}{P.~V. Buividovich},
  \bibinfo{title}{{Numerical Study of Nonperturbative Corrections to the Chiral
  Separation Effect in Quenched Finite-Density QCD}}, \bibinfo{journal}{Phys.
  Rev. Lett.} \bibinfo{volume}{118} (\bibinfo{number}{19})
  (\bibinfo{year}{2017}) \bibinfo{pages}{192003},
  \bibinfo{doi}{\doi{10.1103/PhysRevLett.118.192003}}, \eprint{1611.07263}.

\bibtype{Article}%
\bibitem{Buividovich:2020gnl}
\bibinfo{author}{P.~V. Buividovich}, \bibinfo{author}{D. Smith},
  \bibinfo{author}{L. von Smekal}, \bibinfo{title}{{Numerical study of the
  chiral separation effect in two-color QCD at finite density}},
  \bibinfo{journal}{Phys. Rev. D} \bibinfo{volume}{104} (\bibinfo{number}{1})
  (\bibinfo{year}{2021}) \bibinfo{pages}{014511},
  \bibinfo{doi}{\doi{10.1103/PhysRevD.104.014511}}, \eprint{2012.05184}.

\bibtype{Article}%
\bibitem{Brandt:2024wlw}
\bibinfo{author}{Bastian~B. Brandt}, \bibinfo{author}{Gergely Endr\H{o}di},
  \bibinfo{author}{Eduardo Garnacho-Velasco}, \bibinfo{author}{Gergely
  Mark\'o}, \bibinfo{title}{{On the absence of the chiral magnetic effect in
  equilibrium QCD}}, \bibinfo{journal}{JHEP} \bibinfo{volume}{09}
  (\bibinfo{year}{2024}) \bibinfo{pages}{092},
  \bibinfo{doi}{\doi{10.1007/JHEP09(2024)092}}, \eprint{2405.09484}.

\bibtype{Article}%
\bibitem{Yamamoto:2011gk}
\bibinfo{author}{Arata Yamamoto}, \bibinfo{title}{{Chiral magnetic effect in
  lattice QCD with a chiral chemical potential}}, \bibinfo{journal}{Phys. Rev.
  Lett.} \bibinfo{volume}{107} (\bibinfo{year}{2011}) \bibinfo{pages}{031601},
  \bibinfo{doi}{\doi{10.1103/PhysRevLett.107.031601}}, \eprint{1105.0385}.

\bibtype{Article}%
\bibitem{Yamamoto:2011ks}
\bibinfo{author}{Arata Yamamoto}, \bibinfo{title}{{Lattice study of the chiral
  magnetic effect in a chirally imbalanced matter}}, \bibinfo{journal}{Phys.
  Rev. D} \bibinfo{volume}{84} (\bibinfo{year}{2011}) \bibinfo{pages}{114504},
  \bibinfo{doi}{\doi{10.1103/PhysRevD.84.114504}}, \eprint{1111.4681}.

\bibtype{Article}%
\bibitem{Adhikari:2018cea}
\bibinfo{author}{Prabal Adhikari}, \bibinfo{author}{Jens~O. Andersen},
  \bibinfo{author}{Patrick Kneschke}, \bibinfo{title}{{Pion condensation and
  phase diagram in the Polyakov-loop quark-meson model}},
  \bibinfo{journal}{Phys. Rev. D} \bibinfo{volume}{98} (\bibinfo{number}{7})
  (\bibinfo{year}{2018}) \bibinfo{pages}{074016},
  \bibinfo{doi}{\doi{10.1103/PhysRevD.98.074016}}, \eprint{1805.08599}.

\bibtype{Article}%
\bibitem{Cotter:2012mb}
\bibinfo{author}{Seamus Cotter}, \bibinfo{author}{Pietro Giudice},
  \bibinfo{author}{Simon Hands}, \bibinfo{author}{Jon-Ivar Skullerud},
  \bibinfo{title}{{Towards the phase diagram of dense two-color matter}},
  \bibinfo{journal}{Phys. Rev. D} \bibinfo{volume}{87} (\bibinfo{number}{3})
  (\bibinfo{year}{2013}) \bibinfo{pages}{034507},
  \bibinfo{doi}{\doi{10.1103/PhysRevD.87.034507}}, \eprint{1210.4496}.

\bibtype{Article}%
\bibitem{Brandt:2019hel}
\bibinfo{author}{B.~B. Brandt}, \bibinfo{author}{F. Cuteri},
  \bibinfo{author}{G. Endr{\H{o}}di}, \bibinfo{author}{S. Schmalzbauer},
  \bibinfo{title}{{The Dirac spectrum and the BEC-BCS crossover in QCD at
  nonzero isospin asymmetry}}, \bibinfo{journal}{Particles} \bibinfo{volume}{3}
  (\bibinfo{number}{1}) (\bibinfo{year}{2020}) \bibinfo{pages}{80--86},
  \bibinfo{doi}{\doi{10.3390/particles3010007}}, \eprint{1912.07451}.

\bibtype{Article}%
\bibitem{Kanazawa:2013crb}
\bibinfo{author}{Takuya Kanazawa}, \bibinfo{author}{Tilo Wettig},
  \bibinfo{author}{Naoki Yamamoto}, \bibinfo{title}{{Banks-Casher-type relation
  for the BCS gap at high density}}, \bibinfo{journal}{Eur. Phys. J. A}
  \bibinfo{volume}{49} (\bibinfo{year}{2013}) \bibinfo{pages}{88},
  \bibinfo{doi}{\doi{10.1140/epja/i2013-13088-5}}, \eprint{1211.5332}.

\bibtype{Article}%
\bibitem{Cohen:2015soa}
\bibinfo{author}{Thomas~D. Cohen}, \bibinfo{author}{Srimoyee Sen},
  \bibinfo{title}{{Deconfinement Transition at High Isospin Chemical Potential
  and Low Temperature}}, \bibinfo{journal}{Nucl. Phys. A} \bibinfo{volume}{942}
  (\bibinfo{year}{2015}) \bibinfo{pages}{39--53},
  \bibinfo{doi}{\doi{10.1016/j.nuclphysa.2015.07.018}}, \eprint{1503.00006}.

\bibtype{Article}%
\bibitem{Kogut:2004zg}
\bibinfo{author}{J.~B. Kogut}, \bibinfo{author}{D.~K. Sinclair},
  \bibinfo{title}{{The Finite temperature transition for 2-flavor lattice QCD
  at finite isospin density}}, \bibinfo{journal}{Phys. Rev. D}
  \bibinfo{volume}{70} (\bibinfo{year}{2004}) \bibinfo{pages}{094501},
  \bibinfo{doi}{\doi{10.1103/PhysRevD.70.094501}}, \eprint{hep-lat/0407027}.

\bibtype{Article}%
\bibitem{Astrakhantsev:2019zkr}
\bibinfo{author}{Nikita Astrakhantsev}, \bibinfo{author}{V.~V. Braguta},
  \bibinfo{author}{Massimo D'Elia}, \bibinfo{author}{A.~Yu. Kotov},
  \bibinfo{author}{A.~A. Nikolaev}, \bibinfo{author}{Francesco Sanfilippo},
  \bibinfo{title}{{Lattice study of the electromagnetic conductivity of the
  quark-gluon plasma in an external magnetic field}}, \bibinfo{journal}{Phys.
  Rev. D} \bibinfo{volume}{102} (\bibinfo{number}{5}) (\bibinfo{year}{2020})
  \bibinfo{pages}{054516}, \bibinfo{doi}{\doi{10.1103/PhysRevD.102.054516}},
  \eprint{1910.08516}.

\bibtype{Article}%
\bibitem{Brandt:2025now}
\bibinfo{author}{Bastian~B. Brandt}, \bibinfo{author}{Gergely Endr\H{o}di},
  \bibinfo{author}{Eduardo Garnacho~Velasco}, \bibinfo{author}{Gergely
  Mark\'o}, \bibinfo{author}{A.~Dean~M. Valois},
  \bibinfo{title}{{Out-of-equilibrium Chiral Magnetic Effect via Kubo
  formulas}}, \bibinfo{journal}{PoS} \bibinfo{volume}{LATTICE2024}
  (\bibinfo{year}{2025}) \bibinfo{pages}{196},
  \bibinfo{doi}{\doi{10.22323/1.466.0196}}, \eprint{2502.01155}.

\bibtype{Article}%
\bibitem{Schwarz:2009ii}
\bibinfo{author}{Dominik~J. Schwarz}, \bibinfo{author}{Maik Stuke},
  \bibinfo{title}{{Lepton asymmetry and the cosmic QCD transition}},
  \bibinfo{journal}{JCAP} \bibinfo{volume}{11} (\bibinfo{year}{2009})
  \bibinfo{pages}{025}, \bibinfo{doi}{\doi{10.1088/1475-7516/2009/11/025}},
  \bibinfo{note}{[Erratum: JCAP 10, E01 (2010)]}, \eprint{0906.3434}.

\bibtype{Article}%
\bibitem{Middeldorf-Wygas:2020glx}
\bibinfo{author}{Mandy~M. Middeldorf-Wygas}, \bibinfo{author}{Isabel~M.
  Oldengott}, \bibinfo{author}{Dietrich B{\"o}deker},
  \bibinfo{author}{Dominik~J. Schwarz}, \bibinfo{title}{{Cosmic QCD transition
  for large lepton flavor asymmetries}}, \bibinfo{journal}{Phys. Rev. D}
  \bibinfo{volume}{105} (\bibinfo{number}{12}) (\bibinfo{year}{2022})
  \bibinfo{pages}{123533}, \bibinfo{doi}{\doi{10.1103/PhysRevD.105.123533}},
  \eprint{2009.00036}.

\bibtype{Article}%
\bibitem{Andersen:2018nzq}
\bibinfo{author}{Jens~O. Andersen}, \bibinfo{author}{Patrick Kneschke},
  \bibinfo{title}{{Bose-Einstein condensation and pion stars}}
  (\bibinfo{year}{2018}), \eprint{1807.08951}.

\bibtype{Article}%
\bibitem{Andersen:2022aig}
\bibinfo{author}{Jens~O. Andersen}, \bibinfo{author}{Martin~Kj{\o}llesdal
  Johnsrud}, \bibinfo{title}{{Phases of QCD at nonzero isospin and strangeness
  chemical potentials with application to pion stars}}  (\bibinfo{year}{2022}),
  \eprint{2206.04291}.

\bibtype{Article}%
\bibitem{Stashko:2023gnn}
\bibinfo{author}{O.~S. Stashko}, \bibinfo{author}{O.~V. Savchuk},
  \bibinfo{author}{L.~M. Satarov}, \bibinfo{author}{I.~N. Mishustin},
  \bibinfo{author}{M.~I. Gorenstein}, \bibinfo{author}{V.~I. Zhdanov},
  \bibinfo{title}{{Pion stars embedded in neutrino clouds}},
  \bibinfo{journal}{Phys. Rev. D} \bibinfo{volume}{107} (\bibinfo{number}{11})
  (\bibinfo{year}{2023}) \bibinfo{pages}{114025},
  \bibinfo{doi}{\doi{10.1103/PhysRevD.107.114025}}, \eprint{2303.06190}.

\bibtype{Article}%
\bibitem{Chen:2024cxh}
\bibinfo{author}{Yidian Chen}, \bibinfo{author}{Mingshan Ding},
  \bibinfo{author}{Danning Li}, \bibinfo{author}{Kazem Bitaghsir~Fadafan},
  \bibinfo{author}{Mei Huang}, \bibinfo{title}{{Pion condensation and pion star
  from holographic QCD}}, \bibinfo{journal}{Phys. Rev. D} \bibinfo{volume}{111}
  (\bibinfo{number}{12}) (\bibinfo{year}{2025}) \bibinfo{pages}{126010},
  \bibinfo{doi}{\doi{10.1103/94zt-q2dw}}, \eprint{2408.17080}.

\bibtype{Article}%
\bibitem{Cohen:2003ut}
\bibinfo{author}{Thomas~D. Cohen}, \bibinfo{title}{{QCD inequalities for the
  nucleon mass and the free energy of baryonic matter}},
  \bibinfo{journal}{Phys. Rev. Lett.} \bibinfo{volume}{91}
  (\bibinfo{year}{2003}) \bibinfo{pages}{032002},
  \bibinfo{doi}{\doi{10.1103/PhysRevLett.91.032002}}, \eprint{hep-ph/0304024}.

\bibtype{Article}%
\bibitem{Moore:2023glb}
\bibinfo{author}{Guy~D. Moore}, \bibinfo{author}{Tyler Gorda},
  \bibinfo{title}{{Bounding the QCD Equation of State with the Lattice}},
  \bibinfo{journal}{JHEP} \bibinfo{volume}{12} (\bibinfo{year}{2023})
  \bibinfo{pages}{133}, \bibinfo{doi}{\doi{10.1007/JHEP12(2023)133}},
  \eprint{2309.15149}.

\bibtype{Article}%
\bibitem{Navarrete:2024zgz}
\bibinfo{author}{Pablo Navarrete}, \bibinfo{author}{Risto Paatelainen},
  \bibinfo{author}{Kaapo Sepp{\"a}nen}, \bibinfo{title}{{Perturbative QCD meets
  phase quenching: The pressure of cold quark matter}}, \bibinfo{journal}{Phys.
  Rev. D} \bibinfo{volume}{110} (\bibinfo{number}{9}) (\bibinfo{year}{2024})
  \bibinfo{pages}{094033}, \bibinfo{doi}{\doi{10.1103/PhysRevD.110.094033}},
  \eprint{2403.02180}.

\bibtype{Article}%
\bibitem{Lee:2004hc}
\bibinfo{author}{Dean Lee}, \bibinfo{title}{{Pressure inequalities for nuclear
  and neutron matter}}, \bibinfo{journal}{Phys. Rev. C} \bibinfo{volume}{71}
  (\bibinfo{year}{2005}) \bibinfo{pages}{044001},
  \bibinfo{doi}{\doi{10.1103/PhysRevC.71.044001}}, \eprint{nucl-th/0407101}.

\bibtype{Article}%
\bibitem{Fujimoto:2023unl}
\bibinfo{author}{Yuki Fujimoto}, \bibinfo{author}{Sanjay Reddy},
  \bibinfo{title}{{Bounds on the equation of state from QCD inequalities and
  lattice QCD}}, \bibinfo{journal}{Phys. Rev. D} \bibinfo{volume}{109}
  (\bibinfo{number}{1}) (\bibinfo{year}{2024}) \bibinfo{pages}{014020},
  \bibinfo{doi}{\doi{10.1103/PhysRevD.109.014020}}, \eprint{2310.09427}.

\bibtype{Article}%
\bibitem{Fujimoto:2024pcd}
\bibinfo{author}{Yuki Fujimoto}, \bibinfo{title}{{Interplay between the
  weak-coupling results and the lattice data in dense QCD}}
  (\bibinfo{year}{2024}), \eprint{2408.12514}.

\bibtype{Article}%
\bibitem{Gorda:2025cwu}
\bibinfo{author}{Tyler Gorda}, \bibinfo{author}{Pablo Navarrete},
  \bibinfo{author}{Risto Paatelainen}, \bibinfo{author}{Leon Sandbote},
  \bibinfo{author}{Kaapo Sepp{\"a}nen}, \bibinfo{title}{{A new approach to
  determine the thermodynamics of deconfined matter to high accuracy}}
  (\bibinfo{year}{2025}), \eprint{2511.09627}.

\bibtype{Article}%
\bibitem{Cohen:2025ahp}
\bibinfo{author}{Thomas~D. Cohen}, \bibinfo{title}{{A method to obtain bounds
  on the equation of state of cold nuclear matter from imaginary chemical
  potentials}}  (\bibinfo{year}{2025}), \eprint{2510.07124}.

\bibtype{Article}%
\bibitem{Andersen:2014xxa}
\bibinfo{author}{Jens~O. Andersen}, \bibinfo{author}{William~R. Naylor},
  \bibinfo{author}{Anders Tranberg}, \bibinfo{title}{{Phase diagram of QCD in a
  magnetic field: A review}}, \bibinfo{journal}{Rev. Mod. Phys.}
  \bibinfo{volume}{88} (\bibinfo{year}{2016}) \bibinfo{pages}{025001},
  \bibinfo{doi}{\doi{10.1103/RevModPhys.88.025001}}, \eprint{1411.7176}.

\bibtype{Article}%
\bibitem{Folkestad:2018psc}
\bibinfo{author}{Aasmund Folkestad}, \bibinfo{author}{Jens~O. Andersen},
  \bibinfo{title}{{Thermodynamics and phase diagrams of Polyakov-loop extended
  chiral models}}, \bibinfo{journal}{Phys. Rev. D} \bibinfo{volume}{99}
  (\bibinfo{number}{5}) (\bibinfo{year}{2019}) \bibinfo{pages}{054006},
  \bibinfo{doi}{\doi{10.1103/PhysRevD.99.054006}}, \eprint{1810.10573}.

\bibtype{Article}%
\bibitem{Brandt:2025tkg}
\bibinfo{author}{Bastian~B. Brandt}, \bibinfo{author}{Volodymyr Chelnokov},
  \bibinfo{author}{Gergely Endr\H{o}di}, \bibinfo{author}{Gergely Marko},
  \bibinfo{author}{Daniel Scheid}, \bibinfo{author}{Lorenz von Smekal},
  \bibinfo{title}{{Renormalization group invariant mean-field model for QCD at
  finite isospin density}}, \bibinfo{journal}{Phys. Rev. D}
  \bibinfo{volume}{112} (\bibinfo{number}{5}) (\bibinfo{year}{2025})
  \bibinfo{pages}{054038}, \bibinfo{doi}{\doi{10.1103/fryz-f3vw}},
  \eprint{2502.04025}.

\end{thebibliography*}

\end{document}